\documentclass[11pt]{article}

\usepackage{color}		
\usepackage[dvips]{graphicx}	
\usepackage{amsbsy}		
\usepackage{amsfonts}		
\usepackage{amsmath}		
\usepackage{amssymb}		
\usepackage{upgreek}		
\usepackage{wasysym}
\usepackage{multirow}
\usepackage{mciteplus}
\usepackage{psfrag}
\usepackage{url}
\usepackage{rotate}
\usepackage{cite}
\usepackage{soul}
\usepackage{afterpage}

\include{paperdef}
\graphicspath{{figs/}}

\newcommand*{\figref}[1]{Fig.~\ref{#1}}

\newcommand{\HSM}{H_{\SM}}
\newcommand{\HB}{{\tt HiggsBounds}}
\newcommand{\FH}{{\tt FeynHiggs}}
\newcommand{\SI}{{\tt SuperIso}}
\newcommand{\pMSSM}{pMSSM--7}
\newcommand{\simMH}{126}
\newcommand{\MHexp}{125.7}
\newcommand{\btn}{\ensuremath{\br(B_u \to \tau \nu_\tau)}}
\newcommand{\bmm}{\ensuremath{\br(B_s \to \mu^+\mu^-)}}
\newcommand{\bsg}{\ensuremath{\br(B \to X_s \ga)}}
\newcommand{\gmt}{\ensuremath{(g-2)_\mu}}

\begin{document}
\hfill {\tt DESY 12-200}

\hfill {\tt BONN-TH-2012-028}

\def\thefootnote{\fnsymbol{footnote}}

\begin{center}
\Large\bf\boldmath
MSSM Interpretations of the LHC Discovery:\\[.5em]
 Light or Heavy Higgs?
\unboldmath
\end{center}
\vspace{0.4cm}
\begin{center}
P.~Bechtle$^{1,}$\footnote{Electronic address: bechtle@physik.uni-bonn.de},
S.~Heinemeyer$^{2,}$\footnote{Electronic address: Sven.Heinemeyer@cern.ch},
O.~St{\aa}l$^{3,}$\footnote{Electronic address: oscar.stal@fysik.su.se},
T.~Stefaniak$^{1,}$\footnote{Electronic address: tim@th.physik.uni-bonn.de},
G.~Weiglein$^{4,}$\footnote{Electronic address: Georg.Weiglein@desy.de},
L.~Zeune$^{4,}$\footnote{Electronic address: lisa.zeune@desy.de} \\[0.4cm] 
\vspace{0.4cm}
{\sl$^1$Physikalisches Institut der Universit\"at Bonn\\
 Nu{\ss}allee 12, D-53115 Bonn, Germany}\\[0.2cm]
{\sl$^2$Instituto de F\'isica de Cantabria (CSIC-UC), Santander,  Spain
}\\[0.2cm]
 {\sl $^3$ The Oskar Klein Centre, Department of Physics\\
  Stockholm University, SE-106 91 Stockholm, Sweden}\\[0.2cm]
{\sl $^4$ Deutsches Elektronen-Synchrotron DESY\\
 Notkestra{\ss}e 85, D-22607 Hamburg, Germany}\\[0.2cm]
\end{center}
\vspace{0.2cm}

\renewcommand{\thefootnote}{\arabic{footnote}}
\setcounter{footnote}{0}

\begin{abstract}
A Higgs-like particle with a mass of about $\simMH \gev$ has been
discovered at the LHC. Within the experimental uncertainties, the
measured properties of this new state are compatible with those of the
Higgs boson in the Standard Model (SM). 
While not statistically significant at present, the results show
some interesting patterns of deviations from the SM predictions, 
in particular a higher rate in the $\ga\ga$ decay mode observed
by ATLAS and CMS, and a somewhat smaller rate in the $\tau^+\tau^-$
mode. The LHC
discovery is also compatible with the predictions of the Higgs sector of
the Minimal Supersymmetric Standard Model (MSSM), interpreting the new
state as either the light or the heavy $\cp$-even MSSM Higgs boson.  
Within the framework of the MSSM with seven free parameters (\pMSSM), we
fit the various rates of cross section times branching ratio as measured
by the LHC and Tevatron experiments under the hypotheses of either the light
or the heavy $\cp$-even Higgs boson being the new state around $\simMH \gev$,
with and without the inclusion of further low-energy observables.
We find an overall good quality of the fits, with the best fit
points exhibiting an enhancement of the $\ga\ga$ rate, as well as a
small suppression of the $b \bar b$ and 
$\tau^+\tau^-$ channels with respect to their SM expectations, depending
on the details of the fit.  
For the fits including the whole data set the 
light $\cp$-even Higgs interpretation in the MSSM results in a higher
relative fit probability than the SM fit. On the other hand, we find
that the present data also permit the more exotic interpretation in
terms of the heavy $\cp$-even MSSM Higgs, which could give rise to
experimental signatures of additional Higgs states in the near future.
\end{abstract}
\newpage


\section{Introduction}

One of the most important goals of the Large Hadron Collider (LHC) is
to explore the origin of electroweak symmetry breaking (EWSB).
The spectacular discovery of a Higgs-like particle 
with a mass around $\MHhat\simeq \simMH \gev$, which has been announced
by ATLAS \cite{ATLASdiscovery} and CMS~\cite{CMSdiscovery}, marks a
milestone of an effort that has been ongoing for almost half a century
and opens up a new era of particle physics.  
Both ATLAS and CMS reported a clear excess in the two photon channel, as
well as in the $ZZ^{(*)}$ channel. The discovery is further 
corroborated, though not with high significance, by the
$WW^{(*)}$ channel and by the final Tevatron results~\cite{TCB:2012zzl}.
The combined sensitivity
in each of the LHC experiments reaches more than $5\,\si$. The
observed rate in the $\ga\ga$ channel turns out to be considerably above
the expectation for a Standard Model (SM) Higgs both for
ATLAS~\cite{Atlasnote1291} and CMS~\cite{CMSHig1215}, whereas
the $b\bar b$~and the $\tau^+\tau^-$~channels appear to be
somewhat low in the LHC measurements~\cite{Atlasnote1293,CMSHig1220}.
While those possible deviations from the SM prediction are not
statistically significant at present, so that within the uncertainties
the results are compatible with the SM, if confirmed in the future the
observed patterns could be a first indication of a non-SM nature of the
new state.

Among the most studied candidates for EWSB in the literature 
are the Higgs mechanism within the SM and
within the Minimal Supersymmetric Standard Model (MSSM). Contrary to
the SM, two Higgs doublets are required in the MSSM, resulting in five
physical Higgs boson degrees of freedom. 
Without
explicit $\cp$-violation in the soft supersymmetry-breaking terms
these are the light and heavy $\cp$-even Higgs bosons, $h$ and $H$, the
$\cp$-odd Higgs boson, $A$, and the charged Higgs boson, $H^\pm$.
The Higgs sector of the MSSM can be specified at lowest
order in terms of the $Z$~boson mass, $\MZ$, the $\cp$-odd Higgs
boson mass, $\MA$, and $\tb \equiv v_2/v_1$, the ratio of the
two Higgs vacuum expectation values. 
The masses of the $\cp$-even neutral Higgs bosons and the
charged Higgs boson can be predicted,
including higher-order corrections, in terms of the other MSSM
parameters, see, e.g., 
\cite{Djouadi:2005gj,*Heinemeyer:2004ms,Heinemeyer:2004gx} for reviews.

It was shown that in particular the interpretation of the new
state as the light $\cp$-even Higgs boson of the MSSM is a viable
possibility (called the ``light Higgs case'' in the following). The
implications and phenomenology of this scenario has been studied in a
series of
papers~\cite{Heinemeyer:2011aa,Benbrik:2012rm,Baer:2011ab,*Arbey:2011ab,*Kadastik:2011aa,*Cao:2011sn,*Brummer:2012zc,*Carmi:2012yp,*Ellis:2012aa,*Desai:2012qy,*Cheng:2012np,*Asano:2012rd,*Cao:2012im,*Choudhury:2012tc,*Giardino:2012ww,*Ajaib:2012vc,*Brummer:2012ns,*Evans:2012uf,*Fowlie:2012im,*Buckley:2012em,*Akula:2012kk,*Cao:2012yn,*Hirsch:2012ti,
*Buchmueller:2012hv, *Howe:2012xe, *Wymant:2012zp, *Kang:2012bv,
*Kitahara:2012pb, *Heng:2012at, Carena:2012gp,*Carena:2011aa,Haisch:2012re}. 
On the other hand, it was also
pointed out that the heavy $\cp$-even Higgs boson can have a mass around
$\simMH \gev$~\cite{Heinemeyer:2011aa,Benbrik:2012rm} 
(called the ``heavy Higgs case'')
while maintaining (within the uncertainties) a SM-like behaviour. All 
five MSSM Higgs bosons in this scenario 
would be rather light, and it
would in particular imply the existence of another light Higgs with a
mass below $\simMH \gev$ and suppressed couplings to $W$ and $Z$ bosons.
For a recent discussion of the phenomenology of such a scenario, see
also \cite{Drees:2012fb}.

The question arises whether the MSSM (or another model beyond the SM)
can give a prediction of the production cross sections and decay widths
of the observed Higgs-like state that yields a better 
description of the data than the one
provided by the SM. While at the current level of accuracy no clear
deviation from the SM can be claimed, the situation could change
once the whole dataset of $\sim 20\;\ifb$ to be collected in 2012
is incorporated in the analyses.
The main aim of this paper is to investigate in how much the MSSM can
improve the theoretical description of the current experimental data, and
potentially which parts of the parameter space of the MSSM are favoured
by the current experimental data from the various Higgs search channels.

Because of the large number of free parameters, the MSSM Higgs search
results at 
LEP~\cite{Schael:2006cr}, the Tevatron~\cite{Benjamin:2010xb,*Aaltonen:2012zh}
and the LHC~\cite{ATLASMSSMHiggs,*CMSMSSMHiggs} have been interpreted in
certain benchmark scenarios~\cite{Carena:1999xa,Carena:2002qg,Carena:2005ek}
(of which the $\mhmax$ scenario has been most widely used), where the
MSSM parameters entering via higher-order contributions are set to
specific fixed benchmark values.
However, in order to investigate the potentially favoured regions in the
MSSM parameter space a scan over the relevant SUSY parameters has to be
performed. A complete scan over the in principle more than hundred
free parameters of the MSSM parameter space is neither technically
feasible nor does the available experimental information provide
sufficient sensitivity to simultaneously constrain a large number of
parameters. One therefore needs to focus on a certain subset of
parameters. The most ambitious scans performed up to now were
done~\cite{Arbey:2012na,*Arbey:2012dq} for the phenomenological MSSM
with 19 free parameters (pMSSM--19, see~\cite{AbdusSalam:2011fc}
for details). However, on the one hand it is difficult to sample such a
multi-dimensional parameter space sufficiently well, on the other hand 
it is well known that several of the parameters of the pMSSM--19 hardly
affect Higgs phenomenology. We therefore focus in this paper on a
smaller set of parameters, namely the phenomenological MSSM with the
seven free parameters that we regard as most relevant for the
phenomenology of Higgs and flavour physics (\pMSSM, see below for
details on these parameters). This seven-dimensional parameter space, 
which as we will demonstrate captures most of the allowed Higgs phenomenology
of the MSSM, can be sampled quite well with \order{10^7} scan points.
We will comment in our analysis on the potential impact of varying
further MSSM parameters.

In our analysis we perform fits in the MSSM both for the interpretation
of the LHC signal in terms of the light and the heavy $\cp$-even Higgs
of the MSSM and we compare the fit results with the SM case. We
take into account all available individual search
channels at ATLAS and CMS at $7$~and $8 \tev$ center-of-mass energy that
have been published by the end of July 2012, including also the two
combined Higgs mass values. We furthermore include the
final results of the Tevatron, corresponding to three additional
channels. Besides the direct Higgs search channels we take into account
the Higgs exclusion bounds, limits on the SUSY masses, as well as the
most relevant set of low-energy observables, 
$\br(b \to s \ga)$, $\br(B_s \to \mu^+\mu^-)$, 
$\br(B_u \to \tau \nu_\tau)$, $(g - 2)_\mu$ and the mass of the $W$
boson, $\MW$.  

The paper is organised as follows: Section~2 gives
a summary of the most relevant supersymmetric sectors and parameters.
In section~3 we briefly review the calculations and codes used for the
Higgs sector predictions, as well as for the low-energy
observables. Details on the parameter scan are given.
In section 4 the results of the scan, the best-fit points and the
preferred parameter regions are presented. We briefly discuss the
origin of the most significant deviations of the \pMSSM\ predictions with respect to the SM results.
The conclusions can be found in section~5.


\section{Theoretical framework for \pMSSM}
\label{sect:theory}

In the following we briefly describe the relevant sectors of the MSSM
and define the seven basic parameters for our scan. As a first and
general simplification we restrict ourselves to the MSSM with real
parameters. 
The restriction to seven free/independent parameters is based on the
intention to sample the {\em full} parameter space with \order{10^7}
scan points, while 
ensuring that the most important effects in the MSSM Higgs
sector and flavour physics are covered.

The tree-level values for the predictions of the MSSM Higgs sector quantities
are determined by $\tb$, 
the $\cp$-odd Higgs-boson mass $\MA$, and the $Z$ boson mass $\MZ$. 
These predictions include
the other Higgs boson masses, their couplings to other
MSSM particles, their production cross sections, as well as their decay
properties. Consequently, we choose as free parameters the two
tree-level parameters, 
\begin{align*}
\mbox{(i)}: \quad &\MA, \\
\mbox{(ii)}: \quad &\tb.
\end{align*}

Beyond tree-level, the main correction to the Higgs boson masses
stems from the 
$t/\Stop$ sector, and for large values of $\tb$ also from the 
$b/\Sbot$ sector.
In order to fix our notations, we list the conventions for the input
parameters from the scalar top and scalar bottom sector of the MSSM. The
notation can be taken over to scalar taus  
and scalar tau neutrinos via the substitutions $\Stop \to \Sneut$,
$\Sbot \to \Stau$, taking into account that the MSSM contains only the
scalar superpartners of the left-handed neutrinos. 
Furthermore, the same notation holds for the first and the second
generation of scalar fermions. 
The mass matrices in the basis of the current eigenstates $\StopL,
\StopR$ and $\SbotL, \SbotR$ are given by
\BEA
\label{stopmassmatrix}
{\cal M}^2_{\Stop} &=&
  \ML \MstL^2 + \mt^2 + \CZb (\edz - \frac{2}{3} \sw^2) \MZ^2 &
      \mt \Xt \\
      \mt \Xt &
      \MstR^2 + \mt^2 + \frac{2}{3} \CZb \sw^2 \MZ^2 
  \MR, \\
&& \non \\
\label{sbotmassmatrix}
{\cal M}^2_{\Sbot} &=&
  \ML \MsbL^2 + \mb^2 + \CZb (-\edz + \frac{1}{3} \sw^2) \MZ^2 &
      \mb \Xb \\
      \mb \Xb &
      \MsbR^2 + \mb^2 - \frac{1}{3} \CZb \sw^2 \MZ^2 
  \MR,
\EEA
where 
$\MstL$, $\MstR$, $\MsbL$, and $\MsbR$ are the soft SUSY-breaking
mass parameters in the scalar top and bottom sector, $\mt$ and $\mb$ are
the respective quark masses, $\sw = \sqrt{1 - \MW^2/\MZ^2}$ with
$\MW$ denoting the mass of the $W$~boson, and 
\BE
\mt \Xt = \mt (\At - \mu \CTb) , \quad
\mb\, \Xb = \mb\, (\Ab - \mu \Tb) .
\label{eq:mtlr}
\end{equation}
Here $\At$ ($\Ab$) denotes the trilinear Higgs--stop (Higgs-sbottom)
coupling. The higgsino mass parameter $\mu$, which also appears in
\refeq{eq:mtlr}, is taken as free scan parameter 
\begin{align*}
\mbox{(iii)}: \quad \mu.
\end{align*}
SU(2) gauge invariance leads to the relation $\MstL = \MsbL$. Using a
universal parameter also for the left/right-handed squark soft masses of
the third generation (and similarly for the sleptons), we choose as free
parameters  
\begin{align*}
\mbox{(iv)}&: \quad \msqd := \MstL  (= \MsbL) = \MstR = \MsbR, \\
\mbox{(v)}&: \quad \msld := \MstauL (= \MsneutL) = \MstauR, \\
\mbox{(vi)}&: \quad \Af := \At = \Ab = \Atau.
\end{align*}
For the soft scalar masses of the first two generations, which are much
less relevant for Higgs physics (but can play a relevant role for the
low-energy observables), we choose fixed parameter values as
\begin{align}
\MsqL = \MsqR~(q = c, s, u, d) \; &= \; 1000 \gev, \\
\MslL = \MslR~(l = \mu, \nu_\mu, e, \nu_e) \; &= \; 300 \gev.
\end{align}
The choice for the first and second generation squarks places their
masses roughly at the level currently probed at the LHC. Somewhat
larger values would have a minor impact on our analysis.
The values for the first and second generation slepton mass
parameters were chosen to provide rough agreement with the anomalous
magnetic moment of the muon (see below).

The trilinear Higgs coupling parameter for the first two generations we set to
\begin{align}
A_{c,s,u,c,\mu,e} \; &= \; \Af.
\end{align}
The Higgs sector observables
furthermore depend on the SU(2) gaugino mass parameter, 
\begin{align*}
\mbox{(vii)}: \quad M_2,
\end{align*}
which we take as the final free parameter in our analysis. The other
electroweak gaugino mass parameter, $M_1$, is fixed via the GUT relation 
\BE
M_1 = \frac{5}{3} \frac{\sw^2}{\cw^2} M_2 \approx \frac{1}{2}  \MTwo~.
\end{equation}
The gluino mass parameter, which enters the Higgs mass predictions only from
two-loop order on, is fixed to a value close to the limits from recent
searches at the LHC, 
\begin{align}
M_3 = \mgl = 1000 \gev.
\end{align}
An adjustment of the gluino mass parameter to even higher values is
expected to have a negligible impact on our analysis. 


\section{Model predictions and constraints}
\subsection{Parameter space scanning}

\begin{table}
\centering
\begin{tabular}{rccl}
\hline
Parameter &  Minimum &  Maximum \\
\hline
$\MA$ [GeV]    & 90       & 1000 \\
$\tb$ \phantom{[GeV]}         & 1        & 60  \\
$\mu$ [GeV]    & 200      & 3000 \\
$\msqd$ [GeV]  & 200      & 1500 \\
$\msld$ [GeV]  & 200      & 1500 \\
$\Af$ [GeV]    & -3$\,\msqd$ & 3$\,\msqd$ \\
$M_2$ [GeV]    & 200      & 500 \\
\hline
\end{tabular}
\caption{Ranges used for the free parameters in the \pMSSM\ scan.}
\label{tab:param}
\end{table}
The \pMSSM\ parameter space is sampled by performing random scans (using
uniform distributions) over the seven input parameters in the ranges
given in \refta{tab:param}. The two cases, where either $h$ or $H$
corresponds to the signal at $\MHhat \sim \simMH \gev$, are treated in two
separate scans, and the results are discussed in parallel
below. Each scan starts with \order{10^7} randomly
chosen points with a flat distribution over the parameter ranges.
Dedicated, smaller, sampling is then performed to map the
interesting regions of parameter space.%
\footnote{The reader should keep in mind here (and in the following)
that the point density has no statistical meaning.}%
~In practice, the full parameter ranges from \refta{tab:param} are taken only
for the light Higgs case, while for the heavy Higgs case we limit
$\MA < 200 \gev$ and $\tb < 30$ (still using the full ranges for the other
parameters), which improves the sampling efficiency in the relevant mass
region for $\MH$. In addition to the free parameters listed in
\refta{tab:param}, the remaining parameters are fixed according to
section~\ref{sect:theory}.

Since we are mainly interested in the Higgs sector, we do not exploit
the full possibilities in the low-energy MSSM to vary the soft-breaking
parameters of the first two generations or the gluino
mass 
(we will comment below on a possible impact of this choice on the
MSSM predictions for 
$(g-2)_\mu$). Consequently, it is not relevant to apply LHC exclusion bounds
from supersymmetry searches in our analysis, since these can be avoided
by adjusting the additional parameters to sufficiently high values with
only small effects on the Higgs sector. We do, however, apply the
results from Higgs boson searches, see the next subsection. We also
apply the model-independent limits on sfermion and chargino masses,
typically at the level of $\sim 100 \gev$
from direct searches at LEP (as summarized in the PDG review
\cite{Beringer:1900zz}). Furthermore, we require that the
lightest 
supersymmetric particle (LSP) is the lightest neutralino. 
The top quark pole mass is sampled from a Gaussian distribution with
$\mt=173.2\pm 0.9\gev$, using a cutoff at $\pm 2\,\sigma$. Effects of
other parametric uncertainties from SM quantities are estimated to be
small, and are therefore neglected. 

For the evaluation of the sparticle and Higgs masses we use the code
\FH\ (version 2.9.4)~%
\cite{Heinemeyer:1998np,*Heinemeyer:1998yj,Degrassi:2002fi,Frank:2006yh}.
The residual Higgs mass uncertainty from this calculation (e.g.~from missing
higher orders) is estimated to be around $2\gev$, 
depending on the considered region of parameter space
\cite{Degrassi:2002fi}. 
We are interested in parameter points that give a Higgs mass prediction,
for either $\Mh$ or $\MH$, close to the observed LHC signal. We
therefore constrain the analysis in a first step to points with
$\Mh$ or $\MH$ in the region $121-129 \gev$. To avoid configurations in
parameter space that give an unstable perturbative behavior in the Higgs
mass calculation, we use a criterion based on the $\matr{Z}$-matrix
(as defined in \cite{Frank:2006yh}) and exclude points for which
$\left||Z_{k1}^{\mathrm{2L}}| - |Z_{k1}^{\mathrm{1L}}|\right|/
|Z_{k1}^{\mathrm{1L}}|>0.1$ (see \cite{Benbrik:2012rm} for a similar treatment).
Here $k=1$ (2) is set for a SM-like light (heavy) Higgs, and the
superscripts refer to the same quantity evaluated with 1-loop (1L) or
2-loop (2L) precision. 

To obtain an indication of what the currently favoured regions of the
MSSM parameter space are, we use a simple statistical treatment of the
data where the different observables are taken into account by
calculating, for every parameter point in the scan, a global $\chi^2$
function 
\begin{align}
\chi^2 &= \sum_{i=1}^{n_{\mathrm{LHC}}} \frac{(\mu_i-\hat{\mu}_i)^2}{\sigma_i^2}
         +\sum_{i=1}^{n_{\mathrm{Tev}}} \frac{(\mu_i-\hat{\mu}_i)^2}{\sigma_i^2} 
         +\frac{(M_{h,H}-\MHhat)^2}{\sigma_{\MHhat}^2}
         +\sum_{i=1}^{n_{\mathrm{LEO}}} \frac{(O_i-\hat{O}_i)^2}{\sigma_i^2}.
\label{eq:totchi2}
\end{align}
Quantities with a hat denote experimental measurements, and unhatted
quantities the corresponding model predictions for Higgs signal strength 
modifiers, $\mu_i$, and low-energy
observables (LEO), $O_i$. The different observables entering
\refeq{eq:totchi2}~are described in more detail in the following
sections.
The combined
uncertainties $\sigma_i$ contain the known theory and experimental
uncertainties. Correlations are neglected throughout, since they are
for most cases
not publicly available. The total number of degrees of freedom, $\nu$,
is counted in the naive way as 
$\nu = n_{\mathrm{obs}}-n_{\mathrm{para}}$, where 
$n_{\mathrm{obs}}=n_{\mathrm{LHC}}+n_{\mathrm{Tev}}+1+n_{\mathrm{LEO}}$
(for LHC, Tevatron, the Higgs boson mass, and low-energy observables);
$n_{\mathrm{para}}$ is the number of model parameters. In the SM we have
$n_{\mathrm{para}}=1$ (the Higgs mass), and for both MSSM analyses
$n_{\mathrm{para}}=7$.  


\subsection{Direct Higgs searches}
\label{sect:directHiggs}

We use two kinds of experimental data from direct Higgs searches in our
analysis: exclusion limits at $95\%$ confidence level (CL), and the
signal rates measured 
in different channels for the observed LHC signal around $\MHhat\sim
\simMH \gev$. The exclusion limits from LEP, the Tevatron, and the LHC
(including data published before ICHEP 2012) are taken into account
using \HB\ version 3.8.0 \cite{Bechtle:2008jh,*Bechtle:2011sb}. Since no
$\chi^2$ information is available from these searches, but only the
exclusion limits at a fixed level of confidence, we apply these limits
as hard cuts on the parameter space. It should be noted, 
however, that \HB\ only tests for each parameter point
the model predictions against the single channel with the highest 
expected sensitivity for an exclusion, in order
to ensure a consistent interpretation of the exclusion limit as a $95\%$~CL.

The measured signal strength modifiers, $\hat{\mu}_i$, for the observed
Higgs-like state around $\MHhat\sim \simMH \gev$ are taken into account in
our fit directly in the $\chi^2$ evaluation (see
\refeq{eq:totchi2}). The data for all included channels is given in
\refta{tab:higgschannels}, with the corresponding experimental signal
strengths and their (asymmetric) $1\,\si$ error bands. These rates
provide the main dataset to which we fit the MSSM Higgs sector. In total
we include $37$ observables, where $34$ are from the LHC experiments and
$3$ provide supplementary information from the Tevatron.  The best fit
signal strength modifiers of ATLAS and CMS are given for different Higgs
masses, corresponding to the values measured by the individual
experiments, i.e.~we interpret the experimental discoveries as being
compatible, and due to a single new state. The Tevatron data, which does
not admit a mass measurement from the observed excess
on its own, is evaluated for
$\MHhat = 125 \gev$. All values listed in \refta{tab:higgschannels}~are
extracted directly from the quoted experimental references, with one
exception: ATLAS has not provided a measurement for the signal strength
modifier of $H \to ZZ^{(*)}$ separately for the $7$ and $8$ TeV data,
but only for the combination (the $7$ TeV values are available from a
previous analysis). To compare to our $8$ TeV predictions, these values
are therefore calculated from the $7$~TeV and $7+8$ TeV data under the
assumption of independent Gaussian measurements, following the 
procedure outlined in \cite{Espinosa:2012im}. This should lead to an
uncertainty on the estimated $8$~TeV rate of the same order as the
overall uncertainty from neglecting the (unknown) correlations. 

\afterpage{
\newpage
}
\begin{table}[h!]
\centering
\begin{tabular}{lcr@{.}lr@{.}lr@{.}lc}
\hline
Channel & $\sqrt{s}$~[TeV] & \multicolumn{2}{c}{ $\hat{\mu}_{\mathrm{low}}$} & \multicolumn{2}{c}{$\hat{\mu}$} & \multicolumn{2}{c}{ $\hat{\mu}_{\mathrm{up}}$} & Reference \\
\hline
\multicolumn{9}{c}{ATLAS data at $\MHhat=126.5 \gev$} \\
\hline
$b\bar{b}$              	&7&   -1&646&   0&510&   2&680& \cite{Atlasnote1293}\\
$\tau \tau$     			&7&   -1&550&   0&464&   2&011& \cite{Atlasnote1293}\\
$WW$              			&7&   -0&164&   0&438&   1&103& \cite{Atlasnote1212}\\
$WW$              			 &8&   1&308&   1&920&   2&536  & \cite{Atlasnote1298}\\
$\gamma \gamma$ (inclusive)&7&   1&397&   2&155&   2&903& \cite{Atlasnote1291}\\
$\gamma \gamma$ (inclusive)&8&   1&054&   1&685&   2&326& \cite{Atlasnote1291}\\
$ZZ$              			&7&    0&405&   1&080&   2&177& \cite{Aad:2012an}\\
$ZZ$              			&8&    0&400&   1&049&   1&708& \cite{Atlasnote1292}\\
\hline
\multicolumn{9}{c}{CMS data at $\MHhat=125.0 \gev$} \\
\hline
$b\bar{b}$ (VH) & 7 &            -0&606  &  0&588  &  1&824& \cite{CMSHig1220}\\
$b\bar{b}$ (VH) & 8 &            -0&441  &  0&424  &  1&535& \cite{CMSHig1220}\\
$b\bar{b}$ ($t\bar{t}H$) 	& 7 &   -2&624  & -0&771  &  1&288& \cite{CMSHig1220}\\
$\tau \tau $ (0/1 jet) 		& 7 &   -0&400  &  1&000  &  2&441& \cite{CMSHig1220}\\
$\tau \tau $ (0/1 jet) 		& 8 &    0&588  &  2&153  &  3&635& \cite{CMSHig1220}\\
$\tau \tau $ (VBF) 		& 7 &   -2&912  &  -1&718 & -0&359& \cite{CMSHig1220}\\
$\tau \tau $ (VBF) 		& 8 &   -3&035  &  -1&759 & -0&400& \cite{CMSHig1220}\\
$\tau \tau $ (VH) 			& 7 &    -2&418  &  0&671  &  4&788& \cite{CMSHig1220}\\
$\gamma \gamma$ (Dijet loose) & 8 & -2&660 &-0&626 & 1&409 & \cite{CMSHig1215}\\
$\gamma \gamma$ (Dijet Tight) & 8 & -0&267 & 1&289 & 2&868 & \cite{CMSHig1215}\\
$\gamma \gamma$ (Untagged 3)  & 8 &  2&007 & 3&754 & 5&549 & \cite{CMSHig1215}\\
$\gamma \gamma$ (Untagged 2)  & 8 & -0&195 & 0&930 & 2&080 & \cite{CMSHig1215}\\
$\gamma \gamma$ (Untagged 1)  & 8 &  0&475 & 1&504 & 2&533 & \cite{CMSHig1215}\\
$\gamma \gamma$ (Untagged 0)  & 8 &  0&212 & 1&456 & 2&701 & \cite{CMSHig1215}\\
$\gamma \gamma$ (Dijet)       & 7 &  2&174 & 4&209 & 6&243 & \cite{CMSHig1215}\\
$\gamma \gamma$ (Untagged 3)  & 7 & -0&099 & 1&528 & 3&132 & \cite{CMSHig1215}\\
$\gamma \gamma$ (Untagged 2)  & 7 & -0&434 & 0&715 & 1&887 & \cite{CMSHig1215}\\
$\gamma \gamma$ (Untagged 1)  & 7 & -0&291 & 0&643 & 1&600 & \cite{CMSHig1215}\\
$\gamma \gamma$ (Untagged 0)  & 7 &  1&337 & 3&132 & 4&974 & \cite{CMSHig1215}\\
$WW$ (0/1 jet) & 7 &             -0&029 &  0&588 &  1&206& \cite{CMSHig1220}\\
$WW$ (0/1 jet) & 8 &              0&176 &  0&835 &  1&494& \cite{CMSHig1220}\\
$WW$ (VBF) & 7 &             -3&900 & -1&306 &  0&918& \cite{CMSHig1220}\\
$WW$ (VBF) & 8 &             -0&523 &  1&371 &  3&347& \cite{CMSHig1220}\\
$WW$ (VH) & 7 &              -5&753 & -2&829 &  0&341& \cite{CMSHig1220}\\
$ZZ$ & 7 &                    0&176 &  0&671 &  1&371& \cite{CMSHig1220}\\
$ZZ$ & 8 &                    0&259 &  0&794 &  1&494& \cite{CMSHig1220}\\
\hline
\multicolumn{9}{c}{Tevatron data at $\MHhat=125.0 \gev$} \\
\hline
$b\bar{b}$              & $1.96$ & 1&290 &	1&970 &	2&710  & \cite{TCB:2012zzl}\\
$\gamma \gamma$ & $1.96$ &   1&080&	3&620 &	6&580 & \cite{TCB:2012zzl}\\
$WW$              & $1.96$ &  0&000	&0&320	& 1&450 & \cite{TCB:2012zzl}\\
\hline
\end{tabular}
\caption{Experimentally measured values for the Higgs signal strength modifiers, and their corresponding uncertainties (lower/upper edges of $1\,\sigma$ error bars), in the various channels.} 
\label{tab:higgschannels}
\end{table}

The MSSM predictions for the signal strength modifiers are evaluated
according to  
\begin{align}
\mu_{i}&=\frac{\sum_k \omega_{ik} \sigma_k(pp\to h,H) 
              \times \mathrm{BR}(h,H\rightarrow i)}
              {\sum_k \omega_{ik} \sigma^{\mathrm{SM}}_k(pp\to h,H) 
              \times \mathrm{BR}^{\mathrm{SM}}(h,H\rightarrow i)},
\label{eq:mu}
\end{align}
where $\sigma_k(pp\to h,H)$ denotes the contribution to the Higgs
production cross section from partonic subprocess $k$, evaluated at the
predicted Higgs mass. The production modes considered are gluon-gluon
fusion ($gg$), vector boson fusion (VBF), associated vector boson
production (VH), and associated $t\bar{t}h(H)$ production. The
experimental efficiencies $\omega_{ik}$ have only been published by
ATLAS and CMS for the $\ga\ga$ analysis; for CMS in the case
of the subcategories, and for ATLAS also the inclusive result. We make
use of these numbers when they are available. For all other channels we
have to use the ``naive'' efficiencies deducible from the analysis
description (e.g.~for a VBF-type analysis tagging two forward jets, we
set $\omega=1$ for the VBF cross section, whereas all other modes have
$\omega=0$). In channels where the mass
resolution is not good enough to separate contributions from different
Higgs bosons, we approximate the contributions from $H$ and the
$\cp$-odd Higgs $A$ by adding their signal  
rates incoherently. We do not add the rates of the $\cp$-even Higgs bosons,
whose joint contributions to the signal could also include interference
effects. Our analysis is therefore limited to the case with a single
$\cp$-even Higgs boson close to the observed signal, and we leave a more
detailed treatment of the case when $\Mh\sim\MH\sim 126\gev$ to a dedicated
analysis. Since the $\cp$-odd Higgs does not have tree-level couplings
to vector bosons (and hence 
also a reduced coupling to photons), it gives a negligible
contribution to the channels with vector bosons in the
Higgs production and/or decay. Effectively, the $\cp$-odd Higgs
therefore only plays a role for the inclusive (0/1 jet) $\tau^+\tau^-$
channels. In these channels it can easily dominate over the $H$
contribution for large values of $\tan\beta$.  In the light Higgs
case, we find that the masses of $h$ and $A$ differ by $\MA-\Mh\gtrsim
50\gev $ in the favoured region (see below). Thus we do not take any
contributions to the $h$ rates from the $\cp$-odd Higgs into account.

The cross section predictions entering \refeq{eq:mu} are calculated,
both in the MSSM and the SM, using \FH\ (version 2.9.4)
\cite{Heinemeyer:1998np,*Heinemeyer:1998yj,Degrassi:2002fi,Frank:2006yh}. 
For the SM cross sections the results of the LHC Higgs cross section
working group are
implemented~\cite{Dittmaier:2011ti,*Dittmaier:2012vm,*LHCHXSWG} (where
the $gg$ production cross sections are taken from
\cite{deFlorian:2009hc,*ggHgrazzini}).
The corresponding MSSM production cross-sections are obtained
in the effective-coupling approximation~\cite{Hahn:2006my}. The
$gg$~production cross section follows the description in
\cite{Hahn:2010te}, where results 
of~\cite{Bonciani:2007ex,*Aglietti:2006tp,*Dedes:2002dy,*Dedes:2003km} 
were used.
The decay width evaluation includes a full one-loop correction for
the decay to fermions~\cite{Heinemeyer:2000fa,Williams:2011bu}; see
\cite{Hahn:2010te} for more details on the other channels.

In addition to the signal strength modifiers, we include a $\chi^2$
contribution from the measured Higgs mass $\MHhat$ (which in the MSSM
can correspond to either $\Mh$ or $\MH$). Unlike in the
SM, $\Mh$ and $\MH$ are not free parameters in the MSSM but a prediction of
the theory. Averaging the ATLAS and CMS measurements, we obtain
$\MHhat = \MHexp \gev$. To fully cover the difference between the individual
experimental measurements we assign a conservative uncertainty of
$\si_{\MHhat}^{\mathrm{exp}}=1\gev$. With more data we expect this
experimental uncertainty on the Higgs mass to be significantly
reduced. In addition, we add linearly to this number a theoretical
uncertainty $\si_{\MHhat}^{\mathrm{theo}}=2\gev$~\cite{Degrassi:2002fi}, 
which accounts
for the uncertainty in the MSSM Higgs mass calculation. The total
uncertainty entering \refeq{eq:totchi2} is therefore
$\si_{\MHhat}=3\gev$.


\subsection{Low-energy observables}
\label{subsec:LEO}

In addition to the measurements related to the LHC Higgs signal, we
include several low-energy observables (LEO) in the fit. These are
listed in \refta{tab:flav}, which summarizes the experimental
values%
\footnote{We note that the Belle Collaboration has recently
reported a new measurement of $\br(B_u \to \tau \nu_\tau)$ that is in better
agreement with the SM (and thus naturally with models with two Higgs
doublets)~\cite{Bellebtn}.
While we do not take this new result into account in our overall
fit results, we do comment briefly on its possible effects.}%
~(for the case of $\br(B_s\to \mu^+\mu^-)$ the upper limit) and the
corresponding SM theory predictions (evaluated for
$\MH^{\mathrm{SM}}=\MHexp\gev$ and $m_t=173.2\gev$). The flavour physics
observables are evaluated (both in the SM and the MSSM) using
\SI~(version 3.2)
\cite{Mahmoudi:2007vz,*Mahmoudi:2008tp,*Mahmoudi:2009zz}, which in
particular contains the results for $\br(B\to X_s\gamma)$ based on the
NNLO calculation of \cite{Misiak:2006ab}. 
Our fit includes also the anomalous magnetic moment of
the muon, $a_\mu = \edz \gmt$, which shows a deviation of more than
$3\,\sigma$ between the experimental measurement and the SM prediction.%
\footnote{The most recent evaluation of $a_{\mu}$ in the SM~\cite{Benayoun:2012wc}, taking also $\tau$~data into account, finds an even larger deviation of
more than $4\,\si$.}%
~We use {\tt SuperIso} to calculate
the MSSM contribution $\delta a_\mu$ to the anomalous magnetic moment of
the muon (we have cross-checked those results with \FH\ and found
good agreement) 
including the dominant two-loop
contributions~\cite{Degrassi:1998es,*Heinemeyer:2003dq,*Heinemeyer:2004yq},
see~\cite{Stockinger:2006zn} for a review.

As a final observable we also include the MSSM prediction of the
$W$~boson mass into the fit.
Here the SM prediction shows a $\sim 1.5\, \si$ 
deviation from the latest experimental value, $\MW^{\rm exp} =
80.385 \pm 0.015 \gev$~\cite{Group:2012gb,*LEPEWWG}. Our MSSM evaluation
of $\MW$ is done using \FH, where the full SM
result~\cite{Awramik:2003rn} is supplemented with the leading
corrections from the 
$\Stop/\Sbot$
sector~\cite{Djouadi:1996pa,*Djouadi:1998sq,Heinemeyer:2004gx}. A
comparison with 
the best available MSSM evaluation~\cite{Heinemeyer:2006px,*MWLisa} shows that
corrections larger than $10 \mev$ can be missed if some uncoloured SUSY
particles are light. Consequently, we assign a theory uncertainty of 
$15 \mev$ to our $\MW$ evaluation and conservatively combine it with the
experimental uncertainty linearly. Thus in total we take an uncertainty
of $\pm 30 \mev$ into account. 

\begin{table}
\centering
\begin{tabular}{crlr}
\hline
Observable & \multicolumn{2}{c}{Experimental value}  & SM value  \\
\hline
$\mathrm{BR}(B\to X_s\gamma)$ & $(3.43\pm 0.21\pm 0.07)\times 10^{-4}$ & \cite{Amhis:2012bh} & $(3.08\pm 0.22)\times 10^{-4}$  \\
$\mathrm{BR}(B_s\to \mu^+\mu^-)$ & $< 4.2 \times 10^{-9}$ &\cite{lhcbmumu,*cmsmumu,*atlasmumu} & $(3.55\pm 0.38)\times 10^{-9}$ \\
$\mathrm{BR}(B_u\to \tau \nu_\tau)$ & $(1.66\pm 0.33)\times 10^{-4}$ &\cite{Amhis:2012bh} & $(1.01\pm 0.29)\times 10^{-4}$ \\
$\delta a_{\mu}$ & $(30.2\pm 9.0)\times 10^{-10}$ &\cite{Davier:2010nc,Bennett:2004pv,*Bennett:2006fi} & -- \\
$M_W$ & $(80.385\pm 0.015) \gev$ &\cite{Group:2012gb,*LEPEWWG} & $(80.363 \pm 0.004) \gev$  \\
\hline
\end{tabular}
\caption{The experimental values and (SM) theory predictions for low-energy observables (LEO) used to constrain the MSSM parameter space.}
\label{tab:flav}

\end{table}


\newpage

\begin{figure}[h!]
\centering
\includegraphics[angle=180,width=0.78\columnwidth]{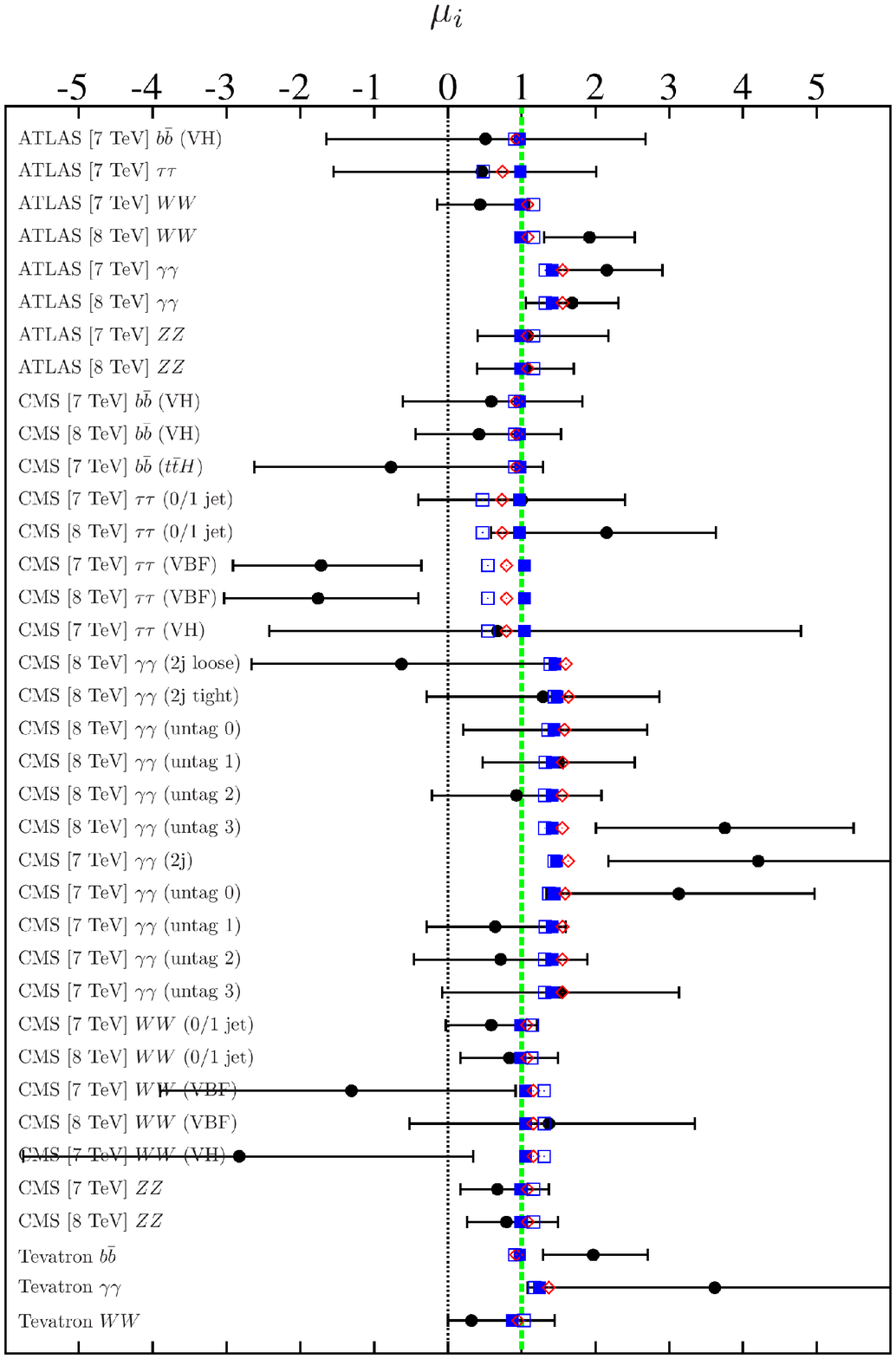}
\caption{Fit results for the signal strength modifiers, $\mu_i$, in the
  case that the light $\cp$-even Higgs is 
    interpreted as the new boson around $\sim \MHexp \gev$ 
(``light Higgs case''). The experimental data is
  shown as black dots (with error bars). The other symbols show best fit
  points, corresponding to the full fit (LHC+Tevatron+LEO) (blue solid
  squares), without the Tevatron data (blue open squares), and without LEO
  (red diamonds).} 
\label{fig:hbestfit}
\vspace{-1em}
\end{figure}

\newpage
\begin{figure}[h!]
\centering
\includegraphics[angle=180,width=0.78\columnwidth]{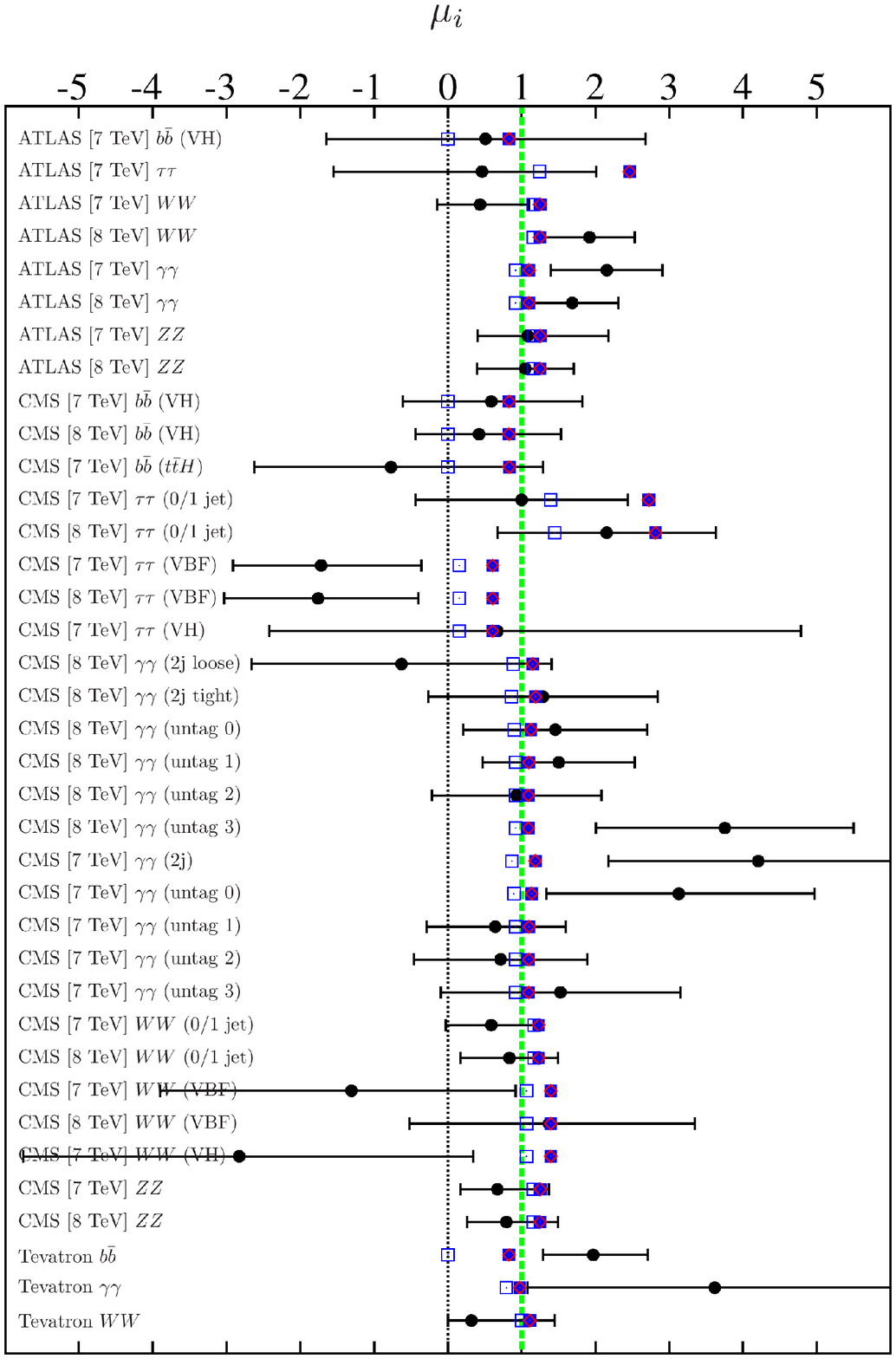}
\caption{Fit results for the signal strength modifiers, $\mu_i$, in
 the case that the heavy $\cp$-even Higgs is
    interpreted as the new boson around $\sim \MHexp \gev$ 
(``heavy Higgs case''). The experimental data is
  shown as black dots (with error bars). The other symbols show best fit
  points, corresponding to the full fit (LHC+Tevatron+LEO) (blue solid
  squares), without the Tevatron data (blue open squares), and without LEO
  (red diamonds).} 
\label{fig:HHbestfit}
\vspace{-1em}
\end{figure}

\begin{table}[t]
\centering
\begin{tabular}{c|ccc|ccc|ccc|ccc}
\hline
 & \multicolumn{3}{c|}{LHC only} & \multicolumn{3}{c|}{LHC+Tevatron} & \multicolumn{3}{|c|}{LHC+LEO} &  \multicolumn{3}{c}{LHC+Tevatron+LEO} \\
Case & $\chi^2/\nu$ & $\chi^2_{\nu}$ & $p$ & $\chi^2/\nu$ & $\chi^2_{\nu}$ & $p$ & $\chi^2/\nu$ & $\chi^2_{\nu}$ & $p$ & $\chi^2/\nu$ & $\chi^2_{\nu}$ & $p$\\
\hline
SM  & $27.6/34$ & 0.81 & 0.77 & $31.0/37$ & 0.84 & 0.74 & $41.6/39$ & 1.07 & 0.36 & $45.3/42$ & 1.08 & 0.34 \\
$h$ & $23.3/28$ & 0.83 & 0.72 & $26.8/31$ & 0.86 & 0.68 & $26.7/33$ & 0.81 & 0.77 & $30.4/36$ & 0.84 & 0.73 \\
$H$ & $26.0/28$ & 0.93 & 0.57 & $33.1/31$ & 1.07 & 0.37 & $35.5/33$ & 1.08 & 0.35 & $42.4/36$ & 1.18 & 0.21\\
\hline
\end{tabular}
\caption{Global $\chi^2$ results with $\nu$ degrees of freedom from the
  fits of the SM and the MSSM with either $h$ or $H$ as the LHC signal, 
  the reduced $\chi^2_\nu \equiv \chi^2/\nu$, 
  and the corresponding
  $p$-values. The number of degrees of freedom are evaluated naively as 
  $\nu = n_{\mathrm{obs}}-n_{\mathrm{param}}$.} 
\label{tab:totchi2}
\end{table}


\section{Results}

In \refta{tab:totchi2} we present the results of our fits in terms
of total $\chi^2$ values (with $\nu$ degrees of freedom), the reduced
$\chi^2_\nu \equiv \chi^2/\nu$, and the corresponding $p$-values. 
Since $\nu$ is derived via the naive counting, the {\em absolute}
numbers of the $p$-values should not be overinterpreted; the 
{\em relative} numbers, however, give a good impression of the 
{\em relative} goodness of the fits.
For each MSSM
intrepretation (the cases of either $h$ or $H$ as the $\MHexp\gev$
signal) we present four different fits: one taking the complete dataset
(LHC+Tevatron+LEO) into account, one where the 
low-energy observables (LEO) are left out, 
one where the Tevatron data are left out, and finally the fit where only
LHC observables are considered.  
When the fit is performed using only the high-energy
collider data, both with and
without the Tevatron results, the obtained $\chi^2$ values of the best
fit points are quite similar between the SM and the two MSSM
interpretations, where the fit in the heavy Higgs case becomes
slightly worse after the inclusion of the Tevatron data. 
When low energy observables are included, the SM and the heavy
Higgs case fits
become somewhat worse. 
In the latter case this can be understood from the potentially larger
contributions of light Higgs bosons to $B$-physics observables. For
the SM fit the reason lies in the fact that the SM 
prediction for $(g-2)_\mu$ differs by more than $3\,\si$ from the
experimental value. Still we find that the
SM provides a good fit to the full dataset, with $p_{\mathrm{SM}}=0.34$.
On the other hand, concerning the MSSM it should be kept in mind
that we did not fit the second generation slepton masses, which could
potentially further improve the $a_\mu$ fit.
For the complete fit, the corresponding $p$-values in the MSSM cases are
$p_h=0.73$ ($p_H=0.21$) for the $h$ ($H$) interpretations,
respectively, which are both acceptable $p$-values.
Overall, the data shows no clear preference
for the MSSM over the SM at this point. While the MSSM fits, in
particular in the light Higgs case, yield lower $\chi^2$ values than the
SM, this comes at the expense of additional parameters, so that the
difference in the $p$-values is rather moderate.
It is interesting to note
that the fit to the heavy Higgs case, while not being as good as in
the light Higgs case, still provides an acceptable
description of all the data, i.e.\ 
the MSSM in the non-decoupling region still provides a
viable solution. This interesting case will be analyzed in more detail
below.

Starting with the best fit for the $h$ case, we show in
\figref{fig:hbestfit} the different best fit points using  all available
data (LHC, Tevatron, LEO) (blue solid squares), leaving out LEO 
(red diamonds) or leaving out the Tevatron data (blue open squares). The
comparison of these three different types of results allows to trace the
origin of trends in the fitted parameters. The experimental data on the
signal strength modifiers in the different channels (as indicated in the
figure) is shown as black dots, with the error bars corresponding to
$\pm 1 \,\si$ uncertainties on $\hat{\mu}$. The values for the best fit
point of the complete fit (LHC, Tevatron, LEO) are also presented in
tabular form in 
\refta{tab:higgschannels_bestfit}. From here we can determine some
characteristics of the best fit point, such as a significantly enhanced
rate in the $\gamma\gamma$ final state and nearly SM rates for the
other channels. Leaving out the Tevatron data a (small) suppression of the
fermionic final states can be observed.
The fitted rates demonstrate that the \pMSSM\ is able to
accomodate
the main trends in the LHC/Tevatron data. Comparing the best fit points
with/without LEO, we find a qualitatively very similar behaviour. 

\begin{table}[h!]
\renewcommand{\arraystretch}{1.05}
\centering
{\small
\begin{tabular}{llc|ccr@{.}l|ccr@{.}l}
\hline
\multicolumn{2}{l}{Channel} & $\sqrt{s}$~[TeV] & $\mu_h$ & $\chi_h^2$ & \multicolumn{2}{c|}{Pull} & $\mu_H$ & $\chi_H^2$ & \multicolumn{2}{c}{Pull}  \\
\hline
ATLAS & $b \bar b$           		&7& 0.98 & 0.05 & 0&22 &         0.83&0.02&0&15         \\
ATLAS &$\tau \tau$                  	&7& 0.98 & 0.11 & 0&33 &         2.46&1.67&1&29              \\
ATLAS & $WW$              			&7& 0.99 & 0.69 & 0&83 & 1.25&1.50&1&22                 \\
ATLAS & $WW$              			&8& 0.99 & 2.31 &-1&52 & 1.25&1.19&-1&09               \\
ATLAS &$\gamma \gamma$				&7& 1.41 & 0.95 &-0&98 & 1.10&1.94&-1&39                   \\
ATLAS &$\gamma \gamma$ 				&8& 1.42 & 0.18 &-0&43 & 1.10&0.87&-0&93                   \\
ATLAS & $ZZ$              			&7& 0.99 & 0.02 &-0&13 & 1.25&0.02&0&16                  \\
ATLAS & $ZZ$              			&8& 0.99 & 0.01 &-0&09 & 1.25&0.09&0&31                 \\
\hline
CMS & $b \bar b$ (VH)                &7& 0.98 & 0.10 & 0&32 &            0.83&0.04&0&19          \\
CMS & $b \bar b$ (VH)                &8& 0.98 & 0.25 & 0&50 &            0.83&0.13&0&36             \\
CMS & $b \bar b$ (ttH)               &7& 0.98 & 0.72 & 0&85 &            0.83&0.61&0&78          \\
CMS & $\tau \tau $ (0/1 jets)        &7& 0.97 & 0.00 &-0&02 &            2.72&1.43&1&20     \\
CMS & $\tau \tau $ (0/1 jets)        &8& 0.97 & 0.57 &-0&76 &            2.81&0.20&0&44     \\
CMS & $\tau \tau $ (VBF)             &7& 1.04 & 4.12 & 2&03 &            0.61&2.92&1&71        \\
CMS & $\tau \tau $ (VBF)             &8& 1.04 & 4.24 & 2&06 &            0.61&3.03&1&74       \\
CMS & $\tau \tau $ (VH)              &7& 1.04 & 0.01 & 0&09 &            0.61&0.00&-0&02        \\
CMS & $\gamma \gamma$ (Dijet loose)  &8& 1.45 & 1.04 & 1&02 &            1.15&0.76&0&87  \\
CMS & $\gamma \gamma$ (Dijet tight)  &8& 1.48 & 0.01 & 0&12 &            1.19&0.00&-0&06 \\
CMS & $\gamma \gamma$ (Untagged 0)   &8& 1.44 & 0.00 &-0&02 &            1.13&0.07&-0&26 \\
CMS & $\gamma \gamma$ (Untagged 1)   &8& 1.42 & 0.01 &-0&09 &            1.10&0.16&-0&39   \\
CMS & $\gamma \gamma$ (Untagged 2)   &8& 1.41 & 0.18 & 0&42 &            1.09&0.02&0&14  \\
CMS & $\gamma \gamma$ (Untagged 3)   &8& 1.41 & 1.80 &-1&34 &            1.09&2.32&-1&52  \\
CMS & $\gamma \gamma$ (Dijet)        &7& 1.48 & 1.80 &-1&34 &            1.19&2.21&-1&49 \\
CMS & $\gamma \gamma$ (Untagged 0)   &7& 1.44 & 0.89 &-0&94 &            1.14&1.24&-1&11 \\
CMS & $\gamma \gamma$ (Untagged 1)   &7& 1.41 & 0.65 & 0&81 &            1.10&0.23&0&48 \\
CMS & $\gamma \gamma$ (Untagged 2)   &7& 1.41 & 0.35 & 0&59 &            1.09&0.10&0&32 \\
CMS & $\gamma \gamma$ (Untagged 3)   &7& 1.41 & 0.01 &-0&07 &            1.09&0.07&-0&27 \\
CMS &  $WW$ (0/1 jets)                &7& 0.98 & 0.40 & 0&64 &            1.23&1.09&1&04                    \\
CMS &  $WW$ (0/1 jets)               &8& 0.98 & 0.05 & 0&22 &            1.23&0.36&0&60                  \\
CMS &  $WW$ (VBF)                    &7& 1.05 & 1.12 & 1&06 &            1.39&1.47&1&21                     \\
CMS &  $WW$ (VBF)                    &8& 1.05 & 0.03 &-0&17 &            1.39&0.00&0&01                   \\
CMS &  $WW$ (VH)                     &7& 1.05 & 1.50 & 1&22 &            1.39&1.78&1&33                     \\
CMS &  $ZZ$                          &7& 0.99 & 0.21 & 0&45 &            1.25&0.69&0&83                        \\
CMS &  $ZZ$                          &8& 0.99 & 0.08 & 0&28 &            1.25&0.43&0&65                     \\
\hline
LHC & \multicolumn{2}{l|}{Higgs mass [GeV]}           & 126.1  & 0.02 & 0&13 &  125.8 & 0.00 & 0&03 \\
\hline
Tevatron & $b \bar b$             &1.96& 0.98 & 2.13 &-1&46 &  0.83&2.82&-1&68           \\
Tevatron & $\gamma \gamma$        &1.96& 1.24 & 0.88 &-0&94 &  0.97&1.08&-1&04           \\
Tevatron &  $WW$                  &1.96& 0.87 & 0.24 & 0&49 &  1.11&0.49&0&70         \\
\hline
LEO & \multicolumn{2}{l|}{BR$(B\to X_s\gamma) \times 10^{4}$}    &  $3.41$ &0.00& -0&03& 4.38 & 2.12 & 1&46  \\
LEO & \multicolumn{2}{l|}{BR($B_s\to \mu^+\mu^-)\times 10^{9}$}  &  $2.79$ &0.00&  0&00& 2.24 & 0.00 & 0&00  \\
LEO & \multicolumn{2}{l|}{BR($B_u\to \tau\nu_\tau)\times 10^{4}$}  &  $0.98$ &2.37& -1&54& 0.80 & 3.78 &-1&94  \\
LEO & \multicolumn{2}{l|}{$\delta a_\mu\times 10^{9}$}           &  $2.58$ &0.24& -0&49 &1.34 & 3.48 &-1&87 \\
LEO & \multicolumn{2}{l|}{$M_W$ [GeV]}                           & 80.379  &0.04& -0&19 &80.383&0.00 &-0&05 \\
\hline
\end{tabular}
}
\caption{Best fit results (for the complete fit) with corresponding $\chi^2$
  contributions and pulls for each observable.}  
\label{tab:higgschannels_bestfit}
\renewcommand{\arraystretch}{1.0}
\end{table}

In \refta{tab:higgschannels_bestfit} we give the details on the
results for the low-energy observables. In the light Higgs case, the only
relevant contribution to the total $\chi^2$ comes from \btn. Using an
experimental value close to the new Belle result~\cite{Bellebtn} would
substantially reduce this $\chi^2$ contribution and lead to an even
better fit. The best-fit value of \bmm\ lies {\em below} the SM
prediction. This feature is indeed found for most of our favoured
region. We have checked that this trend is present already {\em without}
taking the $\chi^2$ contribution of \bmm\ itself
into account, see also
the discussion in~\cite{Haisch:2012re}.

\begin{figure}
\centering
\includegraphics[width=0.24\columnwidth]{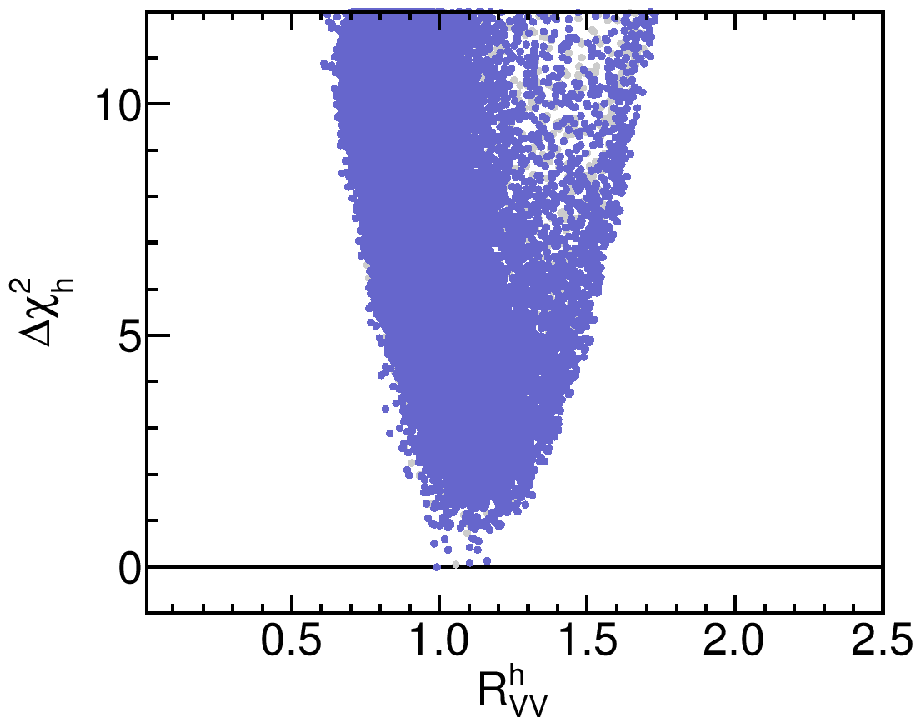}
\includegraphics[width=0.24\columnwidth]{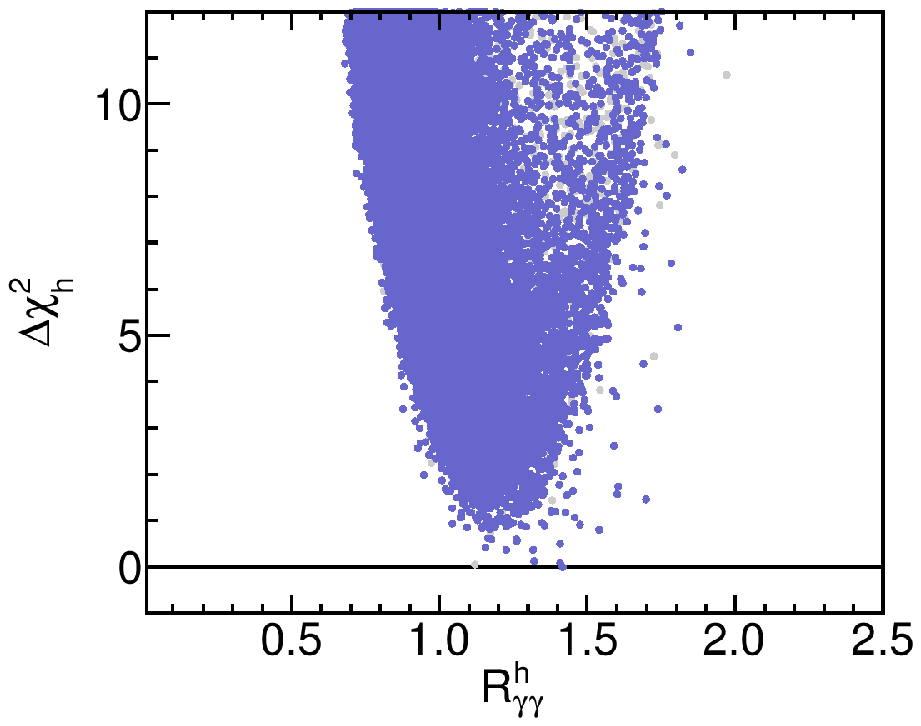}
\includegraphics[width=0.24\columnwidth]{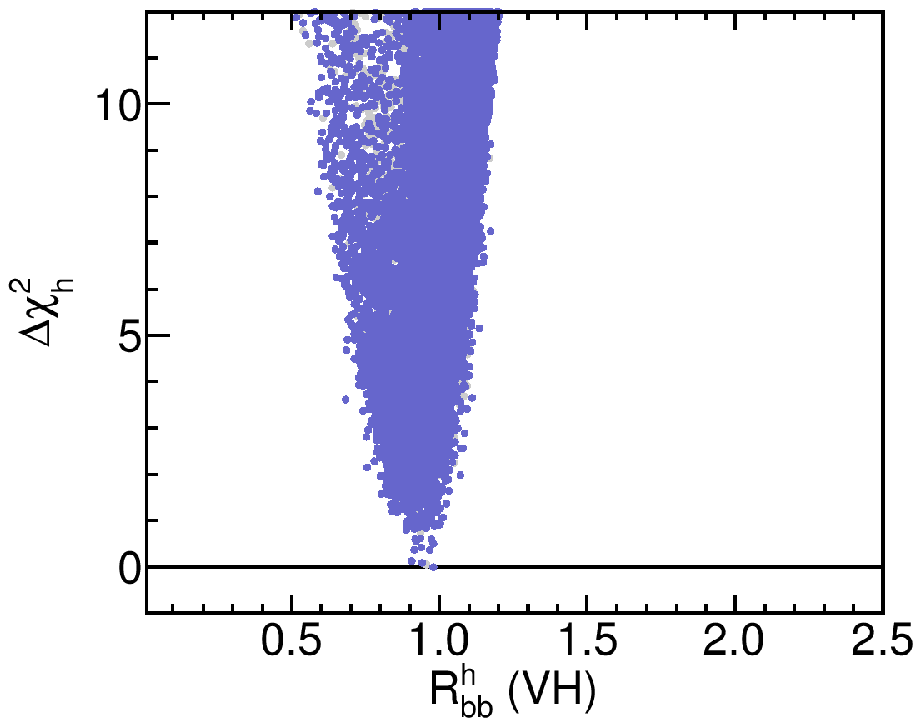}
\includegraphics[width=0.24\columnwidth]{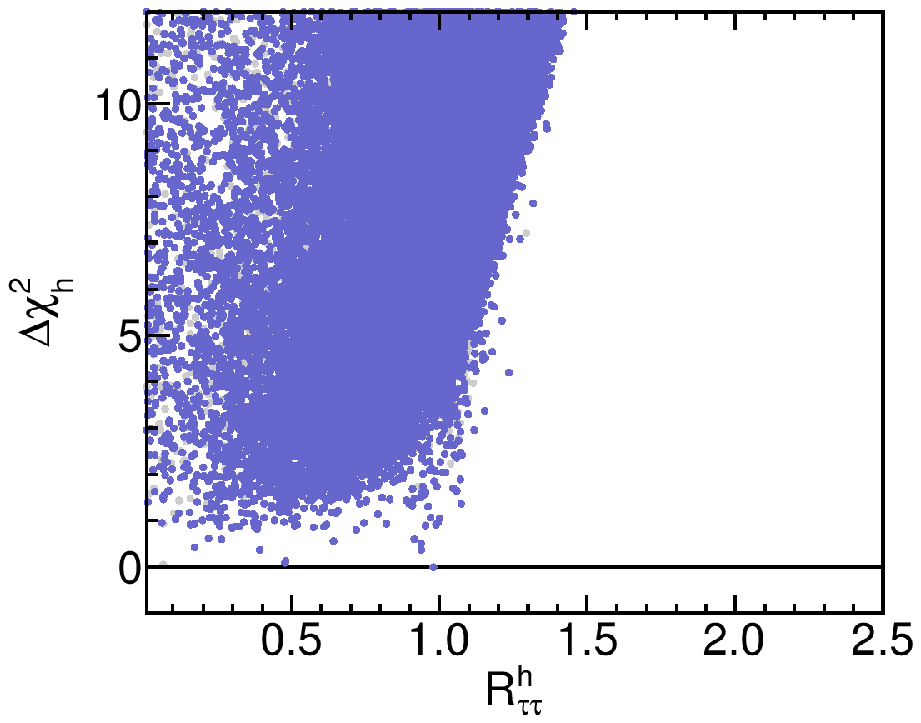}\\
\includegraphics[width=0.24\columnwidth]{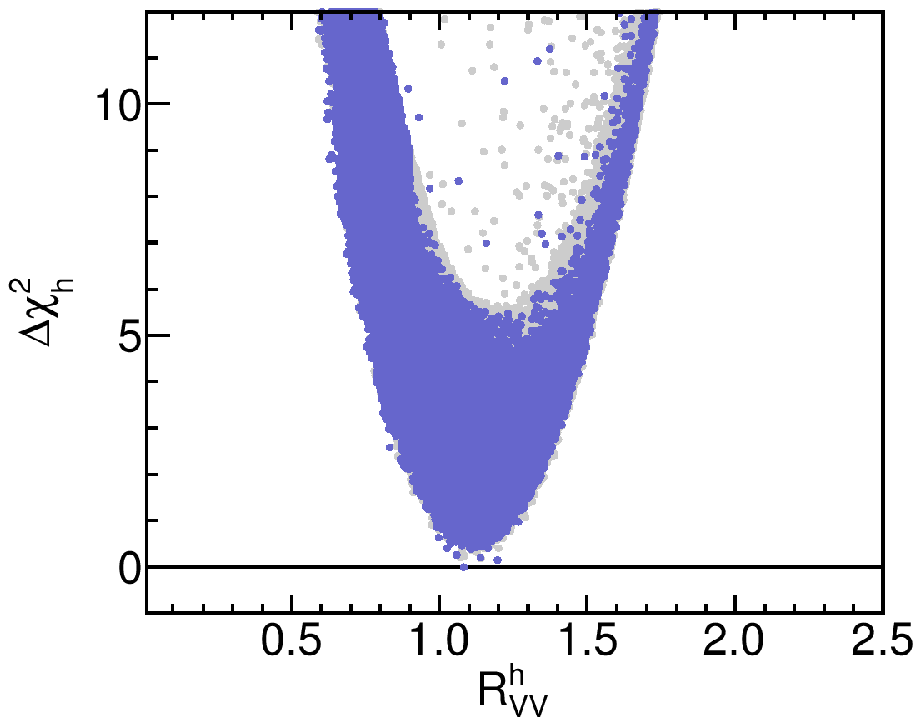}
\includegraphics[width=0.24\columnwidth]{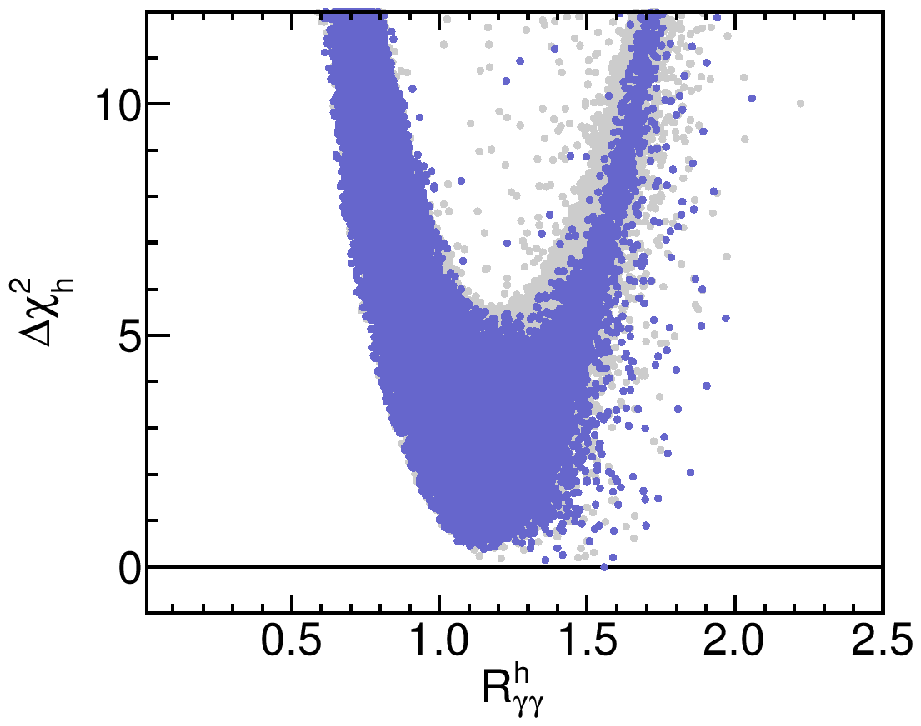}
\includegraphics[width=0.24\columnwidth]{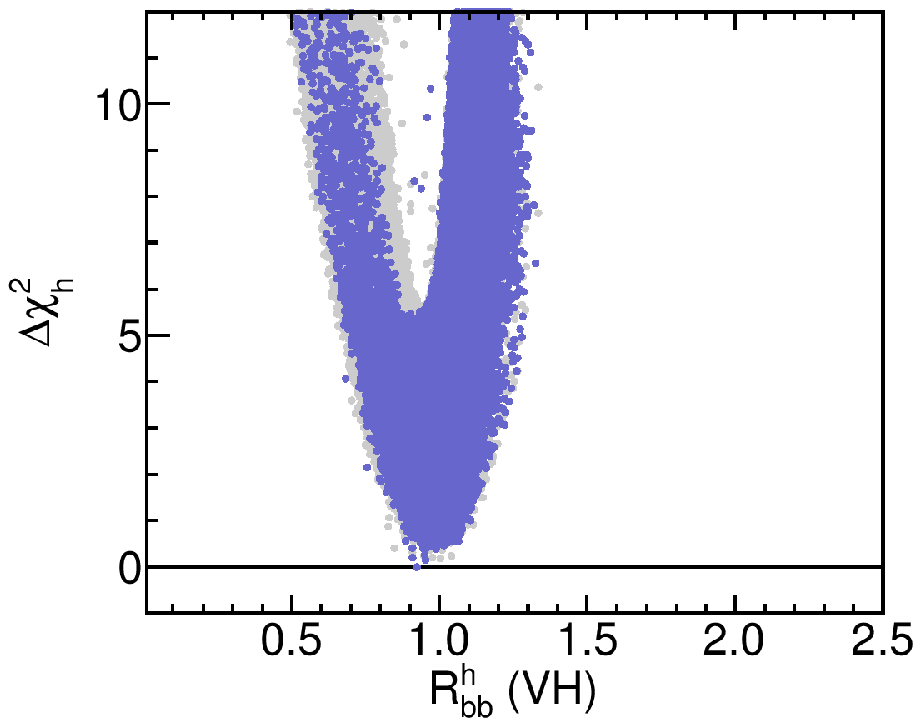}
\includegraphics[width=0.24\columnwidth]{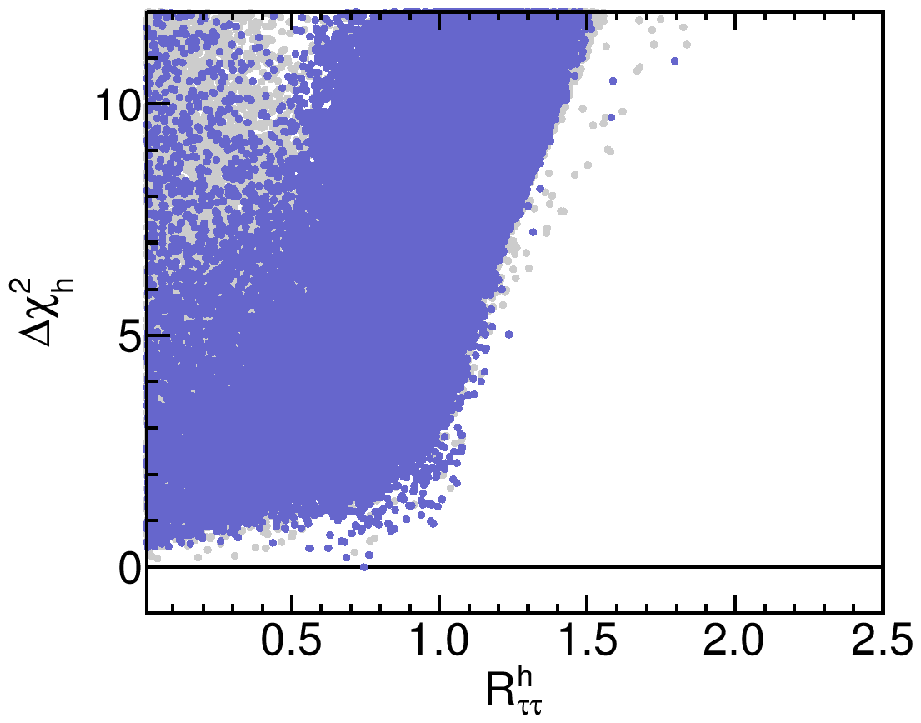}\\
\includegraphics[width=0.24\columnwidth]{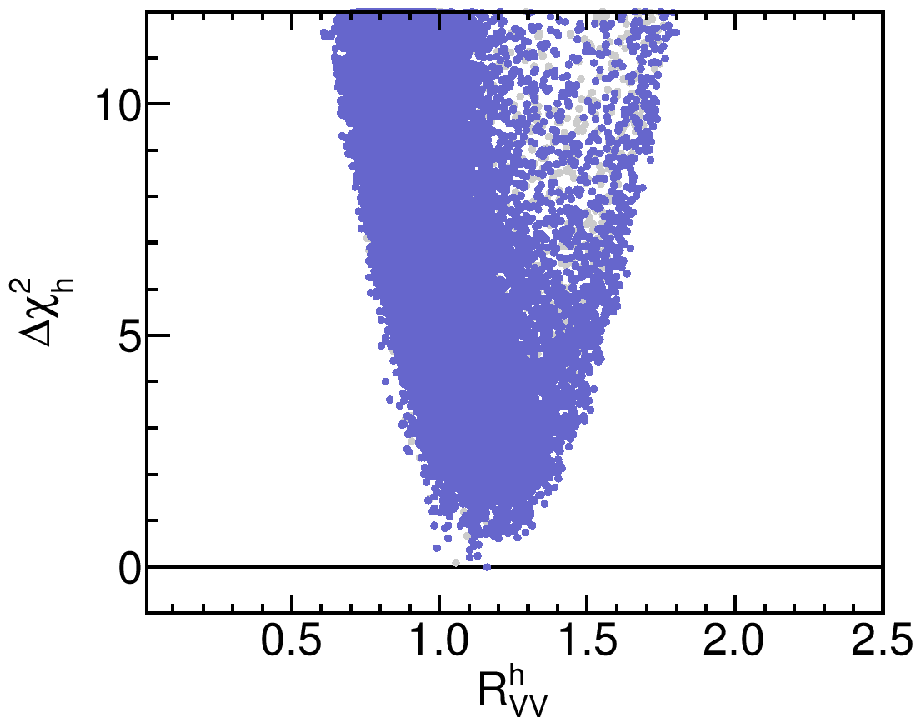}
\includegraphics[width=0.24\columnwidth]{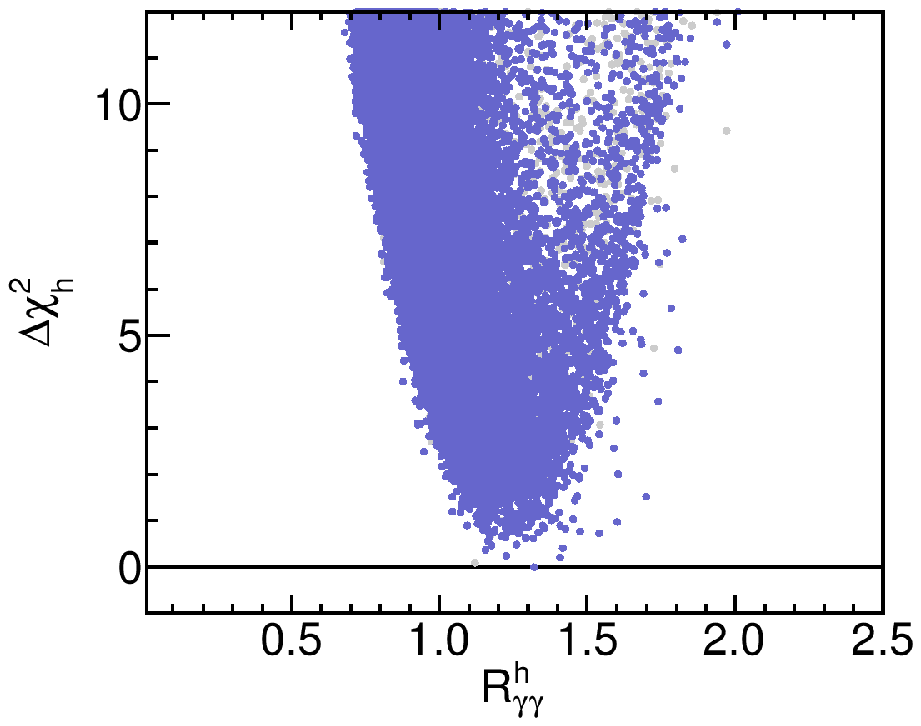}
\includegraphics[width=0.24\columnwidth]{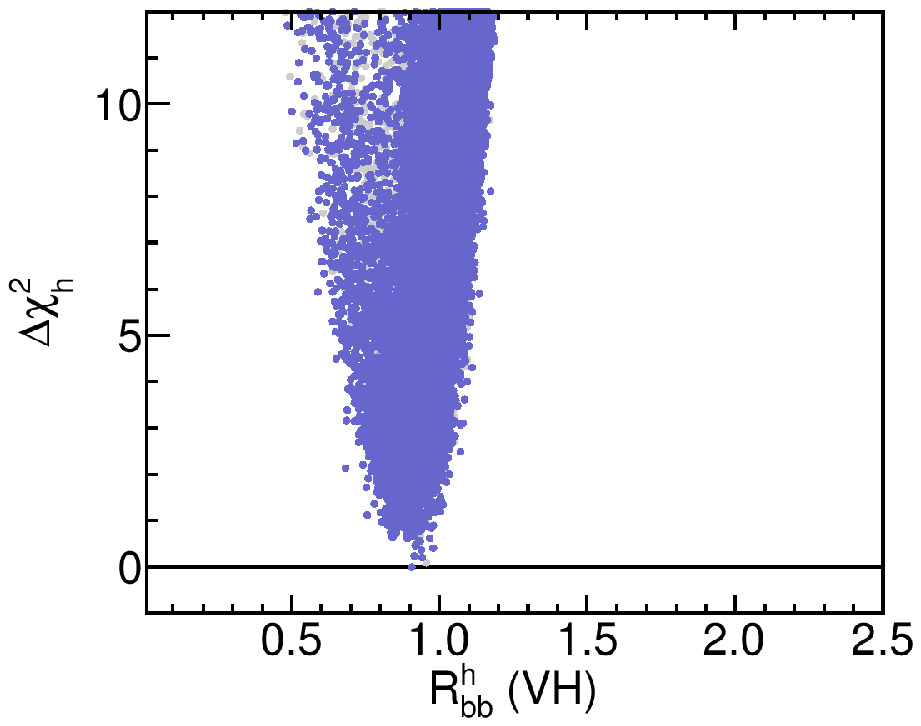}
\includegraphics[width=0.24\columnwidth]{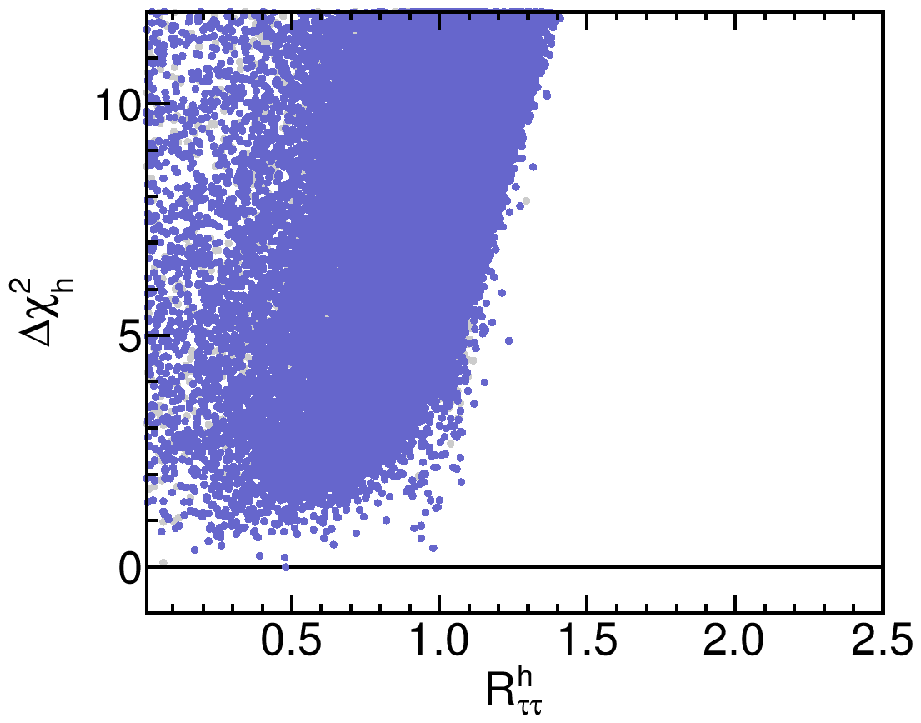}
\caption{Distributions of $\Delta\chi_h^2$  versus the different signal
  rates (defined in the text) for the light Higgs case. The colours show
  all points in the scan (gray), and points that pass the direct Higgs
  search constraints from 
  \HB\ (v.~3.8.0) (blue). The three rows show the distributions (with
  the corresponding minimal $\chi^2$ subtracted) for the complete fit
  (upper row), excluding LEO (middle row), and excluding Tevatron data
  (lower row).}  
\label{fig:totchi2h_rates}
\end{figure}

The best fit points for the heavy Higgs case are presented in
\figref{fig:HHbestfit} (numerical values in
\refta{tab:higgschannels_bestfit}). As the figure shows, the same best
fit point (albeit with different total $\chi^2$) is obtained for the
different cases with/without LEO. 
Leaving out the Tevatron data, however, has a larger qualitative impact on the
results, and rates close to zero are allowed in the $b \bar b$
channel, which will be discussed below in more detail.
Since the alternative hypothesis of a heavy Higgs explanation for the
LHC signal is fitted to the same data as for the $h$ case, and we have
already seen that the overall $\chi^2$ is similar to (although
slightly higher than) the light Higgs 
case, it is perhaps not so surprising to find that comparable rates are
obtained for the best fit in the heavy Higgs case. 
This illustrates again that this interpretation is also a possible
scenario to explain the LHC Higgs signal in the MSSM. The 
more in-depth results below are therefore presented in parallel for the
two separate cases with $h$ or $H$ corresponding to the signal
discovered at the LHC. 

In \refta{tab:higgschannels_bestfit} we also give the results for the
low-energy observables in the heavy Higgs case. One can see that the
relatively small value of the Higgs mass scale in this case leads to
non-negligible $\chi^2$ contributions from \bsg\ and \btn, where the latter
would substantially improve for a value close to the new Belle
result. Also the SUSY 
contribution to $a_\mu$ turns out to be relatively small, giving a
sizable contribution to the total $\chi^2$ (which is however affected by
our choice to keep the slepton mass parameters fixed).
Concerning \bmm\ it
should be noted that, as in the light Higgs case, the preferred value is
{\em below} the SM result, which again holds for most of the favoured
region.

We now turn from the global fit properties and the best fit points to a
more detailed analysis of the scan results. 
\figref{fig:totchi2h_rates} shows distributions of
$\Delta\chi_h^2=\chi_h^2-\chi^2_{h,\mathrm{min}}$ (light Higgs case) for
the different signal rates. The colour coding is as follows: all points
analyzed in the scan (which pass theoretical consistency checks and have
one $\cp$-even Higgs boson in the interval $121\gev < \Mh < 129\gev$)
are shown in gray. The blue points in addition fulfill constraints at
$95\%$~CL from direct Higgs searches applied by \HB\ 3.8.0. The three
rows show the results for the full fit (upper row), excluding LEO (middle row),
and excluding the Tevatron data (lower row). The signal rates are
calculated as the inclusive Higgs production cross section (evaluated at
$\sqrt{s}=8\tev$) times the decay rate, normalized to the SM predictions 
\begin{align}
R^{h,H}_{X} &=\frac{\sum_i \si_i(pp \to h,H) \times \br(h,H\to X)}
                {\sum_i \si^{\SM}_i(pp \to h,H) \times \br^{\SM}(h,H \to X)}.
\label{eq:Rx}
\end{align}
The only final state for which we consider a different observable
than the fully inclusive Higgs
production is $b\bar{b}$, where the sum is only taken over the cross
sections for $(h,H)Z$ and $(h,H)W^\pm$ associated production. As
described above, 
for the inclusive $\tau^+\tau^-$ channels we consider the contribution
of both $H$ 
and $A$ when these are close in mass. To make it clear when this is the
case, we denote the joint (inclusive) rate as
$R^{H/A}_{\tau\tau}$. We also define a common rate for vector boson final states 
$R_{VV} : = R_{WW} = R_{ZZ}$. To keep things simple, we do not include
the experimental efficiencies for the $\gamma\gamma$ channel in  \refeq{eq:Rx}, since the efficiencies are different for the two
experiments. These are however used for the different predictions entering
the fit (as described in \refeq{eq:mu}). 
\figref{fig:totchi2h_rates} allows to investigate the best fit rates in
some more detail (subject to the approximations already
discussed). Uncertainty intervals can be extracted from the range with
$\Delta\chi^2_h < 1$ (corresponding to $68\%$ confidence intervals in the
Gaussian case). The results for the complete fit are  
\begin{align}
R^h_{VV}= 0.99^{+0.22}_{-0.02},\quad 
R^h_{\gamma\gamma}= 1.42^{+0.12}_{-0.38},\quad 
R^h_{bb}= 0.98^{+0.03}_{-0.10},\quad 
R^h_{\tau\tau}= 0.98^{+0.01}_{-0.94}.
\end{align}
For $R^h_{\tau\tau}$ we observe a distribution which is very flat near
the minimum. This indicates a low sensitivity in the fit to constraints
from $\tau^+\tau^-$ final states, and it permits substantially reduced
$\tau^+\tau^-$ rates at a very low additional $\chi^2$
contribution. The light Higgs case could therefore easily explain strongly
reduced $\tau^+\tau^-$ rates, although this is not visible in the best-fit
point.

Results for the heavy Higgs case are shown in
\figref{fig:totchi2HH_rates}. The resulting $\Delta\chi^2_H$
distributions for individual $R_{X}$ are similar to those for $\Delta
\chi^2_h$, except for $R_{\tau\tau}$, where the additional
contribution from the $A$~boson strongly enhances this quantity over
the corresponding result in the light Higgs case.
Extracting the results for the minimal $\chi^2$ in the same
way as for the light Higgs case, we obtain for the complete fit
\begin{align}
R^H_{VV} &= 1.25^{+0.30}_{-0.07},\quad 
R^H_{\gamma\gamma}= 1.10^{+0.18}_{-0.06},\quad 
R^H_{bb}= 0.83^{+0.05}_{-0.12},\quad  
R^{H/A}_{\tau\tau}= 2.54^{+0.31}_{-0.17}.
\end{align}
An interesting behaviour can be observed in the results leaving out
the Tevatron data, as shown in the lower row of
\reffi{fig:totchi2HH_rates}. In that case the best fit points have a
very small value for $R_{bb}^H$. More sampling would be
needed to cover this particular case in detail. Since we focus in the
following on results of the combined fit, we leave this for a dedicated
study. 

\begin{figure}
\centering
\includegraphics[width=0.24\columnwidth]{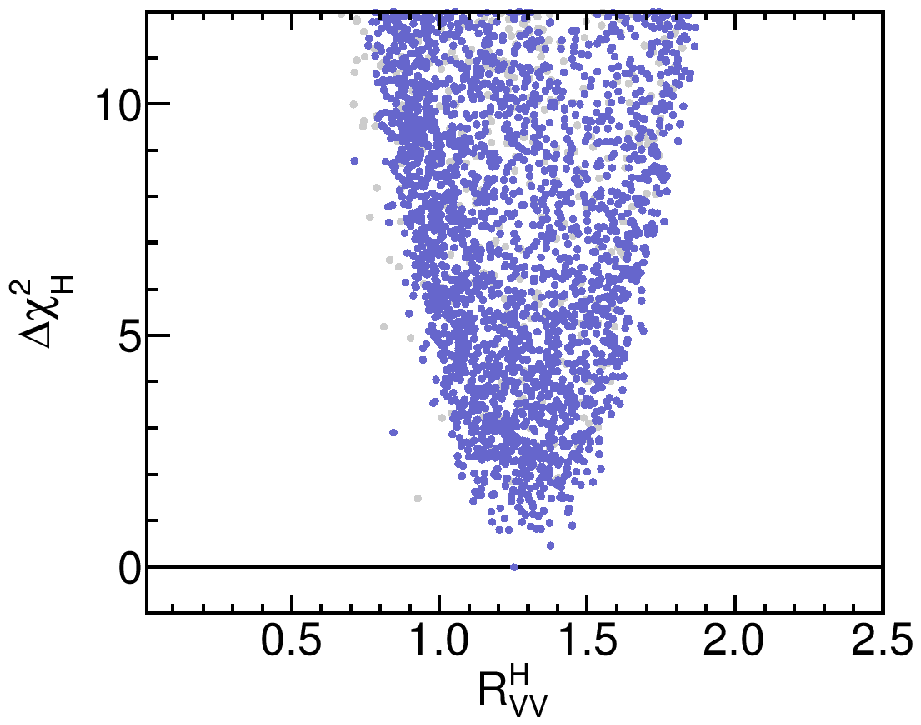}
\includegraphics[width=0.24\columnwidth]{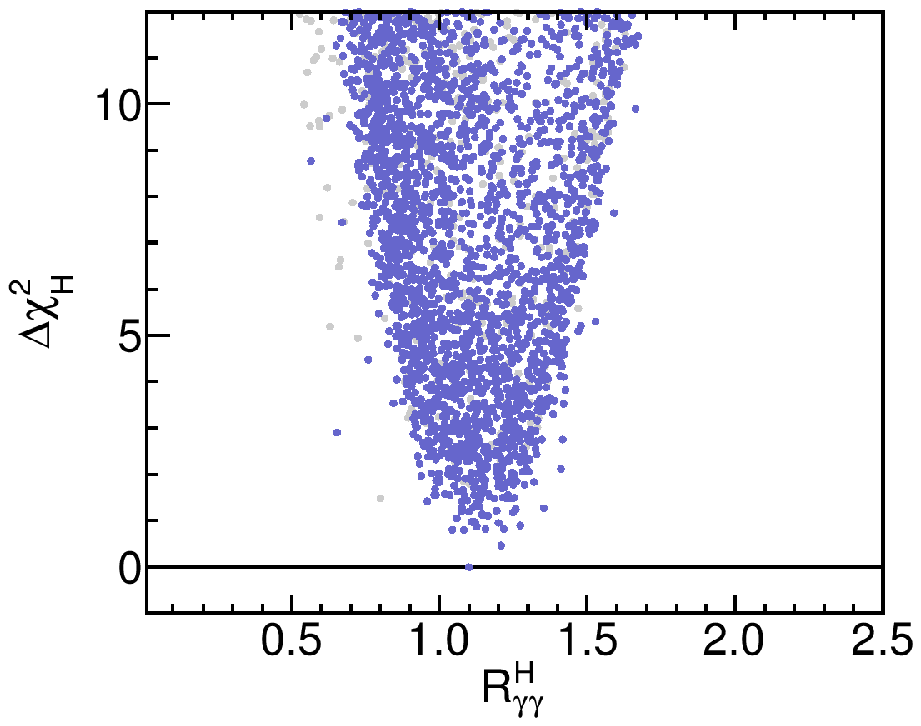}
\includegraphics[width=0.24\columnwidth]{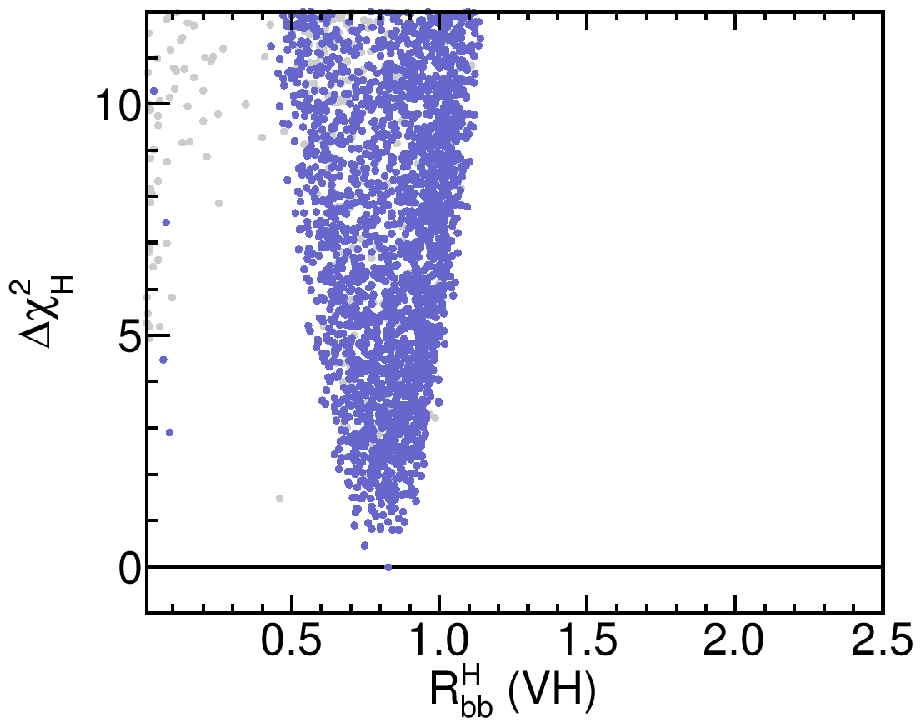}
\includegraphics[width=0.24\columnwidth]{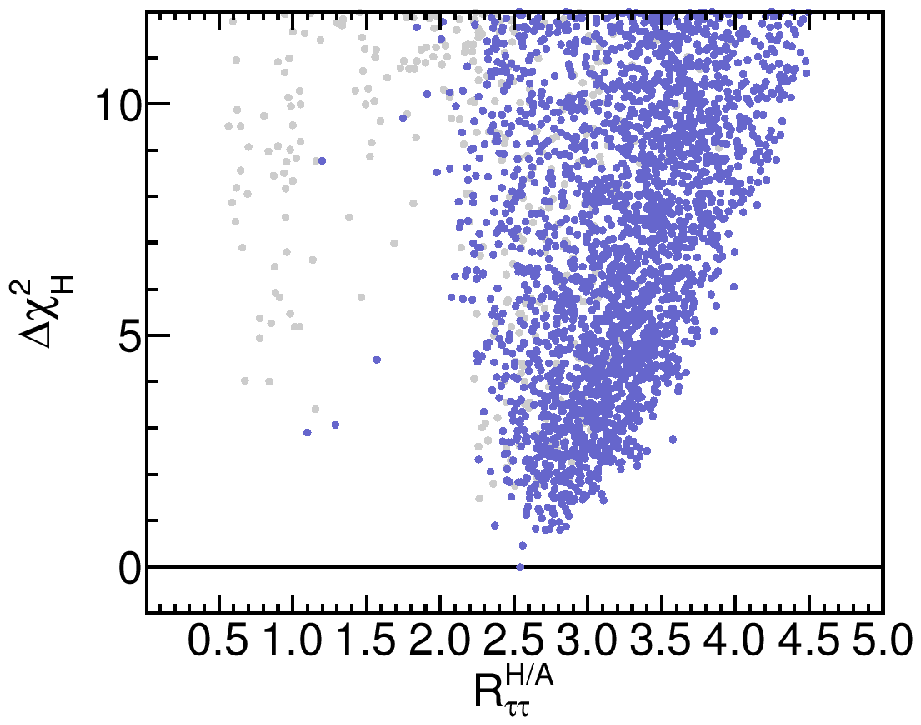}
\includegraphics[width=0.24\columnwidth]{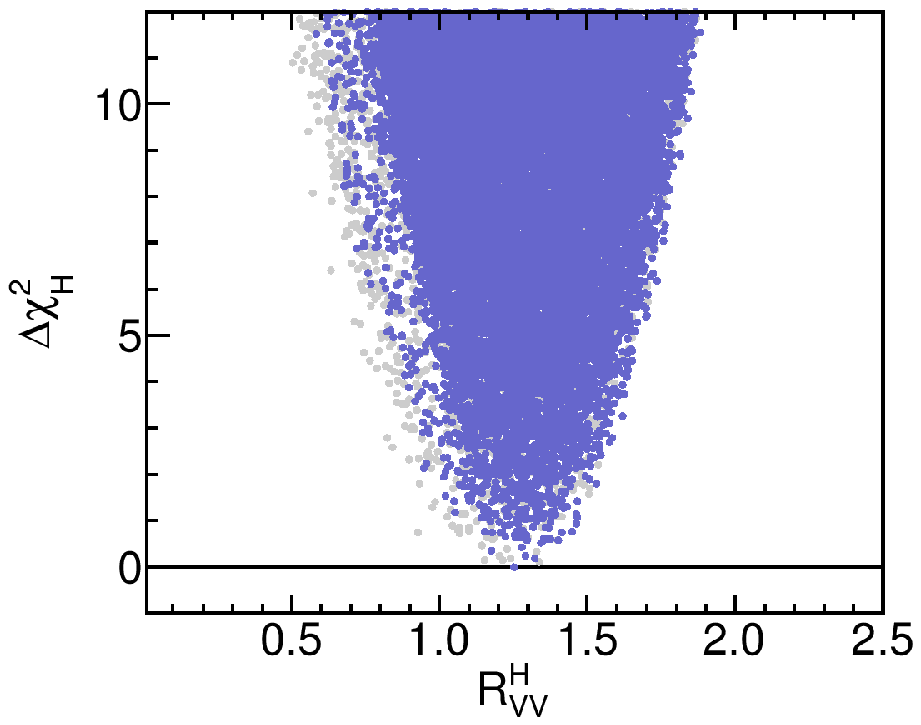}
\includegraphics[width=0.24\columnwidth]{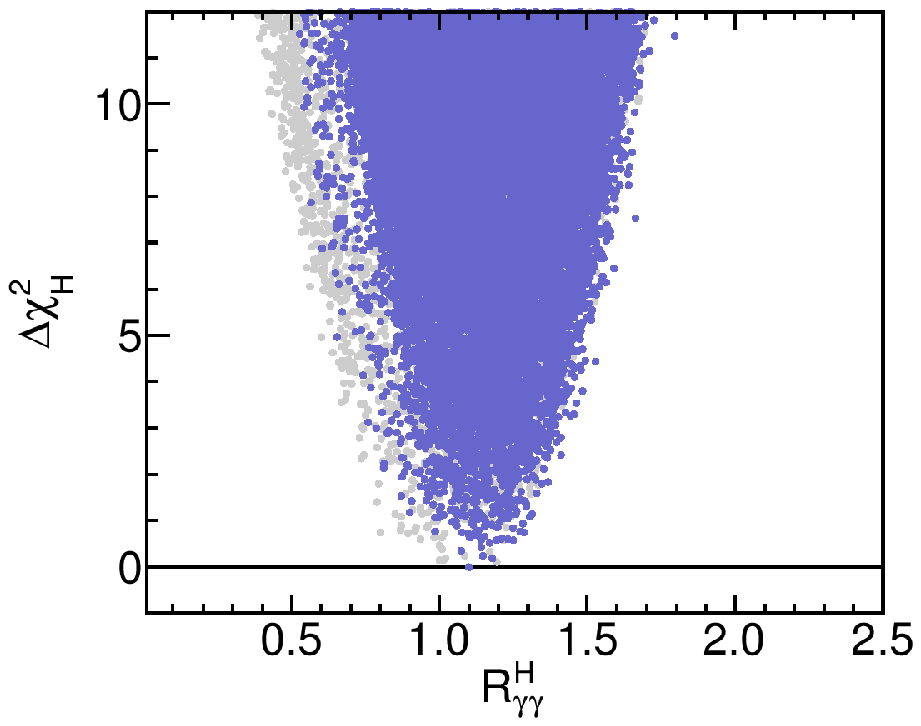}
\includegraphics[width=0.24\columnwidth]{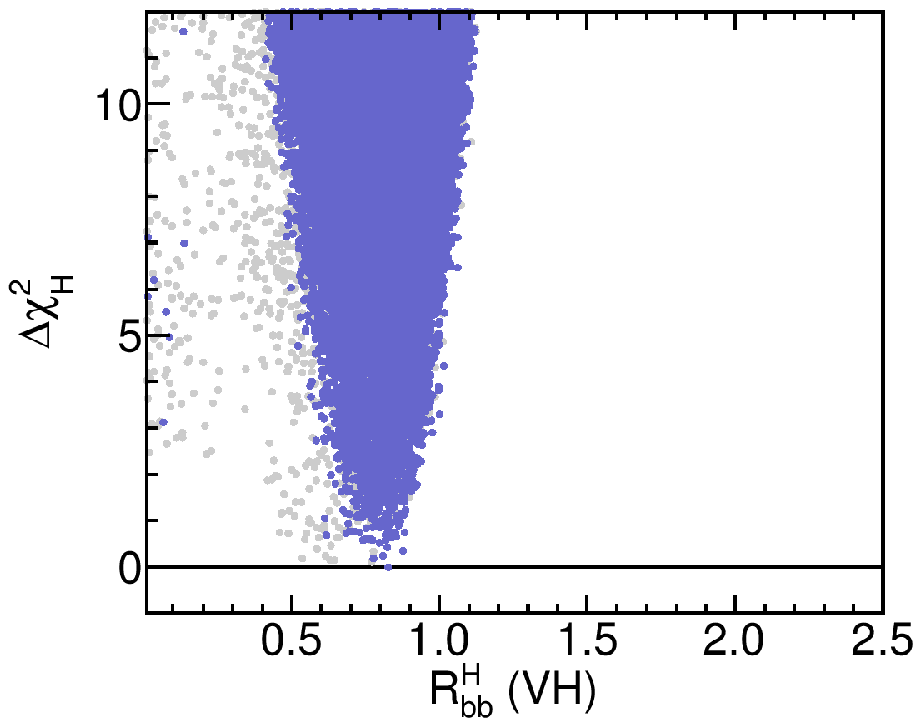}
\includegraphics[width=0.24\columnwidth]{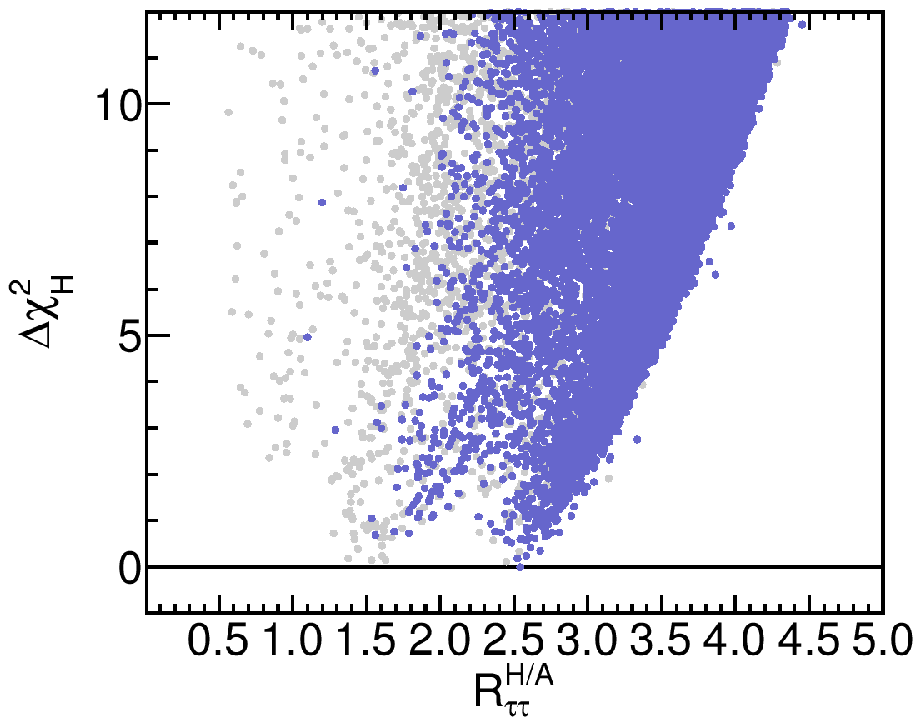}
\includegraphics[width=0.24\columnwidth]{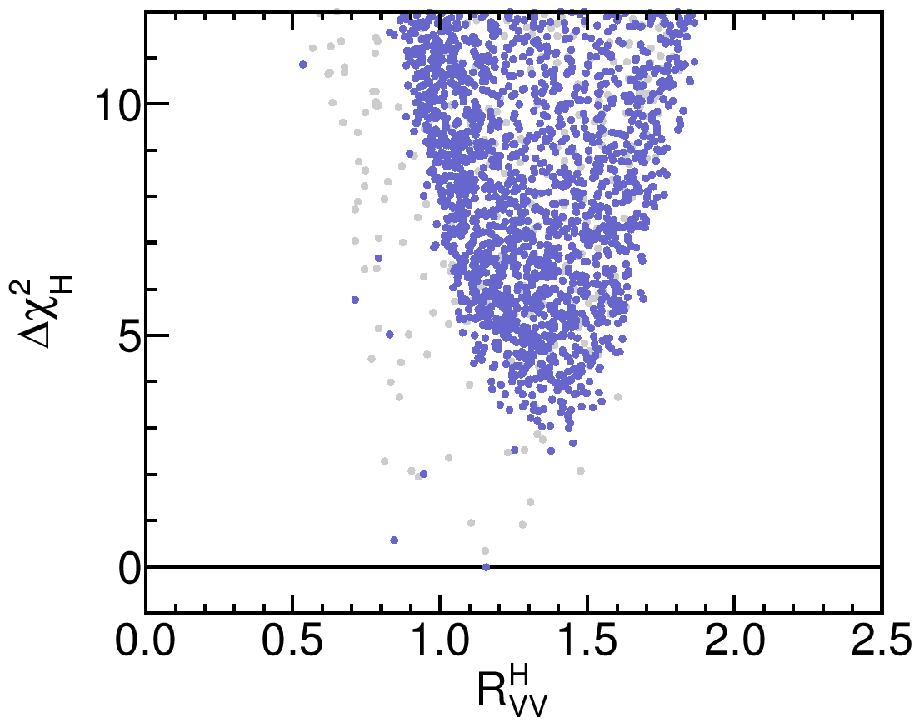}
\includegraphics[width=0.24\columnwidth]{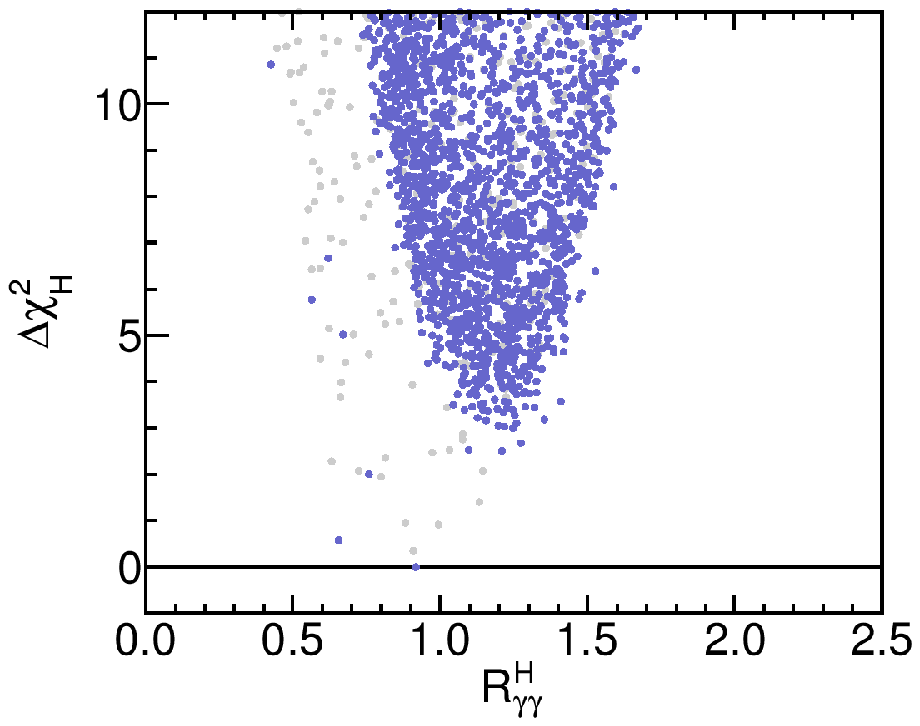}
\includegraphics[width=0.24\columnwidth]{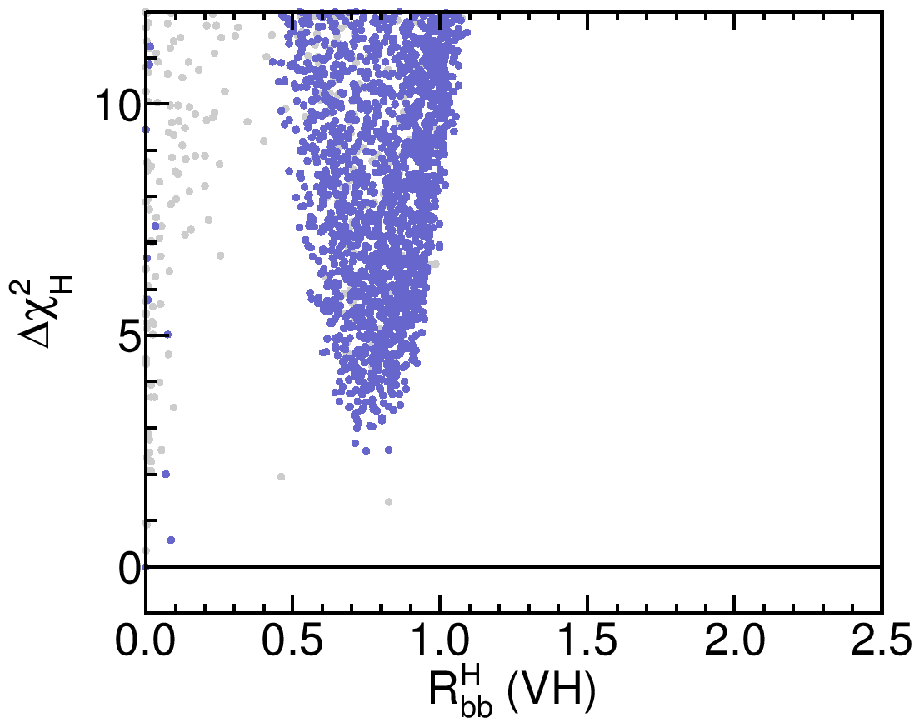}
\includegraphics[width=0.24\columnwidth]{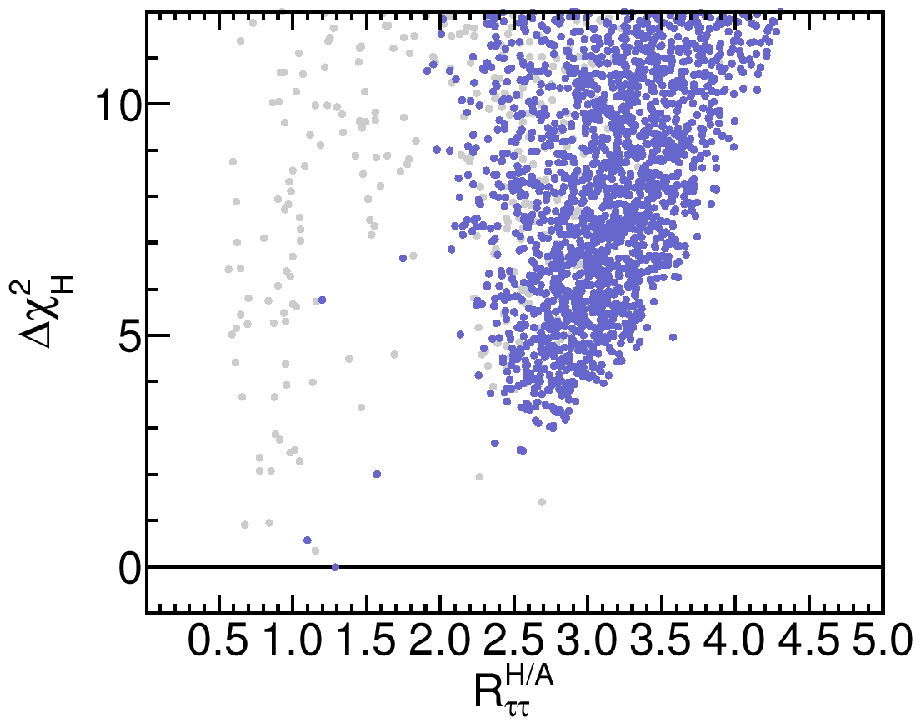}
\caption{$\Delta\chi^2_H$ versus the different signal rates (defined in
  the text) for the heavy Higgs case. Colour coding the same as in
  \figref{fig:totchi2h_rates}. The three rows show the distributions
  (with the corresponding minimal $\chi^2$ subtracted) for the complete
  fit (upper row), excluding LEO (middle row), and excluding Tevatron data
  (lower row).} 
\label{fig:totchi2HH_rates}
\end{figure}

More information about the phenomenology of the \pMSSM\ Higgs sector can
be found from the correlations between the different
rates. This is shown in \figref{fig:hrates_corr} for the light Higgs
case. Compared to the one-dimensional $\chi^2$ distributions of
\figref{fig:totchi2h_rates}, this 
figure introduces two new colours that are used in the following to show
regions close to the minimum 
$\chi^2$. We highlight points for which
$\Delta\chi^2_{h,H}<2.3$ (red) and $\Delta\chi^2_{h,H}<5.99$
(yellow). In the Gaussian limit these correspond to $68\%$ ($95\%$)
confidence regions with two degrees of freedom. 
We shall refer to these
points simply as the \emph{favoured region/points}, or sometimes
\emph{most favoured region/points} when $\Delta\chi^2_{h,H}<2.3$ is
discussed. Here (and in all figures from here on)
we refer to the $\chi^2$ of the complete fit, including LHC, Tevatron
and LEO. The best fit point is indicated in the figures by a black star.
\begin{figure}
\includegraphics[width=0.33\columnwidth]{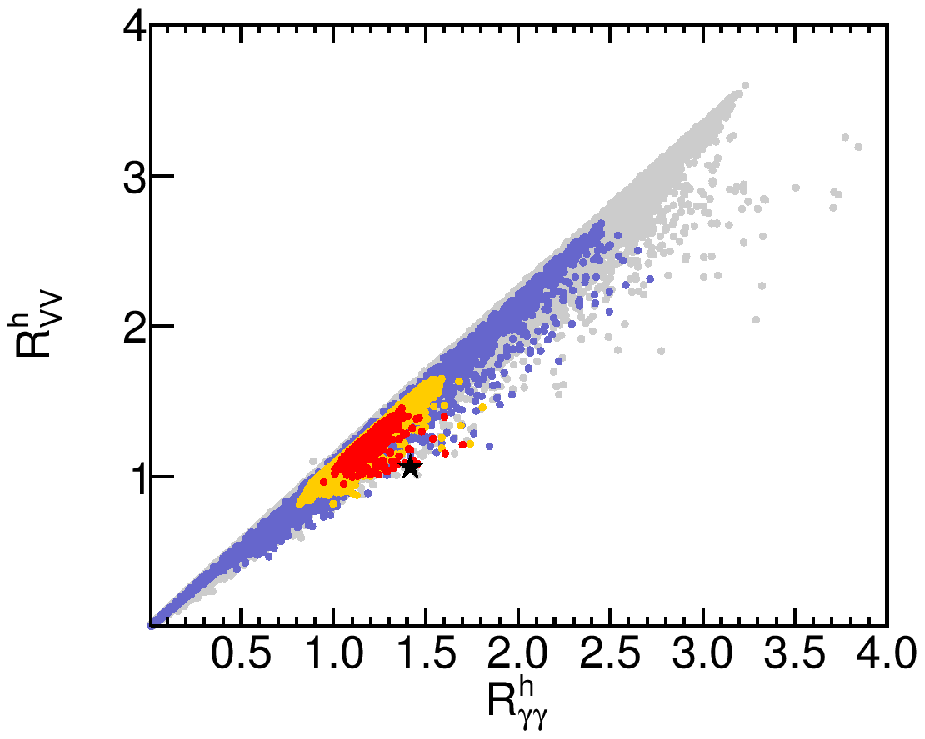}
\includegraphics[width=0.33\columnwidth]{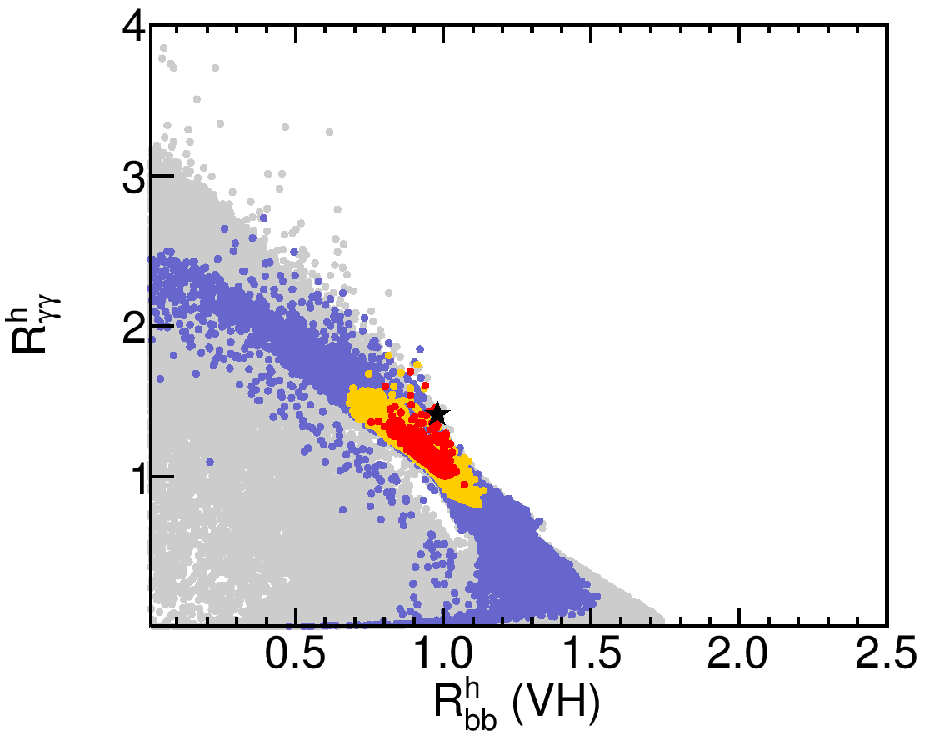}
\includegraphics[width=0.33\columnwidth]{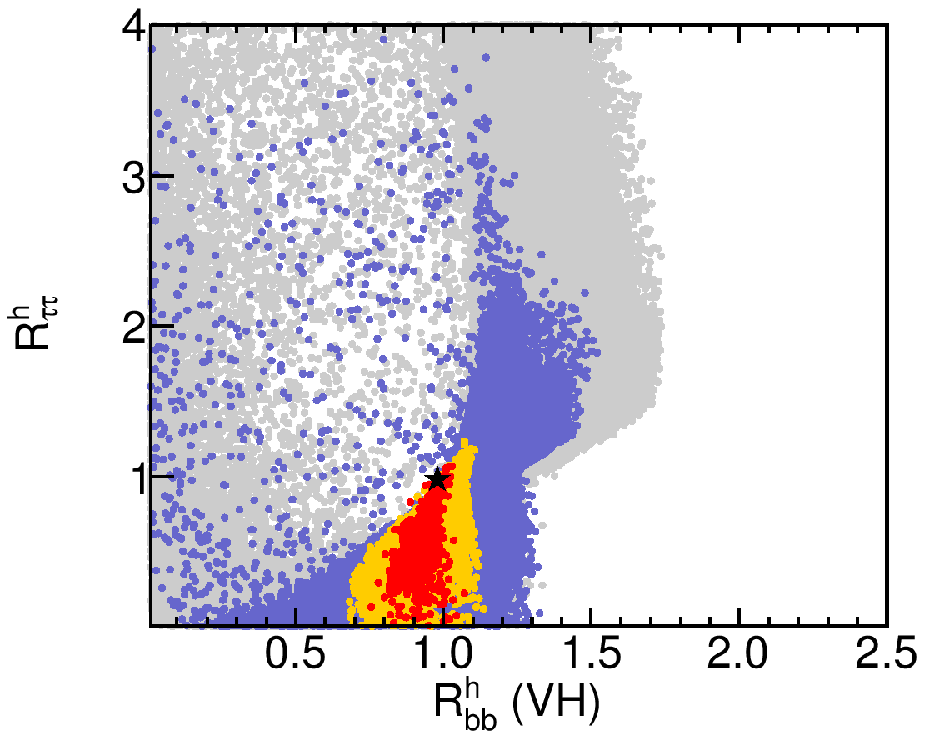}
\caption{Correlations between signal rates for the
  light Higgs case. The colour coding follows that of
  \figref{fig:totchi2h_rates}, with the addition of the favoured regions
  with $\Delta \chi_h^2 < 2.3$ (red) and $\Delta \chi_h^2 < 5.99$
  (yellow). The best fit point is indicated by a black star.} 
\label{fig:hrates_corr}
\end{figure}

\begin{figure}[t]
\includegraphics[width=0.33\columnwidth]{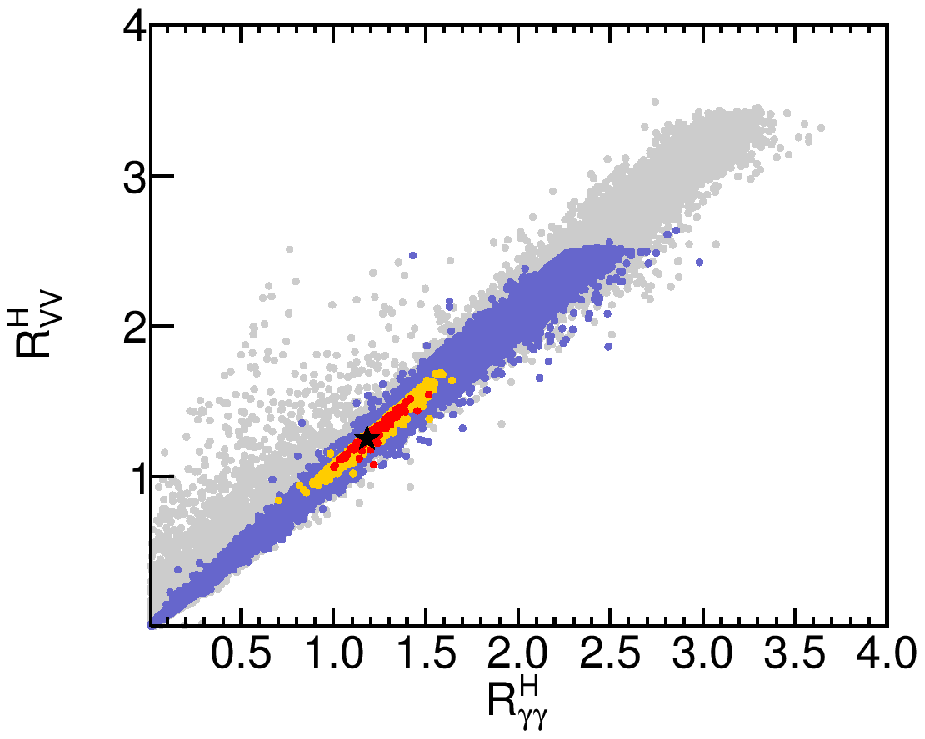}
\includegraphics[width=0.33\columnwidth]{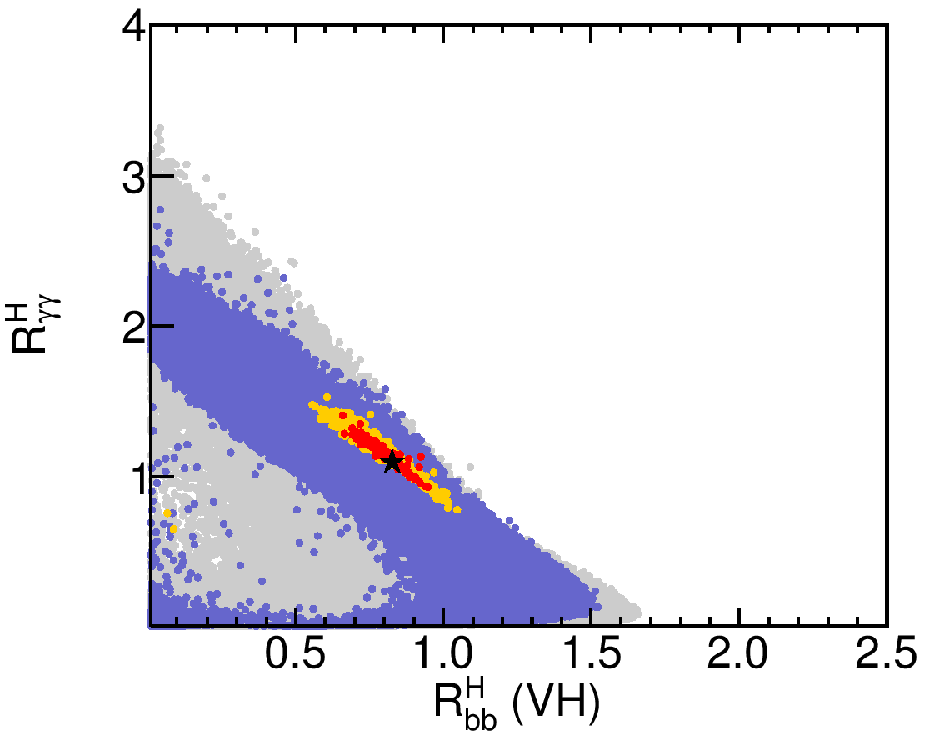}
\includegraphics[width=0.33\columnwidth]{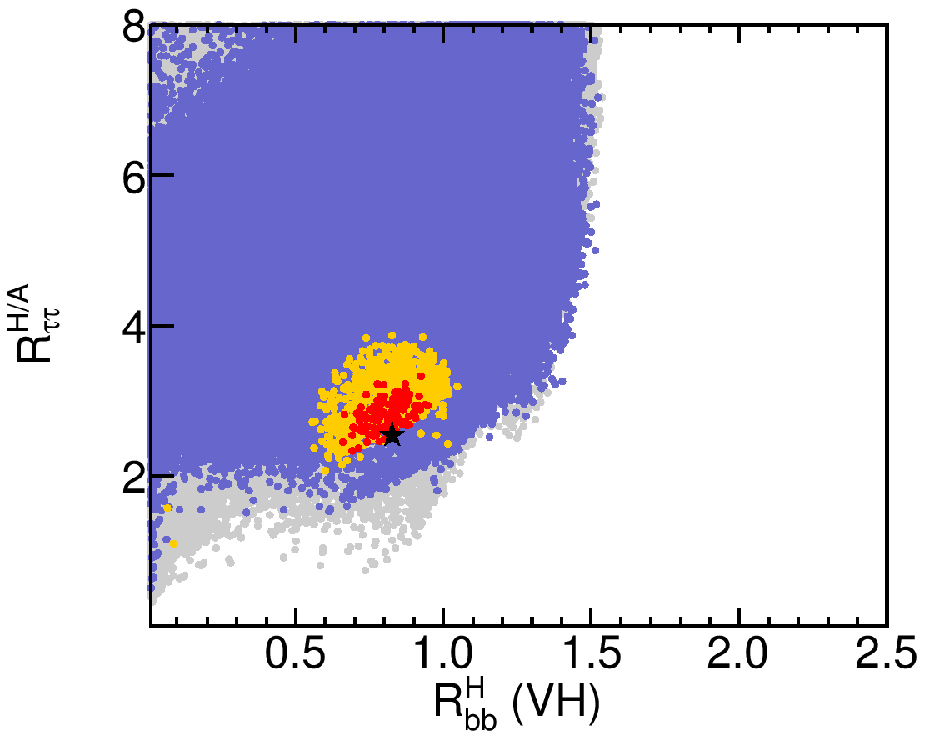}
\caption{Correlations between signal rates in the heavy Higgs
  case. Colours similar to \figref{fig:hrates_corr}, but here
  representing $\Delta \chi_H^2 < 2.3$ (red) and 
  $\Delta \chi_H^2 < 5.99$ (yellow). The black star indicates the best
  fit point for the heavy Higgs case.}  
\label{fig:HHrates_corr}
\end{figure}

The left plot of \figref{fig:hrates_corr} shows the strong, positive,
correlation between $R^h_{VV}$ and $R^h_{\ga\ga}$. In most of the
viable parameter space we find $R^h_{\ga\ga} > R^h_{VV}$. The favoured
region contains points with fully correlated rates in the interval $0.9
\lesssim R^h_{\ga\ga,VV}\lesssim 1.6$, but also solutions with lower degree of
correlation, where a $\ga\ga$ enhancement (up to
$R^h_{\ga\ga}\sim 1.8$) is accompanied by a much smaller (or no) enhancement
of $R^h_{VV}$. In the second plot of \figref{fig:hrates_corr} we
compare the results of $R^h_{\ga\ga}$ and $R^h_{bb}$ (we remind the reader
that the latter rate is calculated using the $VH$ production mode
only). We find an anticorrelation between these two rates. 
This can be understood from the
fact that the $h,H\to b\bar{b}$ decay gives the largest contribution to
the total width for a Higgs boson in this mass range, both in the SM and
(typically) also in the MSSM. A reduction of the $h,H\to b\bar{b}$
partial width is therefore effectively a reduction of the total decay
width, which leads to a simultaneous enhancement of the branching ratios
into the subdominant final states. This has recently been pointed out
\cite{Carena:2012gp,*Carena:2011aa,Benbrik:2012rm} as an important
mechanism to enhance the $\ga\ga$ rate in 
the MSSM. We shall see below how these effects on the Higgs decay widths
affect the parameters in our global fit. The third (right) plot in
\figref{fig:hrates_corr} shows the weak correlation of $R^h_{\tau\tau}$
to $R^h_{bb}$, where in principle any value of $R^h_{\tau\tau}<1$ is
found in the favoured region for $R^h_{bb}\lesssim 1$. Consequently, it
is possible to find a strong reduction of the $\tau^+\tau^-$ mode while
maintaining a SM-like $b \bar b$ rate.

Turning to \figref{fig:HHrates_corr}, we show the rate correlations for
the heavy Higgs case. Similar trends as in the light Higgs case
are visible in the heavy 
Higgs data, with the notable difference in the $\tau^+\tau^-$ rate,
mainly due to the inclusion of the contribution from the $\cp$-odd Higgs
$A$. The favoured regions are found at values for $R_{\tau\tau}^{H/A}$
between~$2$ and~$4$, while $R_{bb}^H$ remains below~$1$.

\figref{fig:effcoup} shows the effective couplings to vector bosons and
gluons%
\footnote{
The effective coupling to gluons is defined via the ratio of
the MSSM and SM decay widths to gluons. The relative vector boson
coupling is obtained from the inclusion of higher-order corrections into
the $\cp$-even Higgs mixing angle, $\al$.} of the MSSM Higgs assigned to
the LHC signal for both the 
light Higgs case (left plot) and heavy Higgs case (right plot). In both
cases the favoured regions have effective Higgs couplings to vector
bosons very close to the SM value ($g^2_{(h,H)VV} = 1$), and the light
(heavy) $\cp$-even Higgs boson behaves SM-like in this production (or
decay) mode (which reflects the general fact that the $VVh$ or
  $VVH$ coupling cannot be larger than the corresponding SM coupling).%
~For the effective coupling to gluons, $g^2_{(h,H)gg}$, the somewhat larger
favoured range $0.7 \lesssim g^2_{(h,H)gg} \lesssim 1.1$ is
obtained. Consequently, 
an overall large enhancement of the Higgs production cross
sections is disfavoured by the fit. The observed (positive) deviations
from the SM in the $VV$ and $\ga\ga$ channels for the best fit points
can therefore be attributed mainly to changes in the decay branching
ratios of the corresponding Higgs states, rather than modifications of
the production rates. 

\begin{figure}
\centering
\includegraphics[width=0.45\columnwidth]{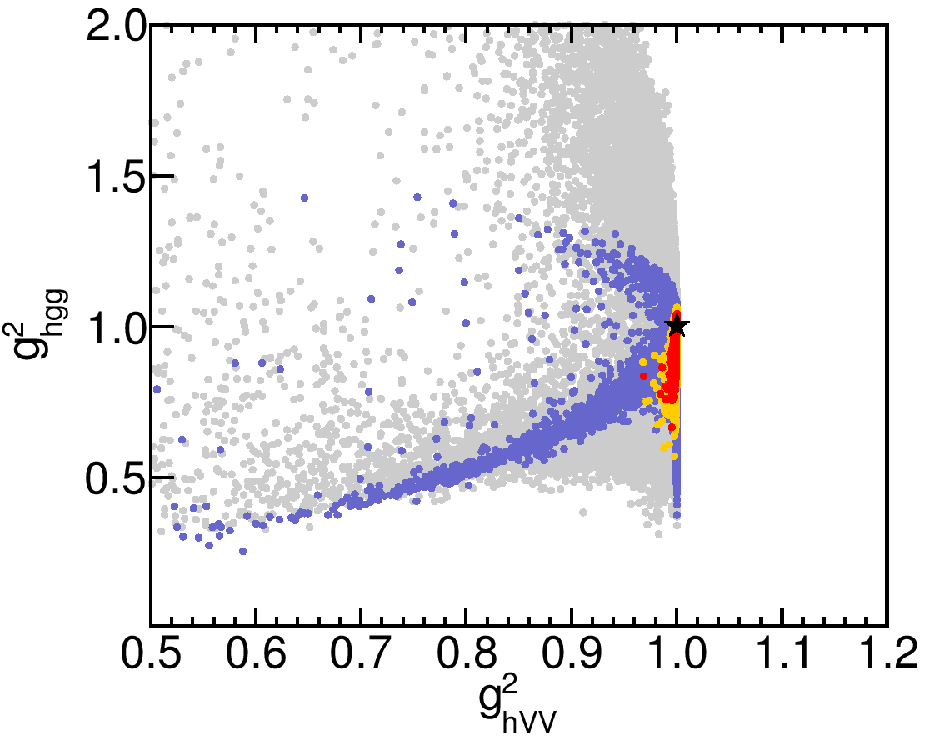}
\includegraphics[width=0.45\columnwidth]{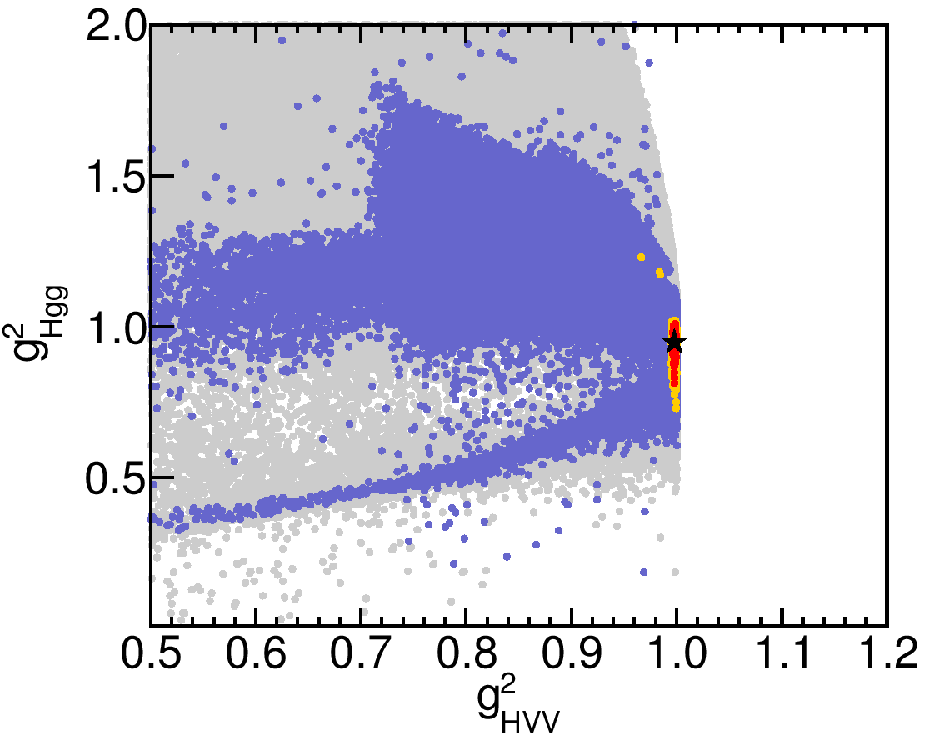}
\caption{Effective couplings to vector bosons and gluon pairs of the
  Higgs boson corresponding to the LHC signal in the light Higgs case
  (left) and heavy Higgs case (right). The colour coding follows that of
  \figref{fig:hrates_corr}.} 
\label{fig:effcoup}
\end{figure}

\begin{figure}
\centering
\includegraphics[width=0.45\columnwidth]{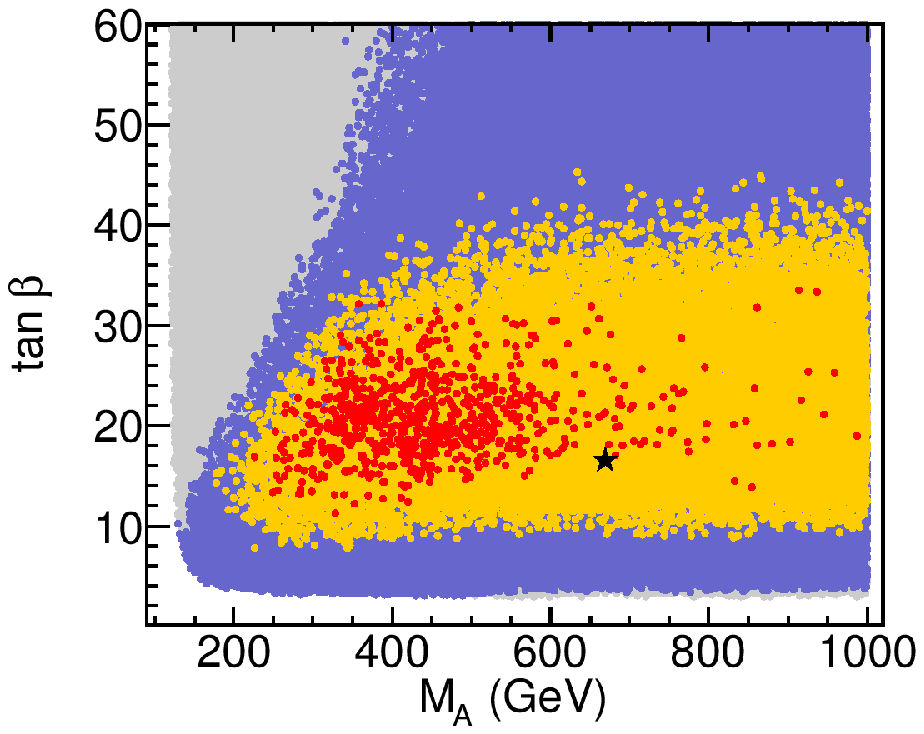}
\includegraphics[width=0.45\columnwidth]{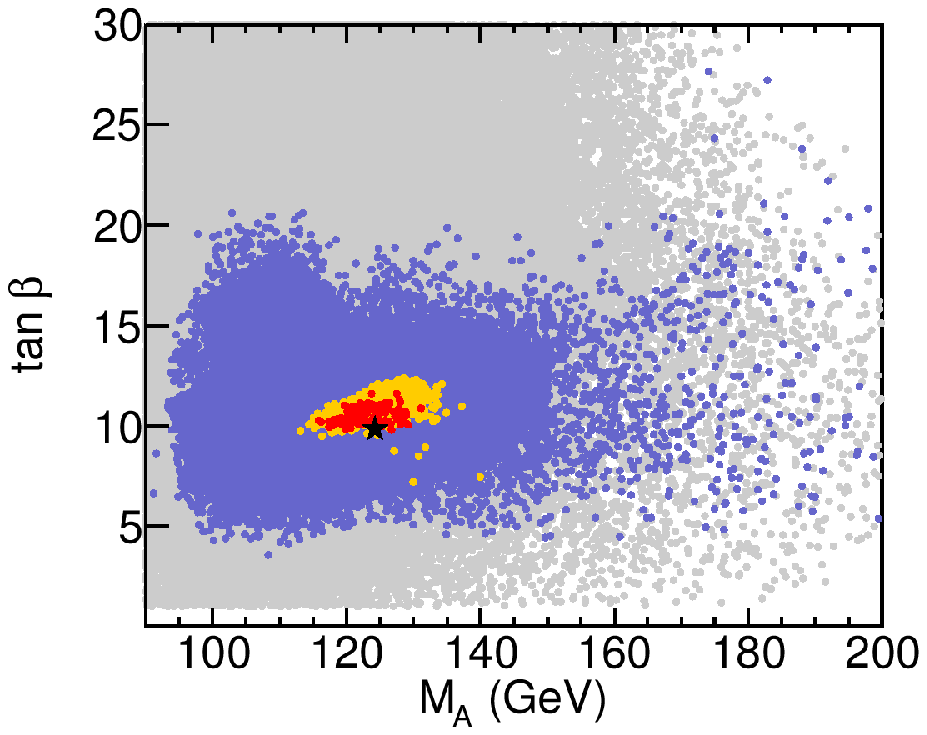}
\caption{Higgs sector tree-level parameters ($\MA$, $\tb$) in the light
  Higgs case (left) and in the heavy Higgs case (right).}
\label{fig:h_TB_MA0}
\end{figure} 

We now briefly discuss what mechanisms can
alter the branching ratios in the manner observed, and what the
consequences are for the favoured regions of MSSM parameter space. 
In \figref{fig:h_TB_MA0}~we show the scan results in the plane of the
Higgs sector tree-level parameters $(\MA,\tb)$, where the results for
the light (heavy) Higgs 
case are shown in the left (right) plot.
In the light Higgs case one can note 
the region at low $\MA$, high $\tb$, which is excluded by direct
MSSM Higgs searches (mainly $H/A\to \tau^+\tau^-$). The excluded region
appears smaller in this plane than the corresponding results published
by the experiments \cite{ATLASMSSMHiggs,*CMSMSSMHiggs}, since their
results are shown only for one particular benchmark scenario (the
so-called $\mhmax$ scenario~\cite{Carena:2002qg}). In an inclusive scan
of the \pMSSM\ parameter space, points  are found where higher order
corrections to the bottom Yukawa coupling 
lead to suppressed production rates for the heavy MSSM Higgs bosons, and
a larger fraction of the parameter space in the $(\MA,\tb)$ plane
therefore opens
up (see the analyses in~\cite{Carena:2005ek,Gennai:2007ys}). 
Sizable branching ratios of $H/A$ to SUSY particles also reduce the
sensitivity of the searches in the $\tau^+\tau^-$ final state.
We see that the regions of very high $\tb \gtrsim 40$, and
also low $\tb \lesssim 8$, are disfavoured by the fit. At high $\tb$
this results from a 
poor fit to $(g-2)_\mu$ and flavour observables, whereas for low $\tb$
the fit to the LHC Higgs observables becomes worse. For
low $\tb$ it also becomes increasingly difficult to fit the relatively
high Higgs mass value ($\MHexp \gev$), although viable solutions down to
$\tb\sim 4$ can be found~\cite{Heinemeyer:2011aa}. Low values of $\MA$ are 
disfavoured by the fit results in the light Higgs case, with the
preferred
region starting at $\MA\gtrsim170\gev$ 
(and the most favoured region at $\MA\gtrsim 230\gev$). Taking the rate
information into account therefore suggests somewhat 
higher mass scales for the MSSM Higgs sector than what is required by
the $\MHhat \sim \MHexp \gev$ Higgs mass measurement alone
\cite{Heinemeyer:2011aa}. 
For the light Higgs case
the lower limits on $\MA$ in the favoured
regions of the fit exclude the possibility of $\MHp < \mt$, where the
charged Higgs can be produced in the decay of the top quark.
On the other hand, the region favoured by the
fit does not show any upper limit for $\MA$, which demonstrates that the
decoupling limit (corresponding to $\MA \gg \MZ$, where the MSSM Higgs 
sector reproduces the predictions for a SM Higgs)
remains a possible scenario. This is to be expected,
given the high quality of the SM fit to the LHC data. 

For the heavy Higgs case, as shown in the right plot of 
\figref{fig:h_TB_MA0}, the situation is very different. Low values
for $\MA$ are preferred, and the favoured region in $(\MA, \tb)$ is much
smaller than for the light Higgs case: $110\gev \lesssim \MA \lesssim
140\gev$ and $7\lesssim\tb\lesssim 13$. Even though the
$H$ can be very
SM-like in this scenario, this situation is very different from the
decoupling limit in the light Higgs case since it implies that all
five MSSM Higgs bosons are light. 
In contrast to the light Higgs case, in this scenario values of the
charged Higgs boson mass only below the top mass ($\MHp < \mt$) are
found, which may offer good prospects for the searches for charged Higgs
bosons in top quark decays. We
therefore show in \figref{fig:HH_MHp_BRtHpb} the results for
$\br(t\to bH^+)$ as a function of $\MHp$. The current upper limit
on this decay mode is 
\order{1\%}~\cite{Aad:2012tj,*CMSHp}, which is close to the maximal
value favoured by 
the fit. With more integrated luminosity, charged Higgs
searches will therefore have an interesting sensitivity to probe
the heavy Higgs scenario at the LHC in the near future.
\begin{figure} 
\centering
\includegraphics[width=0.43\columnwidth]{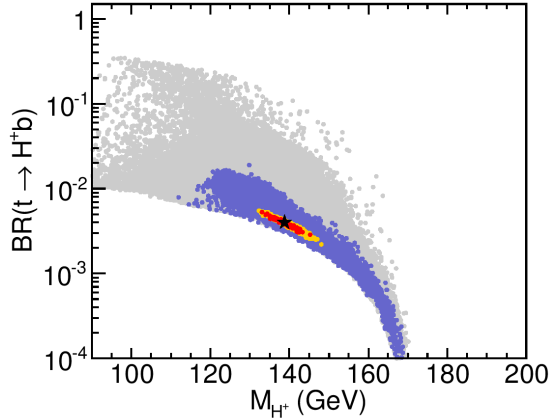}
\caption{Branching ratio of the top quark into a charged Higgs boson and
  a bottom quark in the heavy Higgs case.}  
\label{fig:HH_MHp_BRtHpb}
\end{figure} 

\begin{figure}
\centering
\includegraphics[width=0.45\columnwidth]{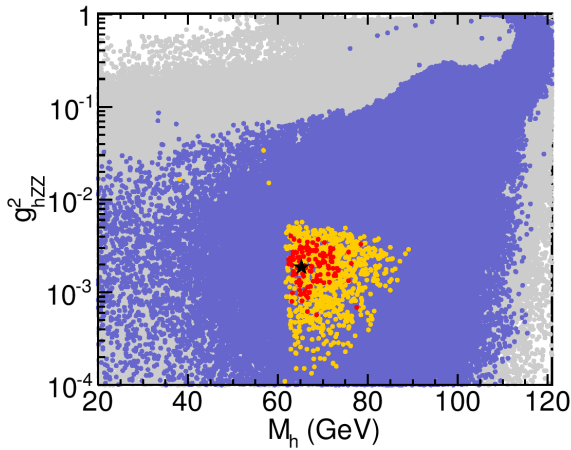}
\caption{Effective coupling squared $g^2_{hZZ}$ of the lightest MSSM Higgs boson
  to a $Z$ boson pair, as a function of the lightest Higgs mass $\Mh$
  in the heavy Higgs case ($\MH\sim \MHexp\gev$).}  
\label{fig:Mh_g2hZZ}
\end{figure} 

While in the heavy Higgs scenario the low preferred values for
$\MA$ typically lead to a situation where $H$, $A$, and $H^\pm$ are
rather close in mass, the lightest Higgs boson, $h$,
can have a significantly lower mass, as illustrated in
\figref{fig:Mh_g2hZZ}. As we see from this figure, points with $\Mh<
90\gev$ have a very small effective coupling to vector bosons,
$g^2_{hZZ}\ll 1$, which explains why such light Higgs bosons are
compatible with the Higgs search limits from LEP.
The bulk of the favoured region is found for $60 \gev \lsim \Mh \lsim 90 \gev$,
with an effective coupling squared to vector bosons at the
sub-percent level. Another feature which is clearly visible in the {\tt
HiggsBounds} allowed points (blue) is the degradation of the limit 
around $\Mh \sim 98 \gev$, which was caused by a slight excess of 
events observed at LEP in that mass region. While a
scenario with $\MH \sim \MHexp$ and $\Mh \sim 98 \gev$ is certainly
possible (see also \cite{Heinemeyer:2011aa,Drees:2012fb}), it is clearly
not favoured by our rate analysis.

\begin{figure}
\centering
\includegraphics[width=0.45\columnwidth]{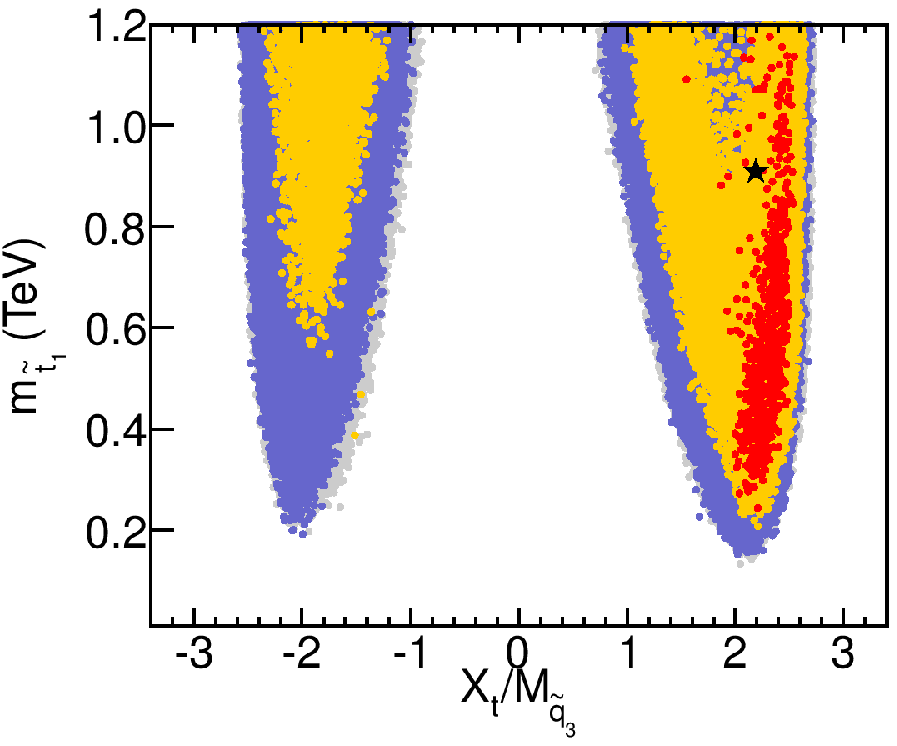}
\includegraphics[width=0.45\columnwidth]{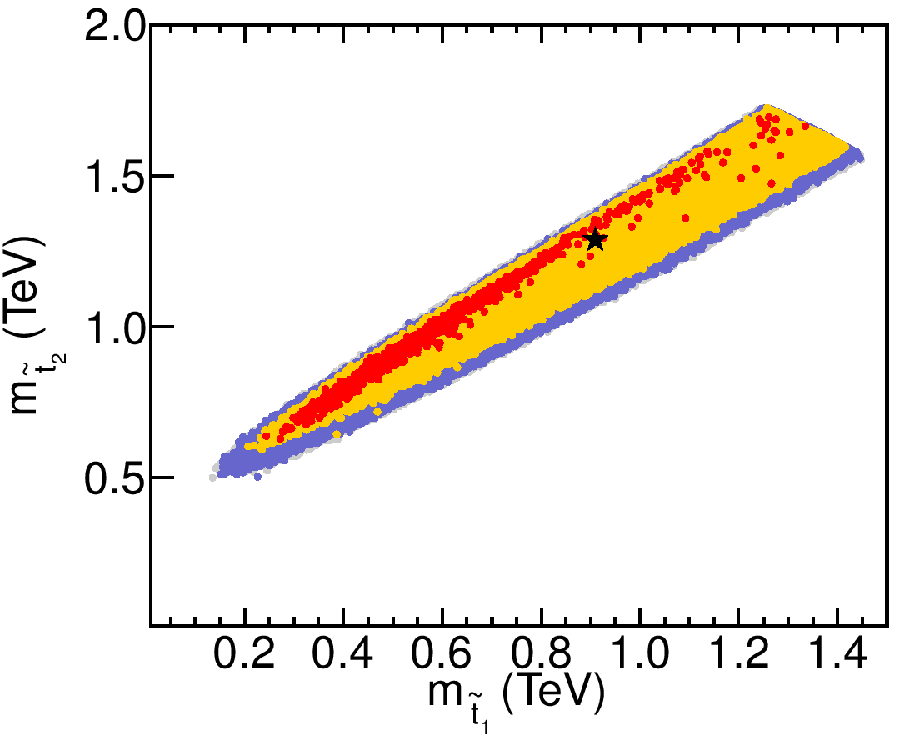}
\caption{Stop mixing parameter $\Xt/\msqd$ vs.\ the light stop mass
  (left), and the light vs.\ heavy stop masses (right) in the light
  Higgs case.} 
\label{fig:h_mstop}
\end{figure} 
\begin{figure}[htb!]
\centering
\includegraphics[width=0.45\columnwidth]{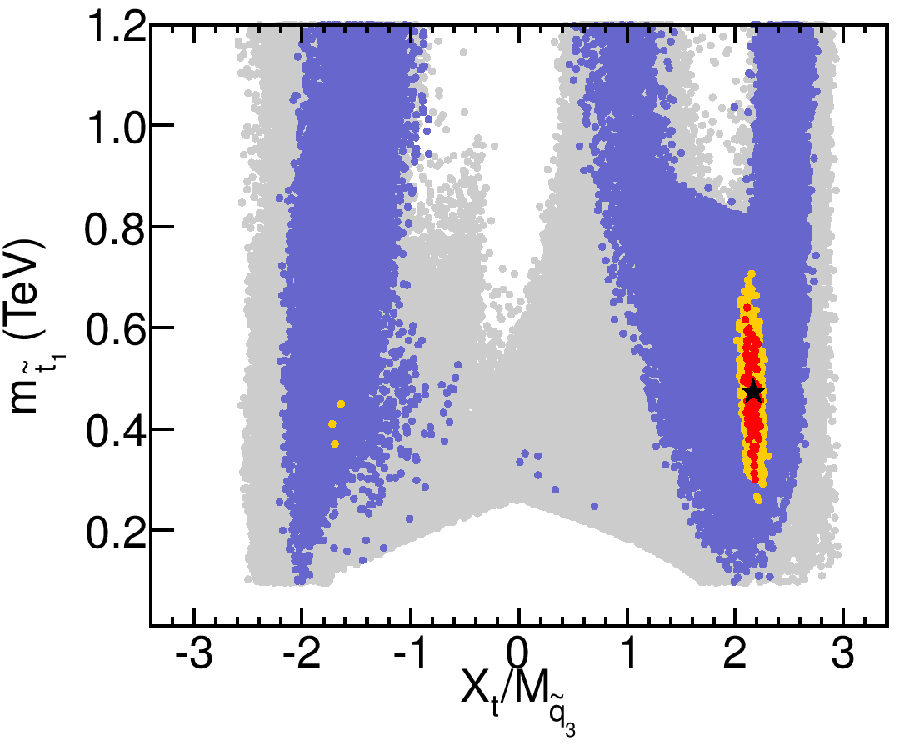}
\includegraphics[width=0.45\columnwidth]{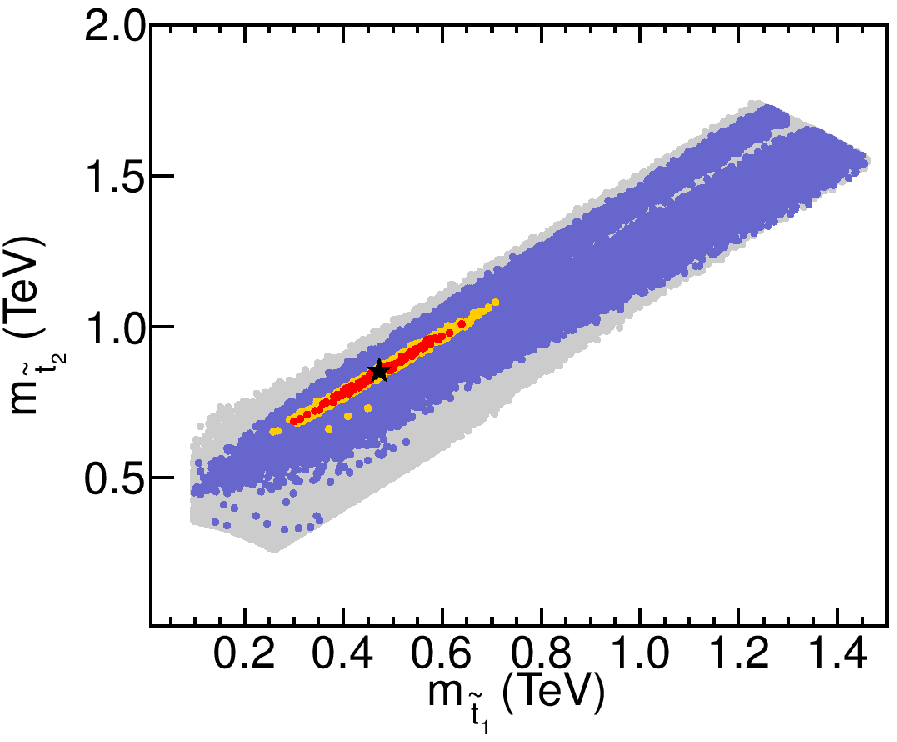}
\caption{Stop mixing parameter $\Xt/\msqd$ vs.\ the light stop mass
  (left), and the light vs.\ heavy stop masses (right) in the heavy
  Higgs case.} 
\label{fig:HH_mstop}
\end{figure}  
The most relevant parameters for higher-order corrections in the MSSM
Higgs sector are the soft SUSY-breaking parameters in the stop sector. 
As it was shown in \cite{Heinemeyer:2011aa}, light scalar top masses
down to $150 \gev$ are in agreement with a light $\cp$-even Higgs mass
around $\sim 125 \gev$, provided the mixing in the scalar top sector is
sufficiently strong. Here we show the corresponding results including
the rate analysis. In \figref{fig:h_mstop} we show 
$\Xt/\msqd$ vs.\ the light stop mass (left plot) and the light
vs.\ the heavy stop mass (right plot) in the light Higgs case.
In the left plot one can see that the case of zero stop mixing in
the MSSM is
excluded by the observation of a light Higgs at $\Mh\sim \MHexp \gev$ 
(unless $\msqd$ is extremely large), and
that values of $|\Xt/\msqd|$ between $\sim 1$ and $\sim 2.5$ must be
realised. For the most favoured region we find $\Xt/\msqd = 2 - 2.5$.
It should be noted here that large values of $|A_t|$ ($|A_t|\gsim
\sqrt{6}\,\msqd$) could potentially  
lead to charge and colour breaking 
minima~\cite{ccb1,*ccb2,*ccb3,*ccb4,*ccb5,*ccb6,*ccb7}. 
We checked that applying a cut at 
$|A_t|\gsim \sqrt{6}\,\msqd$ would still leave
most points of the favoured region.
Concerning the value of the lightest scalar top mass, the overall smallest
values are found at $\mste \sim 200 \gev$, in agreement with
\cite{Heinemeyer:2011aa}.%
\footnote{In this work a Higgs mass measurement of $\MHhat=125\gev$ was
assumed, whereas we now use the average mass $\MHhat=\MHexp\gev$.}%
~Even taking the rate information into
account, the (most) favoured values start at 
$\mste \gtrsim 200 \gev$ for
positive $\Xt$. Such a light $\Stope$ is accompanied by a somewhat
heavier $\Stopz$, as can be seen in the right plot of
\figref{fig:h_mstop}. Still, values of $\mste \sim 200 \gev$ are realised for 
$\mstz \sim 600 \gev$, which would mean that both stop masses are
rather light,
offering interesting possibilities for the LHC. The highest
favoured $\mste$ values we find are $\sim 1.4 \tev$. These are the
maximal values reached in our scan, but from \figref{fig:h_mstop} it
appears plausible that the favoured region extends to larger values of
both stop masses. Such a scenario would be extremely difficult to
access at the LHC. For the intrepretation of these results it is
important to remember that we have assumed a universal value for the
soft mass parameters in the scalar top and bottom sector. Relaxing this
assumption would potentially lead to larger regions of parameter space
in which all applied constraints can be satisfied.

The results for the scalar top masses in the heavy Higgs case look 
in principle similar to the light Higgs case, but with
substantially smaller favoured regions, which are
nearly solely realised for positive $\Xt$ with $\Xt/\msqd = 2$--$2.3$,
as can be seen in 
\figref{fig:HH_mstop}. 
The favoured values of $\mste$ range between $\sim 250 \gev$ and
$\sim 700 \gev$ in this case, whereas the preferred range of the 
heavy stop extends from 
$\mstz\sim 650 \gev$ to $\mstz\sim 1100 \gev$.
\begin{figure}
\centering
\includegraphics[width=0.45\columnwidth]{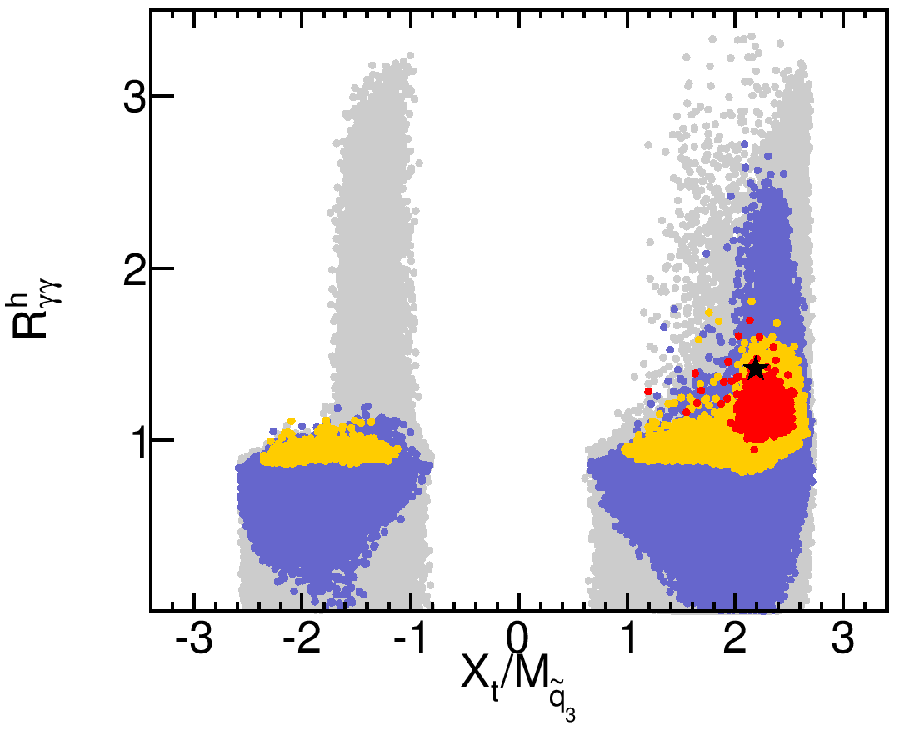}
\includegraphics[width=0.45\columnwidth]{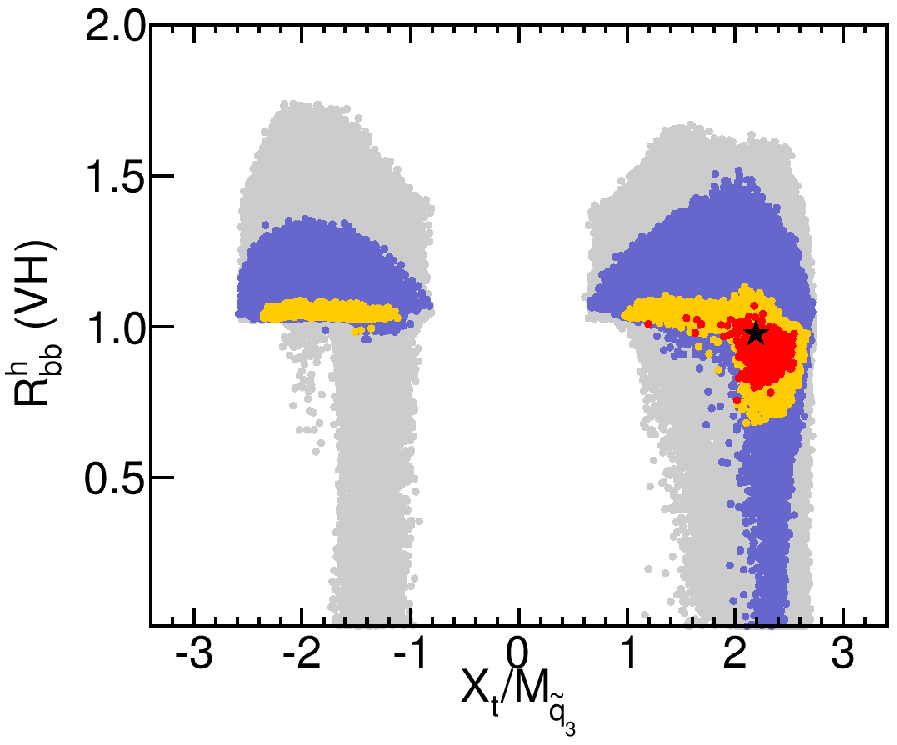}
\caption{Dependence of the rates $R^h_{\ga\ga}$ and $R^h_{bb}$
  (VH) on the stop mixing parameter $\Xt/\msqd$ for the light Higgs case.} 
\label{fig:h_Xt_Rgaga}
\end{figure} 
We now turn to the analysis of rates as a function of the underlying
MSSM parameters. This comparison allows to analyse the various
mechanisms that are responsible for the observed differences in the
decay rates with respect to the SM values. 
In \figref{fig:h_Xt_Rgaga} we analyse the correlation between the ratio
$\Xt/\msqd$ and $R_{\ga\ga}^h$ (left) or $R_{bb}^h$ (VH) (right) 
in the light Higgs case.
It can be seen that the enhancement in the $\ga\ga$ channel is only
substantial for $\Xt/\msqd \gsim 2$, where values of up to 
$R_{\ga\ga}^h \sim 1.7$ can be
reached in the favoured region. Such an enhancement can have two
sources: a suppression of $\Ga(h \to b \bar b)$, as the by far largest
contribution to the total width, or a direct enhancement of 
$\Ga(h \to \ga\ga)$. That the first mechanism is indeed responsible for a
substantial part of the scenarios with an enhancement of $R_{\ga\ga}^h$
can be seen in the right plot 
of \figref{fig:h_Xt_Rgaga}, which together with the middle plot of
\figref{fig:hrates_corr} illustrates that the enhancement in the $\ga\ga$
channel in the favoured regions is accompanied by some suppression of
the $b \bar b$ channel. This suppression/enhancement is realised for
large values of $\Xt/\msqd$.  

Such a suppression of the $b \bar b$ channel can happen in two different
ways. In the MSSM, the (effective) coupling $g_{hb\bar{b}}$ is given by,
\begin{equation}
\begin{split}
\frac{g_{hb\bar{b}}}{g_{\HSM b\bar{b}}} = \frac{1}{1+\De_b} \left(
-\frac{\sin \aeff}{\cos \beta}+\De_b \frac{\cos \aeff}{\sin
\beta}\right), 
\label{eq:hbbdeltamb}
\end{split}
\end{equation}
where $\alpha$ is the mixing angle in the $\cp$-even Higgs sector. 
Higher-order contributions from Higgs propagator corrections can
approximately be included via the introduction of an effective 
mixing angle, corresponding to the replacement  
$\al \to \aeff$~\cite{Heinemeyer:2000fa} (in our numerical analysis we treat
propagator-type corrections of the external Higgs bosons in a more
complete way, which is based on wave function normalisation factors that
form the $\matr{Z}$~matrix~\cite{Frank:2006yh}). 
A suppression of the $h\to b \bar b$ channel thus occurs for scenarios
with small $\aeff$. 
Furthermore, genuine corrections to the $hb \bar b$ vertex enter
\refeq{eq:hbbdeltamb} via the quantity
$\db \propto \mu\tb$~\cite{Hempfling:1993kv,*Hall:1993gn,*Carena:1994bv,Carena:1999py},\footnote{
The dominant contributions to $\db$ beyond one-loop order are the QCD
corrections, given 
in~\cite{Noth:2008tw}. Those two-loop
contributions are not included in our analysis, 
but their numerical effect is approximated by using a scale of 
$\mt$ for the evaluation of the
one-loop expression.}
  see~\cite{Benbrik:2012rm} for a more detailed discussion of the 
possible mechanisms giving rise to suppression of the $h\to b \bar b$
channel.
\begin{figure}
\centering
\includegraphics[width=0.45\columnwidth]{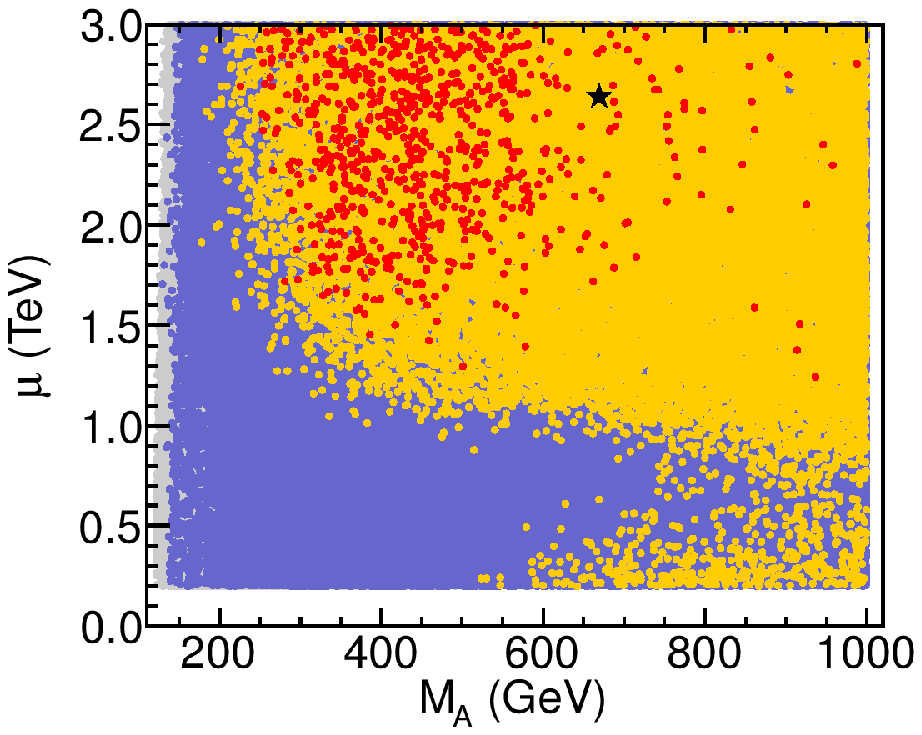}
\includegraphics[width=0.45\columnwidth]{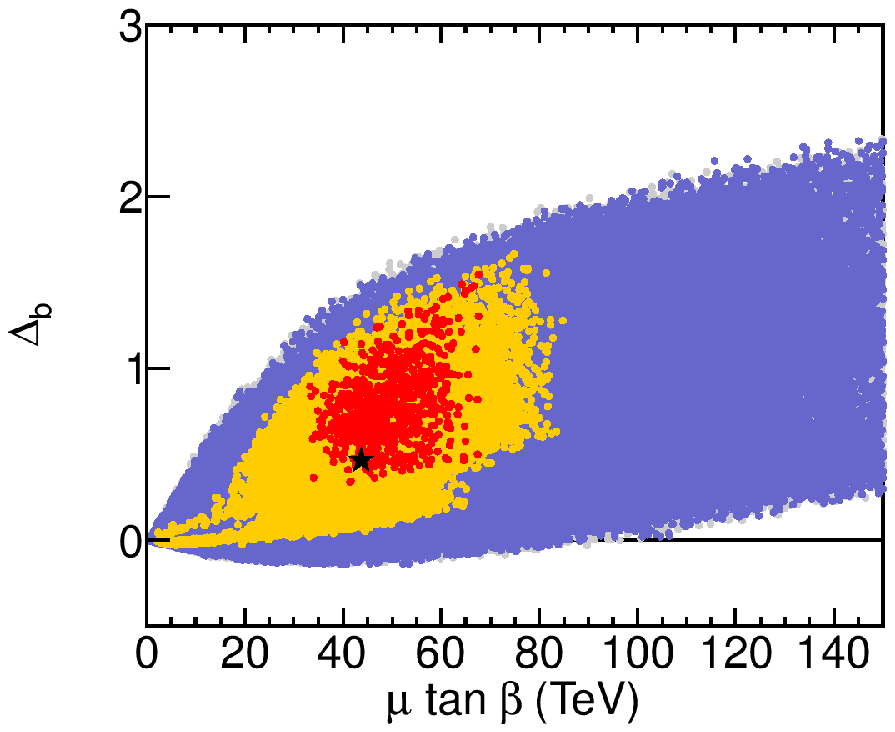}
\caption{Correlation of the $\mu$ parameter to the value of $\MA$ (left), 
and dependence of $\db$ corrections on $\mu\tb$
  (right), both in the light Higgs case.} 
\label{fig:h_Deltab}
\end{figure} 
\begin{figure}
\centering
\includegraphics[width=0.45\columnwidth]{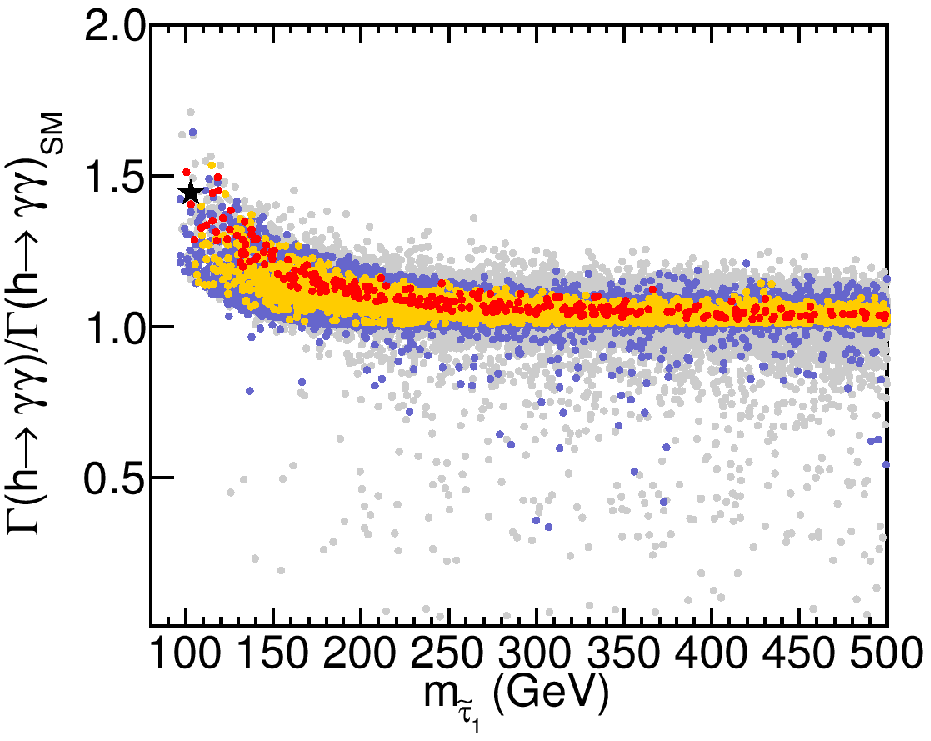}
\includegraphics[width=0.45\columnwidth]{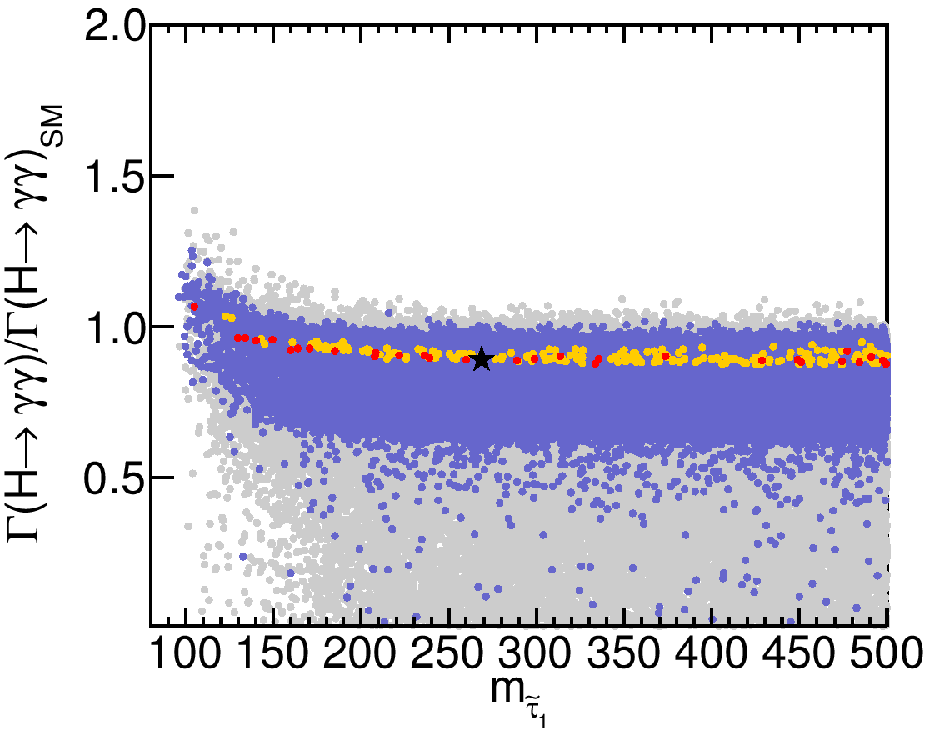}
\caption{Enhancement of the $h\to\ga\ga$ partial width in the presence
  of light staus for the light Higgs case (left) and heavy Higgs case (right).} 
\label{fig:h_mstau}
\end{figure} 

While the loop-corrected coupling $g_{hb\bar{b}}$, as given in
\refeq{eq:hbbdeltamb}, approaches the SM coupling in the
decoupling limit ($\MA \gg \MZ$), a suppression of $g_{hb\bar{b}}$ is 
possible for $\MA$ not too large provided that $\Delta_b$ is numerically
sizable and positive.   
We analyze this in \figref{fig:h_Deltab}. The left plot in this figure
shows that the most favoured regions are obtained for 
$\mu>1 \tev$, and that the combination of small $\mu$ and
$\MA<500\gev$ is disfavoured. The corresponding $\db$ values  are shown
in the right plot as a function of 
$\mu\tb$. The most favoured regions here are found
in the range $0.3 \lsim \db \lsim 1.5$, for correspondingly large values
of $\mu\tb\sim 30$--$70\tev$. Note that the large values for the $\db$
corrections do not pose problems with perturbativity, since they tend to
reduce the bottom Yukawa coupling. 
It should be noted that the $\Delta_b$ corrections in
\refeq{eq:hbbdeltamb} have another important effect: while in the
absence of those contributions a small value of $\aeff$ would give rise
to a simultaneous suppression of the Higgs couplings to $b \bar b$ and
to $\tau^+\tau^-$, the $\Delta_b$ corrections differ from the
corresponding contributions to the $g_{h\tau^+\tau^-}$ coupling. This
implies in particular that the $g_{h\tau^+\tau^-}$ coupling may be 
suppressed while the $g_{hb\bar{b}}$ coupling remains unsuppressed 
(and vice versa), see the discussion of 
\reffi{fig:hrates_corr} above.

\begin{figure}
\centering
\includegraphics[width=0.45\columnwidth]{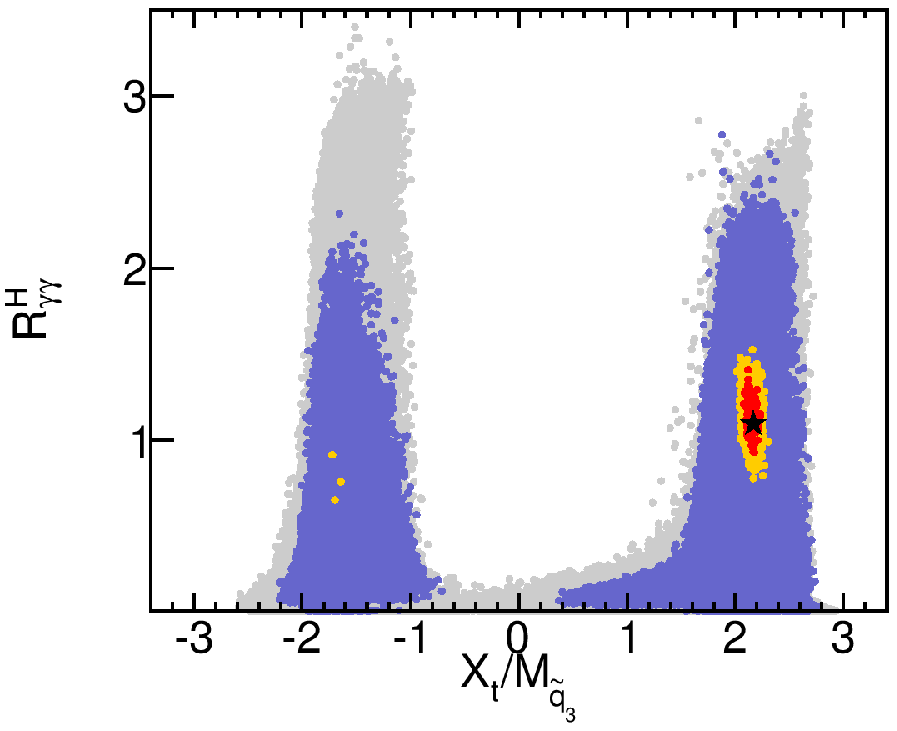}
\includegraphics[width=0.45\columnwidth]{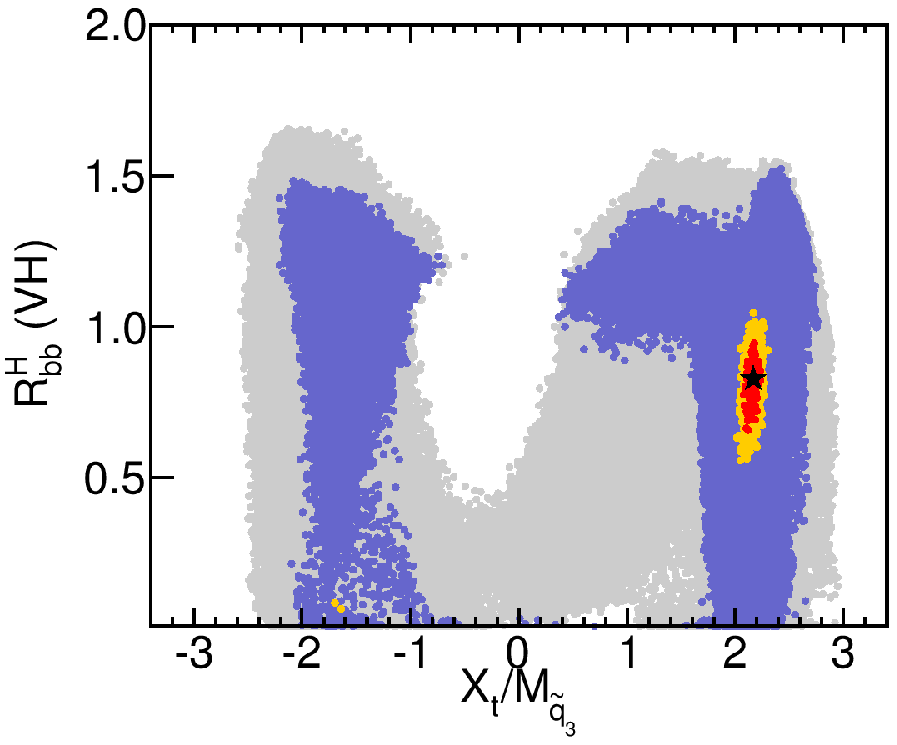}
\caption{Dependence of the rates $R^H_{\ga\ga}$ and $R^H_{bb}$ (VH)
on the stop mixing parameter $X_t/\msqd$ for the heavy Higgs case.} 
\label{fig:HH_Xt_Rgaga}
\end{figure} 
\begin{figure}[h!]
\centering
\includegraphics[width=0.45\columnwidth]{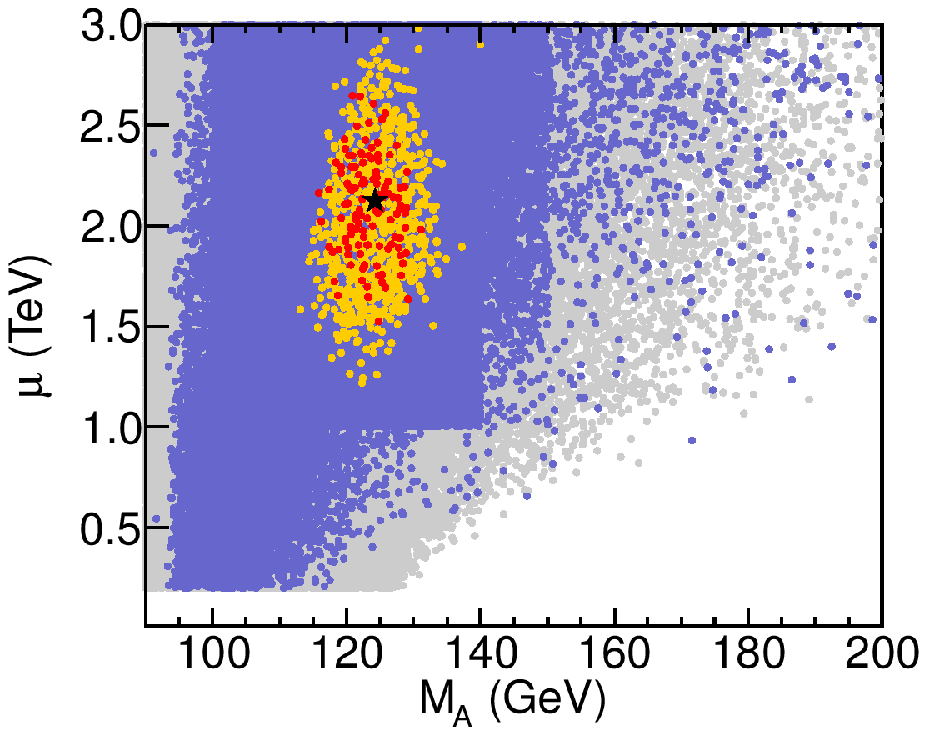}
\includegraphics[width=0.45\columnwidth]{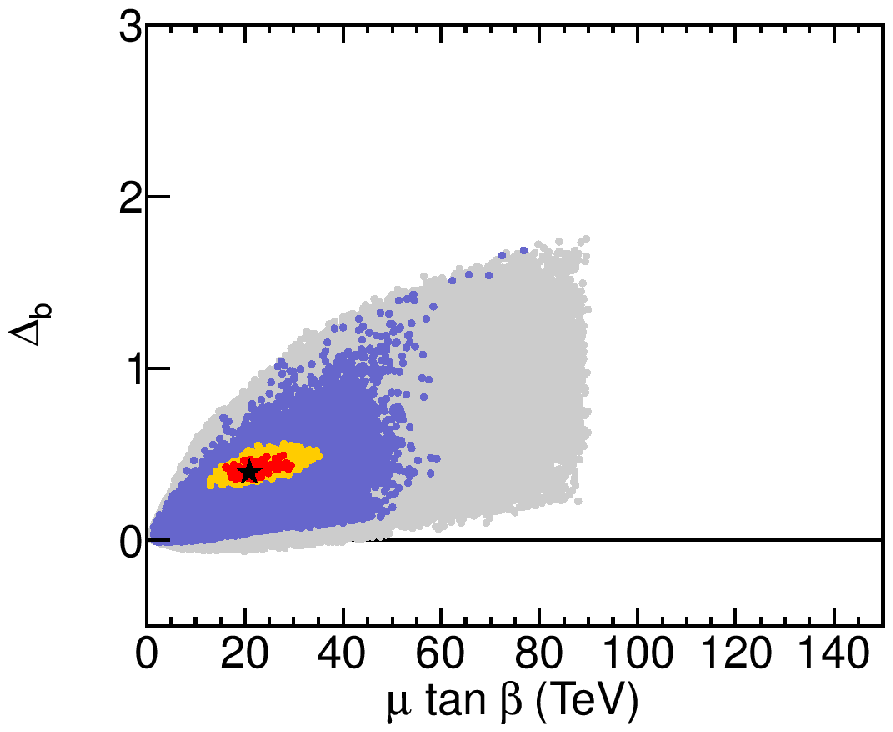}
\caption{Correlation of the $\mu$ parameter to the value of $\MA$ (left), and 
dependence of $\db$ corrections on $\mu\tb$ (right), both in the heavy
Higgs case.}  
\label{fig:HH_Deltab}
\end{figure} 

For the second mechanism, direct enhancement of the $\Ga(h \to \ga\ga)$
width, it is known that other SUSY particles can play an important
role. One possibility that has been discussed recently is to have very
light scalar taus~\cite{Carena:2012gp,*Carena:2011aa}. 
The effect of light scalar
taus can also be observed in our analysis, as can be seen in 
\figref{fig:h_mstau}. Here we show 
$\Ga(h,H \to \ga\ga)/\Ga(h,H \to \ga\ga)_{\SM}$ as a function of
$\mstaue$. In the light Higgs case, shown in the left plot, 
for $\mstaue\sim 100\gev$ the enhancement over the SM width
reaches $50\%$ in the favoured region. 
Even lower values of $\mstaue$ (which are allowed regarding the limits
from direct searches, see \cite{Beringer:1900zz}) are forbidden in our scan
from the requirement that the LSP is the lightest neutralino,
together with the lower limit of $M_2 \ge 200 \gev$ and the GUT
relation between $M_1$ and $M_2$. Relaxing
these assumptions would allow for a larger enhancement of $\Ga(h \to
\ga\ga)/\Ga(h \to \ga\ga)_{\SM}$, as is clear from the sharp rise of
this rate seen in \figref{fig:h_mstau} for low $\mstaue$. 
For $\mstaue \gsim 300 \gev$ a decoupling to the SM rate is
observed. Through the contributions of light scalar taus it is thus
possible to accommodate enhanced values of $R_{\ga\ga}^{h}$, while
maintaining $R_{bb}^{h}$ and $R_{VV}^{h}$ at the SM level (as also
observed in our results, cf.~\figref{fig:hrates_corr}). 
In the heavy Higgs case, on the other hand, as shown in the right
plot of \reffi{fig:h_mstau}, the favoured region is
located close to one, and light staus do not contribute to a possible
enhancement of $R_{\gamma\gamma}^H$.

Similarly to the light Higgs case, we investigate the dependence of the
rates on the stop 
sector parameters for the heavy Higgs case. The results are shown in
\reffi{fig:HH_Xt_Rgaga}. As in \reffi{fig:HH_mstop}, the favoured
regions are given for large and positive $\Xt/\msqd$, where we find 
$0.8 \lsim R_{\ga\ga}^H \lsim 1.6$ and a corresponding suppression
of $0.6 \lsim R_{bb}^H \lsim 1.0$. The $\db$
corrections, that enter analoguously to \refeq{eq:hbbdeltamb}, can also
in this case be largely responsible for the 
suppression of the $R_{b \bar b}^H$ rate, as we show in \reffi{fig:HH_Deltab}.
Here one can see that in the heavy Higgs scenario only values of
$\db$ between $\sim 0.3$ and $\sim 0.6$ are favoured, which are
realised for  
$10 \tev \lsim \mu\tb \lsim 35 \tev$, i.e.\ smaller values than in the
light Higgs case. 

To summarize the discussion on favoured MSSM parameter regions, we
list
in \refta{tab:fitpara} the parameter values for the best fit points (for
the complete fit). We also give the parameter ranges corresponding to
$\Delta \chi_{h,H}^2 < 1$. For several of the parameters this range
extends to the 
limits of our scanned interval. Cases like this have been indicated in
\refta{tab:fitpara} with parentheses around the corresponding
numbers. One can see that in most cases the ranges, even evaluated for
$\De\chi^2_{h,H} < 1$, are quite wide. One exception is $\tb$, which is 
relatively tightly constrained (at least at the level of
$\De\chi^2_{h,H} < 1$) in the light Higgs case, and even more so in the
heavy Higgs case. In the latter case, as discussed above, also the
masses of the additional Higgs bosons are relatively tightly
constrained, offering some important information on how this scenario
can be further explored (see the next section). 
More precise experimental data would be
needed to achieve tighter constraints on the other fitted
parameters, which enter the MSSM Higgs phenomenology via loop
corrections. The fact that even in the more ``exotic'' scenario, where
the signal at $\sim \MHexp\gev$ is interpreted in terms of the heavier
$\cp$-even Higgs of the MSSM, the values of individual SUSY parameters
are only moderately constrained by the fit illustrates that a reasonably
good description of the data can be achieved without the need of tuning
certain parameters to specific values. This is of course even more the
case for the interpretation in terms of the light $\cp$-even Higgs.

\begin{table} 
\centering
\begin{tabular}{c|ccc|ccc}
\hline
 & \multicolumn{3}{|c|}{Light Higgs case} 
 & \multicolumn{3}{|c}{Heavy Higgs case}  \\
 Parameter & & Best fit & & & Best fit & \\
\hline
$\MA$ [GeV]     & 300   & 669  & 860    & 120.5 & 124.2 & 128.0 \\
$\tb$           & 15    & 16.5 & 26     & 9.7   & 9.8   & 10.8 \\
$\mu$ [GeV]     & 1900  & 2640 & (3000) & 1899  & 2120  & 2350 \\
$\msqd$ [GeV]   & 450   & 1100 & (1500) & 580   & 670   & 740 \\
$\msld$ [GeV]   & 250   & 285  & (1500) & (200) & 323   & (1500) \\
$\Af$ [GeV]     & 1100  & 2569 & 3600   & 1450 & 1668  & 1840 \\
$M_2$ [GeV]     & (200) & 201  & 450    & (200) & 304   &  370 \\
\hline
$\Mh$ [GeV]     & 122.2 & 126.1 & 127.1 & 63.0  & 65.3  & 72.0  \\
$\MH$ [GeV]     & 280   & 665   & 860   & 123.9 & 125.8 & 126.4 \\
$\MHp$ [GeV]    & 310   & 673   & 860   & 136.5 & 138.8 & 141.5 \\
\hline
\end{tabular}
\caption{Best fit parameter values (in the respective middle column) and
ranges for $\Delta\chi_{h,H}^2<1$. Values in parentheses indicate that the
limit of the scan range has been reached.}  
\label{tab:fitpara}
\end{table}


\section{Conclusions and Outlook}

We have analyzed the compatibility of the Minimal Supersymmetric
Standard Model (MSSM) with the recent discovery at the LHC of a
Higgs-like state at $\MHhat\sim \MHexp \gev$. To this end we have studied
the low-energy (phenomenological) \pMSSM\ parameter space, where 
we allowed the seven parameters most relevant for Higgs and flavour
phenomenology to vary freely:  the 
$\cp$-odd Higgs boson mass, $\MA$, the ratio of the two vacuum
expectation values, $\tb$, a common soft SUSY-breaking parameter for the
scalar top- and bottom quarks, $\msqd$, a soft SUSY-breaking parameter
for the scalar tau and neutrino sector, $\msld$, a common trilinear
coupling for the third generation, $\Af$, the 
higgsino mass parameter, $\mu$, as well as the $\rm{SU}(2)$ gaugino mass
parameter, $M_2$. The $\rm{U}(1)$ gaugino mass parameter $M_1$ was fixed from
the value of $M_2$ using the GUT relation.  
The other parameters have been set to fixed values as to be 
generically in agreement with recent SUSY searches at the LHC and with 
low-energy observables such as $(g-2)_\mu$. 

A random parameter scan over the seven free parameters with
$\mathcal{O}(10^7)$ scan points has been performed. For each scan point,
a $\chi^2$ function was evaluated, taking into account the measured
rates in $37$ individual Higgs search channels from ATLAS, CMS, and the
Tevatron, the best-fit mass values of the LHC experiments, as well as the
following low-energy observables: 
$\br(B \to X_s \ga)$, $\br(B_s \to \mu^+\mu^-)$, 
$\br(B_u \to \tau \nu_\tau)$, $(g - 2)_\mu$ and $\MW$.

As a starting point we find that the SM yields a good fit to the data,
with a $\chi^2$ per degree of freedom (dof) around unity. The precise
value depends on whether low-energy observables and/or the Tevatron 
data are included in the fit. Turning to the MSSM, we find that the
\pMSSM\ provides an excellent fit to 
the Higgs data in the case that the light $\cp$-even Higgs 
is interpreted as the new state at $\sim \MHexp \gev$. In the case that
the heavy $\cp$-even Higgs boson is interpreted as the newly discovered state
the fit is still acceptable, but somewhat worse than in the light Higgs
case once Tevatron and low-energy data is included. 
The two MSSM best-fit points have a total $\chi^2/$dof of $30.4/36$
($42.4/36$) for the light (heavy) Higgs case, respectively, after the
inclusion of LHC, Tevatron and low-energy data.
This translates into $p$-values of $73\%$ and $21\%$, respectively. 
The corresponding SM value for $\chi^2/\nu$ is $45.3/42$, resulting
in a fit probability of $34\%$.
The largest $\chi^2$ contribution in the SM comes from the inclusion of
$(g-2)_\mu$, which shows a more than $3\,\si$ deviation from the SM
prediction. Regarding the comparison of 
the results for the light Higgs case and the heavy Higgs case in the MSSM 
it should be noted that a sizable part of the additional $\chi^2$
contribution in the heavy Higgs case 
results from the \btn\ measurement and from $(g-2)_\mu$. The agreement
between theory and experiment (both for the MSSM and the SM) for \btn\
would improve with the inclusion of the new Belle measurement.
The $\chi^2$ contribution arising from $(g-2)_\mu$ for the heavy Higgs
case of the MSSM could potentially be improved if in addition to the
seven parameters that are varied in our fit also the second generation
slepton parameters would be treated as free fit parameters, which would
essentially select the slepton mass parameters yielding the lowest 
$\chi^2$ value from $(g-2)_\mu$ for each point in parameter space
without affecting the other phenomenology.
Thus, while the best description of the data is achieved if the new
state at $\sim \MHexp \gev$ is interpreted as the light $\cp$-even Higgs
boson of the MSSM, the more ``exotic'' interpretation in terms of the
heavier $\cp$-even Higgs of the MSSM is also permitted by the data, even
if the results from the Higgs searches at the LHC are supplemented with 
results from the Tevatron Higgs searches and with results from flavour
physics and electroweak precision data. The latter interpretation would
imply that also the other four Higgs bosons of the MSSM would be rather
light, giving rise to exciting prospects for the searches for non
SM-like Higgses.

In the case of the light $\cp$-even Higgs with $\Mh \sim \MHexp \gev$, we
find for the best-fit point in the full fit an enhancement of production
times branching ratio for the 
$\ga\ga$ channels of about $40\%$ with respect to the SM prediction. 
The rates for the gauge boson channels that we obtain are similar
as in the SM, and the same holds for the fermionic
channels ($b \bar b$ and $\tau^+\tau^-$).
While the fit results for the $\ga\ga$, $VV$, and $b \bar b$ rates show a
clear $\chi^2$ minimum, the $\tau^+\tau^-$ channel has a very broad 
distribution close to the minimum, and no strong preference
can be attributed to the actual best-fit value.

In the case of the heavy $\cp$-even Higgs with $\MH \sim \MHexp \gev$ we
find for the best fit point a somewhat smaller enhancement of the
$\ga\ga$ channel, an enhancement of the gauge boson channels,
and a suppression of the $\tau^+\tau^-$ (VBF) and $b \bar b$
channels, whereas the $\tau^+\tau^-$ inclusive channel is enhanced
due to the contribution of the \cp-odd Higgs boson.

For the light Higgs case, as well as for the heavy Higgs case, the rates
in the $\ga\ga$ and $VV$ channels are strongly correlated,
however in most cases with the possibility of a stronger enhancement (or
smaller suppression) in the $\ga\ga$ channel. Between the $\ga\ga$ channel and
the $b \bar b$ channel an anti-correlation can be observed. This shows
that the fit within the MSSM favours at least over a part of the
preferred region a scenario where a $\ga\ga$ enhancement is
caused by a
suppression of the $b \bar b$ channel. 
Our results furthermore show that the $\tau^+\tau^-$ channel can be
strongly suppressed, while the $b \bar b$ channel can remain close to
the SM strength.
Since we find that the $gg\to h,H$ production channels in
the favoured regions are not substantially enhanced, the observed 
enhancement of the $\ga\ga$ channel is not caused by larger
production cross sections. A suppression of the $b \bar b$ channel in
the light and the heavy Higgs case can be caused by a large value of
$\db$, which can reach values exceeding unity in the favoured
region.
In the light Higgs case, the $\ga\ga$ channel can also be enhanced by the
contribution of light scalar taus to the decay process. In the 
case where the lightest scalar tau mass is as low as about
$100\gev$, we find an enhancement of up to $50\%$ from this mechanism.

For the scalar top masses, we find that the favoured regions start at
$\mste \sim 200 \gev$ and $\mstz \sim 600 \gev$ in the light Higgs
case. They extend up to $\sim 1.4 \tev$ and $\sim 1.6 \tev$,
respectively, which are the maximal values accessible in our scan. The
mixing in the scalar top 
sector must exceed $|\Xt/\msqd| \sim 1$, where the most favoured regions
have $\Xt/\msqd = 2 - 2.5$. Similar values for the lower
bounds on the scalar top masses
are found in the heavy Higgs case. However, for this case we find that
the favoured regions are also bounded from above by (roughly) 
$\mstz \lesssim 1 \tev$. 

As is evident from our analysis (as demonstrated e.g.~by
Figs.~\ref{fig:hbestfit} and 
\ref{fig:HHbestfit}), the fitted rates in the MSSM interpretations are
not significantly different from the SM predictions, using
the current experimental and theoretical uncertainties. Therefore,  
if no other new states beyond the current Higgs candidate are discovered
in the near future, the question arises how much the precision of the
current measurements would need to be improved in order to distinguish
the MSSM from the SM, based on precisely measured rates alone.  
In order to obtain a rough estimate for the answer to this
question, we set the hypothetical future central values of
the measurements to the MSSM best fit point in the light Higgs
case, so that we can investigate the impact of prospective future
experimental results in a scenario where this particular 
MSSM point is actually
realized in nature. Then, we scale the full uncertainties of the signal
rates measured in the LHC8 
(i.e., the channels measured at the LHC with $\sqrt{s} = 8 \tev$)
channels by a global scale factor 
and infer the value of the scale factor
 at which the
deviation of the SM prediction from the assumed future 
measurements with higher precision reaches a
significant level. We scale only the LHC8 results (and not LHC7,
Tevatron, or LEO) to account for those channels for which an actual
improvement of precision is expected in the future (at $\sqrt{s}=8$\,TeV
or higher). To reach $2\,\si$ ($3\,\si$) significance for
rejecting the SM, we find that a scale factor of $21\,\%$
($18\,\%$) is required. Reducing the uncertainties to 
$\sim 20\,\%$ of their current values for all channels (i.e., a
factor five improvement compared to the present situation) appears to be 
reasonable as a long-term goal for the
LHC~\cite{HratesATLAS-HL-LHC,*HratesCMS-HL-LHC}. For the 
necessary improvement of the statistical, systematical and theoretical
precision to go significantly beyond this number, as needed in this
case to go towards $5\,\si$ sensitivity for SM exclusion, would require
something close to the expected experimental sensitivity of the
International Linear Collider (ILC), see \cite{Brau:2012hv} and
references therein.  
The corresponding numbers in the heavy Higgs case are $31\%$ ($26\%$), 
which would be needed to reach a $2\,\si$ ($3\,\si$) rejection of the SM via
Higgs rate measurements alone.

While distinguishing the fitted MSSM interpretations from the SM by
virtue of improving the precision of the rate measurements alone
may be
difficult, promising paths to establishing the presence of
physics beyond the SM
in the Higgs sector are on the one hand the search for
additional (non SM-like) Higgs states and on the other hand
the precise measurement of further Higgs properties. Concerning the 
latter, a
promising example would be to measure the Higgs $\cp$ properties in the
$\tau^+\tau^-$ final state (assuming that the existence of this decay mode
will be confirmed for the observed signal). In the heavy Higgs case, as
mentioned in \refse{sect:directHiggs}, the contributions from~$H$
and~$A$ add up in channels with poor  
mass resolution when no $g_{HVV}$ coupling is involved. 
This is the case for the inclusive production mode with decays
into $\tau^+\tau^-$. There, an analysis of the $\tau$ spin
correlations~\cite{Berge:2008wi,*Berge:2011ij} (and references therein)
could show significant deviations from the SM prediction of $\cp = +1$,
due to the presence of the $A$ production mode. In contrast, in
the light Higgs case, the discovered state at $\MHhat\sim \MHexp\gev$ would
be measured with exact $\cp = +1$.

The potential discovery of additional Higgs like states would be the
clearest way to distinguish the MSSM $h$ and $H$ interpretations
from each other. As discussed above, see in particular
Tab.~\ref{tab:fitpara},
the masses of the yet undiscovered Higgs bosons are expected
in different ranges for the $h$ and $H$ interpretations.
However, discovering the lighter $\cp$-even state $h$ in the 
heavy Higgs case appears to be difficult at the LHC due to its low mass
and heavily suppressed coupling to vector bosons, $g_{hVV}^2$. 
At the ILC, however, the $hA$ and $H^+H^-$
production modes would be straightforward to measure in the heavy Higgs
case. On the other hand, in the light Higgs case all other states could
be beyond the mass reach within the current fit uncertainties. Here, the
ILC measurements would have to rely on precise coupling measurements to
distinguish the $h$ and 
$H$ interpretations directly~(see \cite{Brau:2012hv} and references therein).
In that case also a combined interpretation of LHC and ILC data might
be very valuable~\cite{Weiglein:2004hn,Desch:2004cu}.

In the heavy Higgs interpretation, the LHC searches for MSSM Higgs
bosons in the $\tau^+\tau^-$ final states will soon have sensitivity to
start probing the region of small $\MA$ and moderate $\tb$ that is favoured in
this scenario. Furthermore we find that the charged Higgs
boson in this case
should be lighter than $m_t$, so that at the production from
top quark decays at the LHC would be kinematically possible.
The favoured values of $\br{(t \to H^+b)}$ are just below the current
experimental bounds. Still, an improvement of the limits by one order of
magnitude would be required to fully cover this possibility.  

New data from the ATLAS and CMS Higgs boson searches is rapidly
emerging. It will be particularly important to investigate on the
one hand potential
deviations of the rates from the SM predictions and on the other hand
the outcome of searches for additional non SM-like Higgses.
Confronting these results with predictions in the MSSM
will show whether this model, whose unambigous prediction of 
a light (and potentially SM-like) Higgs boson seems to be well supported
by the data,
will continue to provide a viable description of nature also in the
future.


\section*{Acknowledgments}
This work has been supported by the Collaborative Research Center SFB676
of the DFG, ``Particles, Strings, and the Early Universe''. It has also been 
partially funded by the Helmholtz Alliance ``Physics at the Terascale''.
The work of S.H.\ was partially supported by CICYT (grant FPA
2010--22163-C02-01) and by the Spanish MICINN's Consolider-Ingenio 2010
Programme under grant MultiDark CSD2009-00064. The research of O.S.\ is
supported by the Swedish Research Council (VR) through the Oskar Klein Centre.


\bibliography{hifi}
\bibliographystyle{JHEP}

\end{document}